\documentclass[prb,twocolumn,showpacs]{revtex4-1}

\usepackage{amsmath}
\usepackage{enumerate}
\usepackage{amssymb}
\usepackage{bm}
\usepackage{graphicx}
\usepackage{wrapfig}
\usepackage{setspace}
\usepackage[hypertex]{hyperref}
\usepackage{color}
\usepackage{cancel}
\usepackage{verbatim}

\newcommand{\SLE}[1]{SLE$\,_{#1}$}
\newcommand{\SLEr}[2]{SLE(${#1},{#2}$)}
\renewcommand{\>}{\rangle}
\newcommand{\<}{\langle}

\renewcommand \L{\mathcal{L}}

\newcommand{\al}{\alpha}

\newcommand{\de}{\delta}

\DeclareMathOperator{\re}{Re}
\DeclareMathOperator{\im}{Im}

\begin{document}

\title{Quantum Hall transitions: An exact theory based on conformal restriction}

\author{E. Bettelheim}
\affiliation{Department of Physics, Hebrew University, Jerusalem, Israel}
\author{I. A. Gruzberg}
\affiliation{The James Franck Institute and the Department of Physics,
The University of Chicago, Chicago, IL 60637, USA}
\author{A. W. W. Ludwig}
\affiliation{Department of Physics, University of California, Santa Barbara}

\date{October 25, 2012}

\begin{abstract}

We revisit the problem of the plateau transition in the integer quantum Hall effect. Here we develop an analytical approach for this transition, and for other two-dimensional disordered systems, based on the theory of ``{\it conformal restriction}''. This is a mathematical theory that was recently developed within the context of the {\it Schramm-Loewner evolution} which describes the ``stochastic geometry'' of fractal curves and other stochastic geometrical fractal objects in two-dimensional space. Observables elucidating the connection with the plateau transition include the so-called {\it point-contact conductances} (PCCs) between points on the boundary of the sample, described within the language of the Chalker-Coddington network model for the transition. We show that the disorder-averaged PCCs are characterized by  a {\it classical} probability distribution for certain geometric objects in the plane (which we call {\it pictures}), occurring with positive statistical weights, that satisfy the crucial so-called restriction property with respect to changes in the shape of the sample with {\it absorbing boundaries}; physically, these are boundaries connected to {\it ideal leads}. At the transition point, these geometrical objects ({\it pictures}) become fractals. Upon combining this restriction property with the expected conformal invariance at the transition point, we employ the mathematical theory of ``conformal restriction measures'' to relate the disorder-averaged PCCs to correlation functions of (Virasoro) primary operators in a conformal field theory (of central charge $c=0$). We show how this can be used to calculate these functions in a number of geometries with various boundary conditions. Since our results employ only the conformal restriction property, they are equally applicable to a number of other critical disordered electronic systems in two spatial dimensions, including for example the spin quantum Hall effect, the thermal metal phase in symmetry class D, and classical diffusion in two dimensions in a perpendicular magnetic field. For most of these systems, we also predict exact values of critical exponents related to the spatial behavior of various disorder-averaged PCCs.

\end{abstract}


\maketitle

\section{Introduction}

Effects of static, randomly placed impurities (disorder) are central to our understanding of transport properties of electronic solids. Indeed, building on Anderson's seminal work, \cite{Anderson58} an immense amount of research activity has emerged over the past few decades on models of disordered electronic solids, in particular of non-interacting electronic systems, the subject now generically known as the Anderson localization. \cite{AL50} About a decade and a half ago, the field of Anderson localization has received a tremendous boost through the work of Zirnbauer, \cite{Zirnbauer96} and Altland and Zirnbauer \cite{Altland97}  (AZ) which provided a very general classification scheme of the behavior of non-interacting fermions subject to static disorder potentials. Their work showed that universal behavior emerging on length scales much longer than the mean free path must be in one of only 10 possible symmetry classes, which depend solely on the behavior of the Hamiltonian under generic symmetries (time-reversal, particle-hole, chiral). \cite{ReviewSymmetryClasses} These 10  symmetry classes are in one-to-one correspondence with the 10 types of symmetric (constant curvature) Riemannian spaces in the classification scheme of the mathematician Cartan.

Electronic disordered systems exhibit, in a variety of symmetry classes and spatial dimensions, second order quantum phase transitions between insulating and conducting phases, which are examples of Anderson (localization) transitions. (For a recent review of Anderson transitions, see e.g. Ref. \onlinecite{Evers08}).  Other examples of Anderson (localization) transitions are quantum Hall plateau transitions between insulating phases with different topological order and different quantized values of a Hall conductance. A famous example is the integer quantum Hall (IQH) plateau transition observed in two-dimensional (2D) semiconductor devices subject to strong magnetic fields. The nature of the critical state at and the critical phenomena near the IQH transition are at the focus of intense experimental \cite{Amsterdam-group, Tsui-group, Amado10, Saeed11, Huang12, Shen12} and theoretical research. \cite{Zirnbauer99, tsvelik, LeClair, Pruisken, Obuse08b, Evers08b, Slevin09, Burmistrov10, Amado11, stabilitymap}

In spite of much effort over several decades, an analytical treatment of most of the critical conducting states in disordered electronic systems, including in particular that of the mentioned IQH transition, has been elusive (although some proposals \cite{Zirnbauer99, tsvelik, LeClair} have been put forward, but see Refs.  \onlinecite{Obuse08b} and \onlinecite{Evers08b}). A notable exception is the so-called spin quantum Hall (SQH) plateau transition, \cite{Kagalovsky99, Senthil1999b} which is similar to the IQH transition, but in a different symmetry class (class C in the AZ classification). In this case an exact mapping to the classical problem of bond percolation  is available. \cite{Gruzberg99} Through (variants of) this mapping, exact expressions for various disorder-averaged observables and critical exponents for the SQH transition were obtained. \cite{Gruzberg99, Cardy00, Beamond02, Mirlin03, Subramaniam08, Bondesan11}

The universal (critical) properties of Anderson transitions can be formulated in terms of so-called network models. The prime example of a network model is the celebrated Chalker-Coddington network model \cite{Chalker88} describing IQH transition. A similar network for the SQH transition \cite{Kagalovsky99} was the starting point for the mapping to percolation \cite{Gruzberg99} mentioned above. While a network model formulation exists for systems in all 10  (AZ) symmetry classes, a particularly rich behavior is seen in symmetry class D in two dimensions, \cite{Chalker01, Gruzberg01, ReadLudwig00, Mildenberger07} comprising, for example, a fermionic representation of the two-dimensional short-range Ising spin glass (Ising exchange couplings with random signs). The phase diagram of a generic network model in symmetry class D contains three phases: an insulator, the so-called thermal quantum Hall state, and a metal with continuously varying (thermal) Hall conductivity.

Conventional critical statistical mechanics models are known to possess conformal invariance. \cite{Polyakov:1970} Implications of this invariance are most powerful in two dimensions, where it allows us to apply methods of 2D conformal field theory (in short, CFT) \cite{BPZ, YellowBook} to study critical phenomena in such models. As in any field theory description, basic objects of study in CFT are correlation functions of local observables. One of the more important characteristics in any CFT is the so-called central charge $c$. This parameter is related to the way a critical system responds to changes in its geometry.

It is widely believed that Anderson transitions in two dimensions also possess conformal symmetry (and there is numerical evidence to support this belief in certain cases \cite{Obuse07, Obuse08, Obuse10}). However, in this case one is usually interested in correlation functions (density of states, conductivities, etc.) averaged over all disorder realizations. Taking such averages is complicated since the partition function of a disordered system undergoes statistical fluctuations from one realization of disorder to another. \cite{footnoteFluctuationPartitionFct} One way to handle this difficulty is to apply the supersymmetry method where two types of fields (bosonic and fermionic) are introduced in the theory. The outcome is a theory whose partition function is unity, $Z=1$, independent of the  particular disorder realization, as well as of the shape and the size of the system. This implies the vanishing of the central charge for a CFT describing an Anderson transition in two dimensions (see e.g. Ref. \onlinecite{GurarieLudwig2005Review} for a recent review).

Recently, another approach to the study of two-dimensional critical systems has appeared. The approach uses methods of probability theory and conformal maps, and can be called the ``stochastic conformal geometry approach''. The focus of stochastic geometry is to directly describe randomly fluctuating geometric objects in scale-invariant (i.e.,, critical) systems: regions in space of fractal dimension (often referred to as ``clusters'') and their boundaries (often referred to as ``cluster boundaries'') which form fractal curves. In a seminal paper, \cite{Schramm1999} Oded Schramm has introduced a one-parameter family of random processes, since then called the Schramm-Loewner evolutions (SLE), which describe growth processes of random fractal conformally invariant curves. Conformal invariance in this case is understood precisely as a statement about probability measures on curves. Since their original discovery, the SLE processes have been studied in depth, have been related to traditional CFT, and generalized in several ways. Many reviews of this beautiful theory exist by now, and we recommend Refs. \onlinecite{Werner-review, Lawler-book, Kager-Nienhuis-review, Cardy-review, BB-review, IAG-review, RBGW07} for more details.

The curves described by SLE are unique candidates for (scaling limits of) cluster boundaries in 2D critical statistical mechanics systems. A one-parameter family of SLE processes conventionally denoted by SLE$_\kappa$, which was discovered by Schramm, fully exhausts all possible ensembles of SLE curves. The real parameter $\kappa$ of the SLE$_\kappa$ family is related to the central charge of the CFT describing the critical system by
\begin{align}
c = \frac{(3\kappa - 8)(6 - \kappa)}{2\kappa}.
\end{align}
Since we are interested in theories with $c=0$, the values $\kappa = 8/3$ and $\kappa = 6$ play a special role for us. The SLE$_{8/3}$ process describes the scaling limit of 2D self-avoiding random walks (SAW) or polymers, and the SLE$_6$ process describes the percolation hulls. These two types of critical curves possess special properties called locality (for $\kappa = 6$) and restriction (for $\kappa = 8/3$). It is the restriction property that is intimately related to CFTs with $c=0$. It turns out that the notion of `conformal restriction' (i.e., the presence of the restriction property in a conformally invariant 2D system) can be extended to certain two-dimensional sets (`clusters'). In fact, there is a one-parameter family of conformal restriction measures, supported on such sets, which are fully characterized by a real number $h$ called the {\it restriction exponent}. In terms of CFT this exponent is the scaling dimension of a certain boundary primary operator. The sets  (`clusters') that are described by conformal restriction all have boundaries (`cluster boundaries') which are fractal curves that happen to be variants of SLE$_{8/3}$, called SLE$(8/3,\rho)$. Here the parameter $\rho$ is related to the exponent $h$ mentioned above by
\begin{align}\label{hfunctionofrho}
h(\rho) &= \frac{(3\rho+10)(2+\rho)}{32}, & \rho(h) = \frac{2}{3} \sqrt{24 h + 1} - \frac{8}{3}.
\end{align}
The theory of conformal restriction, the related theory of (multiple) SLE$(\kappa, \rho)$ processes, and their connections with CFT are the subject of Refs. \onlinecite{Lawler-book} and \onlinecite{Friedrich02, Friedrich03, Friedrich04, LSW-conformal-restriction, Werner-restriction-review, BBK2005, Kytola, Graham, Dubedat-1, Dubedat-2, Dubedat-3}, and we will review relevant results later in the paper.

In this paper we propose to make use of the theory of conformal restriction to study quantum Hall transitions and other 2D disordered electronic systems. A connection between a 2D disordered electron system and the theory of conformal restriction  can be established by studying the so-called point contact conductance (PCC), that is the conductance between two infinitely narrow leads introduced and studied in Ref. \onlinecite{Janssen99}. Loosely speaking, in a given microscopic model the disorder-averaged PCC is represented as a sum of contributions from paths that the current follows between the point contacts, in the sense of the Feynman path integral (or sum) for a quantum mechanical amplitude. When the contacts are placed at the {\it boundary} of a disordered conductor, we obtain the so-called {\it boundary} PCC. As we will explicitly show, the current paths, when studied in the presence of {\it absorbing boundaries}, \cite{RefCommentAbsorbingBoundaries} satisfy the restriction property on the lattice (i.e., at the  discrete, as opposed to the continuum level). Assuming that the discrete (lattice) model has a continuum limit at its critical point, we expect the continuum analogs of the current paths to satisfy the (continuum) restriction property. Furthermore, upon  making the assumption of conformal invariance, we conclude that scaling limits of the current paths can be described by conformal restriction measures. In the following we will keep these assumptions in mind without stressing the difference between discrete and continuous settings.

The fact that the continuum limits of current paths satisfy the restriction property turns out to imply immediately that, in the language of CFT, the point contacts are points of insertions of (Virasoro) {\it primary} conformal boundary operators. The connection between current paths and restriction measures opens up the possibility to obtain analytical results for disorder-averaged PCC's at the IQH critical point with a variety of boundary conditions. In addition, we show how our results naturally apply to the SQH transition, where the current paths are percolation hulls which, in the continuum limit, are known to be described by SLE and are rigorously known to satisfy the restriction property.

Problems of classical diffusion and transport in two dimensions in a strong perpendicular magnetic field \cite{RG1980, ML1993, KY1994, XRS1997} also admit a description in terms of conformal restriction. This is especially clear in the classical limit of the Chalker-Coddington model considered in Ref. \onlinecite{XRS1997}, where it was shown that in the continuum limit conductances of various kinds can be obtained by solving Laplace's equation with tilted (oblique) boundary conditions. The tilt angle is the Hall angle. This setting is naturally related to Brownian motions reflected at an angle (related to the Hall angle) upon hitting a reflecting boundary. Such reflected Brownian motions in fact underlie microscopic constructions of arbitrary restriction measures. \cite{LSW-conformal-restriction} A field theory formulation of classical high-field transport was given in the form of a Gaussian model which is the linearized version of Pruisken's (replica) sigma model for the IQH effect. \cite{XRS1997}

The same Gaussian model field theory results from linearization of a {\it different} nonlinear sigma model that describes thermal transport of quasiparticles in disordered superconductors in class D in 2D.\cite{Senthil2000, Read-Green2000, Bocquet2000} The perturbative renormalization group flow in this model is towards weak coupling, and in a finite system of size $L$ one can linearize the nonlinear sigma model to obtain the Gaussian model with a coupling constant of order $(\ln L)^{-1}$. In this limit quasiparticle transport is essentially classical with thermal conductivities (divided by temperature, and in the corresponding units) growing logarithmically with length scale, $\sigma_{xx} \sim \ln L$, while $\sigma_{xy}$ is arbitrary. Thus, our results obtained from the general theory of conformal restriction apply to this system as well.

Before we proceed with a detailed derivation of our results, we briefly summarize them here. The main results that apply to all systems that we have mentioned above are as follows:

(1) Disorder-averaged PCCs within microscopic models are mapped to {\it classical} statistical mechanics problems with {\it positive, albeit in some cases nonlocal, weights}.

(2) The so-obtained weights are {\it intrinsic}, which means that they are specific to certain geometric objects, and depend only on the shape and the structure of these objects, while they are independent of the shape of the rest of the system and of the boundary conditions. These weights also satisfy the crucial {\it restriction property} with respect to deformations of {\it absorbing} boundaries. More details regarding the meaning of intrinsic weights and the significance of the boundary conditions will  become clearer in the sequel.

(3) Upon assuming conformal invariance we find that current insertions through point contacts on a boundary are (Virasoro) {\it primary} CFT operators. The dimensions of these operators are known exactly in some cases, and numerically in others. Other operators related to changes in boundary conditions are also shown to be primary. This immediately allows us to use global conformal invariance to determine disorder-averaged PCCs that reduce to two- and three-point functions of primary operators [see Eqs. (\ref{2-point-g-absorbing}), (\ref{2-point-g-mixed}), (\ref{2-point-g-reflecting}), (\ref{delta-g}), and (\ref{g-reflecting-segment})]. This also sets the stage for future work  addressing the computation  of PCCs that reduce to the more complex four-point functions.

The rest of the paper is organized as follows. In Sec. \ref{sec:c=0}, we explain the conformal restriction property. We also explain in general terms how the graphical representation of boundary PCCs in terms of Feynman paths satisfies restriction with respect to absorbing boundaries. In Sec. \ref{restriction in models}, for each of the models mentioned above (IQH, SQH, diffusion in strong magnetic field, and the metal in class D), we provide a detailed derivation of the relation between the disorder-averaged PCCs and classical weights satisfying the restriction property. Specifically, in Sec. \ref{restrictionINChalker} we will explicitly show how the construction outlined in Sec. \ref{sec:c=0} works for the disorder-averaged PCCs in the Chalker-Coddington model for the integer quantum Hall plateau transition. \cite{Chalker88} We do the same for the network model for the SQH transition\cite{Kagalovsky99} through the mapping to classical percolation\cite{Gruzberg99} in Sec. \ref{restriction in SQH}, then for the classical limit of the Chalker-Coddington (CC) model in Sec. \ref{restriction in classical CC}, and for the metal in class D in Sec. \ref{restriction in class D}. Section \ref{restriction theory} is devoted to a presentation of the theory of conformal restriction and multiple SLEs. Sec. \ref{sec:CFT} sets up some useful notation and explains the relation of conformal restriction and SLEs to CFT in the so-called Coulomb gas formalism. In Sec. \ref{sec:calculation1}, we make use of the conformal restriction theory to obtain certain information on the transport behavior of the systems of interest. We establish the functional forms of disorder-averaged PCCs in several geometries. We also compute the conformal weights (scaling dimensions) of some of the relevant primary operators. Some of these weights turn out to be {\it superuniversal} in the sense that they are fully determined by conformal restriction alone, and do not depend on the particular symmetry class of the model (see Table \ref{DimensionsTable} for a summary). In Sec. \ref{sec:conclusions} we discuss our results with the view on possible extensions and generalizations. Appendices provide some relevant background information from graph theory, and details of some calculations.

\section{Conformal restriction and models with $c=0$}
\label{sec:c=0}

We begin this section by describing the conformal restriction property. Then we explain how current paths contributing to boundary PCCs at Anderson critical points naturally satisfy this property with respect to {\it absorbing} boundaries (which, we recall, describe ideal leads attached to the boundaries).

\subsection{Conformal restriction property}
\label{subsec:conf-restriction}

\begin{figure}[t]
\centering
\includegraphics[width=0.45\columnwidth]{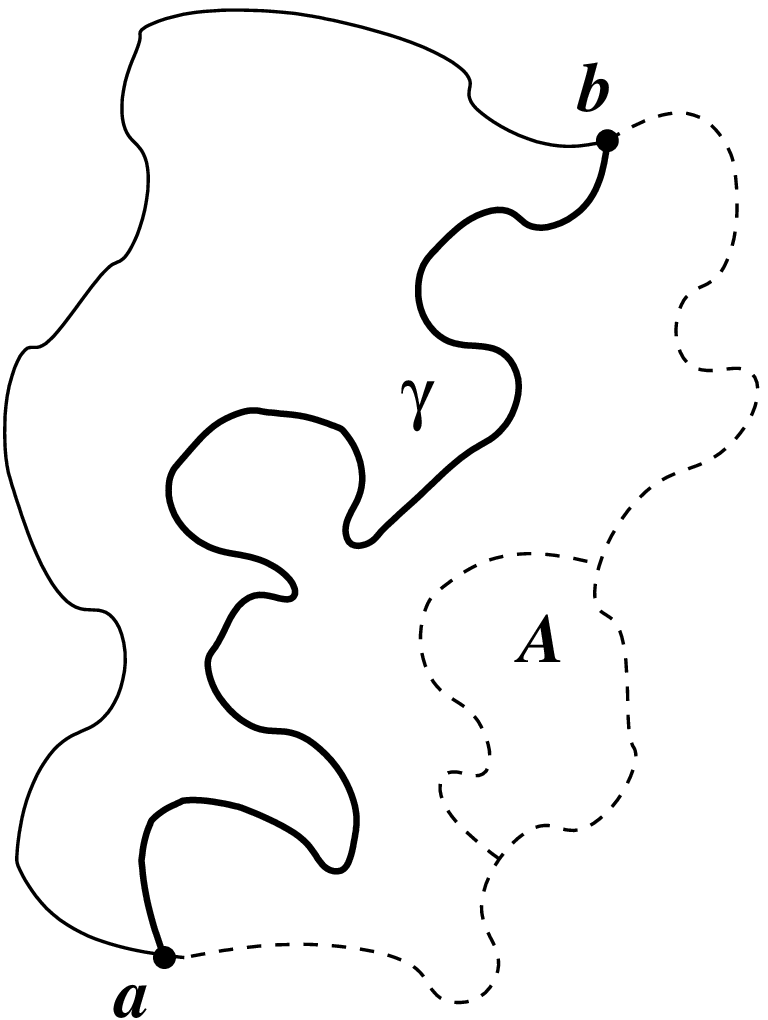}
\hfill
\includegraphics[width=0.45\columnwidth]{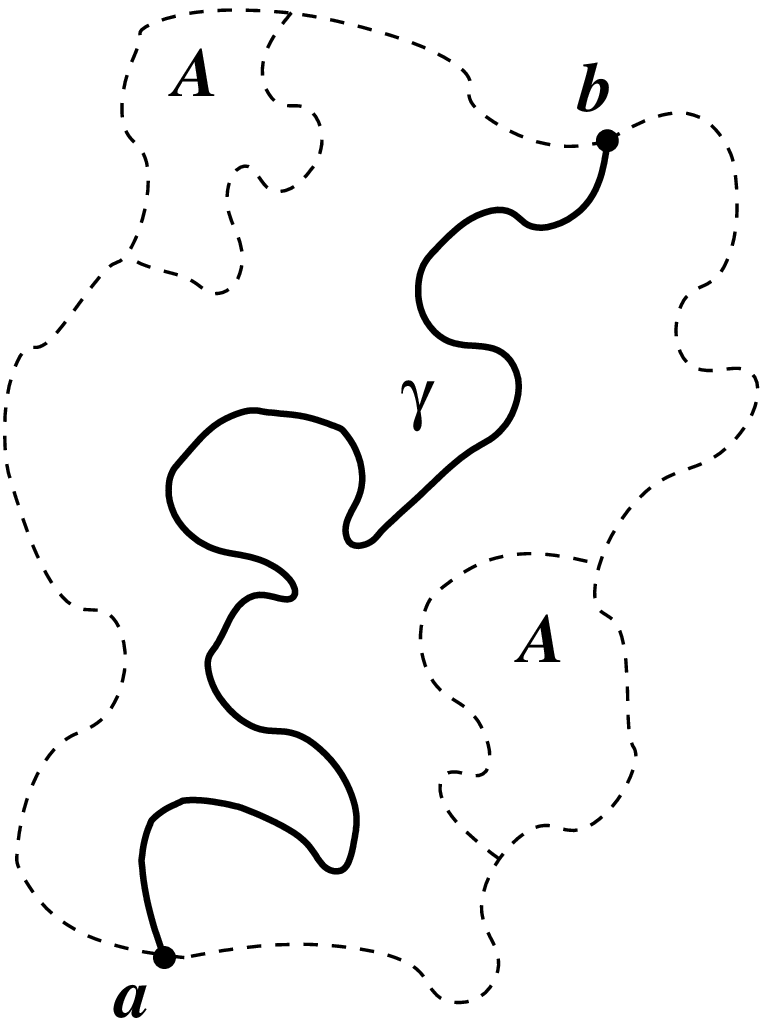}
\caption{Restriction property for curves. A curve $\gamma$ in $D$ is restricted not to enter the sets $A$. The portions of the boundary where the set $A$ can be attached are shown by dashed lines. They correspond to {\it absorbing} boundaries in the physical models we consider. Left: one-sided restriction. Right: two-sided restriction.}
\label{restriction-curves}
\end{figure}

Consider a statistical ensemble of curves defined in a simply-connected domain $D$ of the complex plane. All these curves start at a fixed point $a$ on the boundary of $D$ and end at another fixed boundary point $b$ (see Fig. \ref{restriction-curves}). The ensemble is specified by a finite measure on the curves. The measure can be normalized to be a probability measure, but it is more natural and convenient to think about un-normalized weights associated with curves in the ensemble, similar to Boltzmann weights of configurations in statistical mechanics.

Next, consider a set $A$ such that the topology of the sub-domain $D \!\setminus\! A$ is the same as that of $D$. This means that $A$ is ``attached'' to the boundary of $D$, so that $D \!\setminus\! A$ is simply-connected, and that the points $a$ and $b$ belong to the parts of the boundary that are common between $D$ and $D \!\setminus\! A$. Notice that we allow for sets $A$ that have more than one connected component.

The original ensemble of random curves can be used to define two new ensembles of curves in the sub-domain $D \!\setminus\! A$. The first one is obtained by restriction: it is the ensemble of curves in $D$ conditioned not to intersect $A$. In other words, of all the curves in the original ensemble we keep only those that do not enter $A$. To a curve $\gamma \in D \!\setminus\! A$ this definition assigns the same weight in the new ensemble that this curve has in the original ensemble. The second way to define a new ensemble in the sub-domain is to choose a conformal map $\Phi$ from $D \!\setminus\! A$ to $D$ that fixes the points $a$ and $b$ [$\Phi(a) = a$, $\Phi(b) = b$], and to any curve $\gamma \in D \!\setminus\! A$ assign the weight of its image $\Phi(\gamma)$ in the original ensemble. This is called the conformal transport of the probability measure.

Now the original ensemble is said to satisfy the conformal restriction property if both ways of defining a new ensemble in the sub-domain $D \!\setminus\! A$ lead to the same {\it  probability}  measure on curves  for {\it any} set $A$ of the type described above. Note that the equivalence is at the level of probabilities and not statistical weights.

If in this construction we use sets $A$ that can only border the boundary of $D$ on one arc from $a$ to $b$, say, the one that goes counterclockwise, then we have the so-called one-sided restriction (see the left panel in Fig. \ref{restriction-curves}). If different connected components of $A$ can be attached to either of the arcs of the boundary, we have the two-sided restriction (see the right panel in Fig. \ref{restriction-curves}). Notice that this is a stronger property since any two-sided restriction measure automatically satisfies the one-sided restriction, but the opposite is not necessarily true.

\begin{figure}[t]
\centering
\includegraphics[width=0.45\columnwidth]{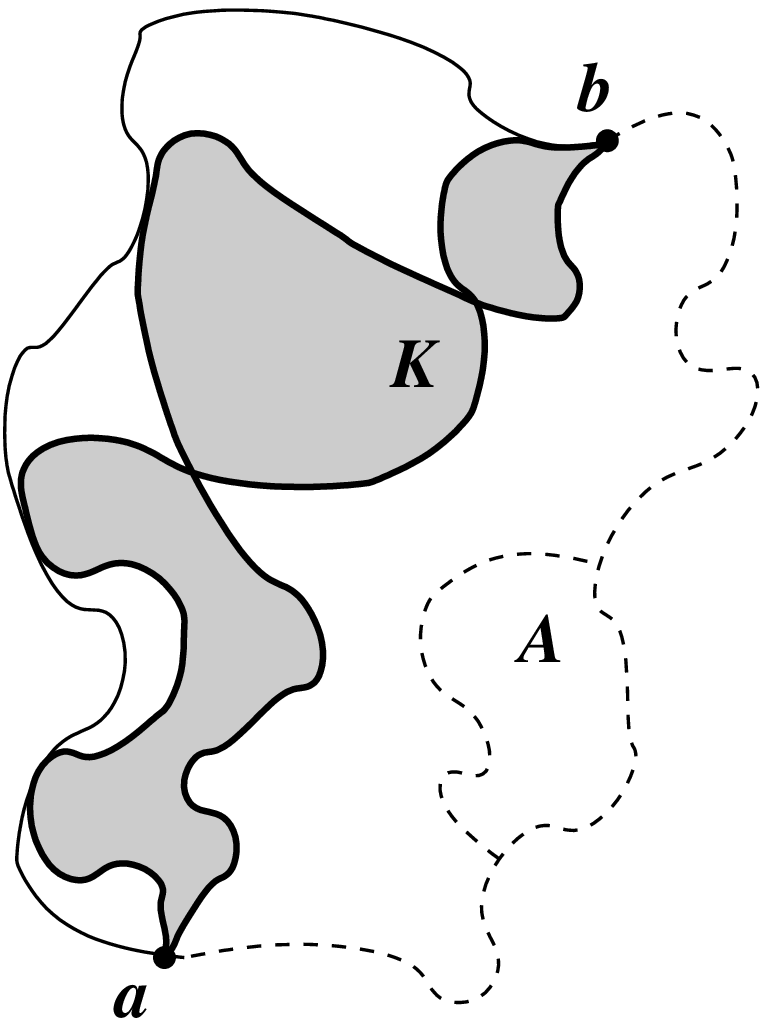}
\hfill
\includegraphics[width=0.45\columnwidth]{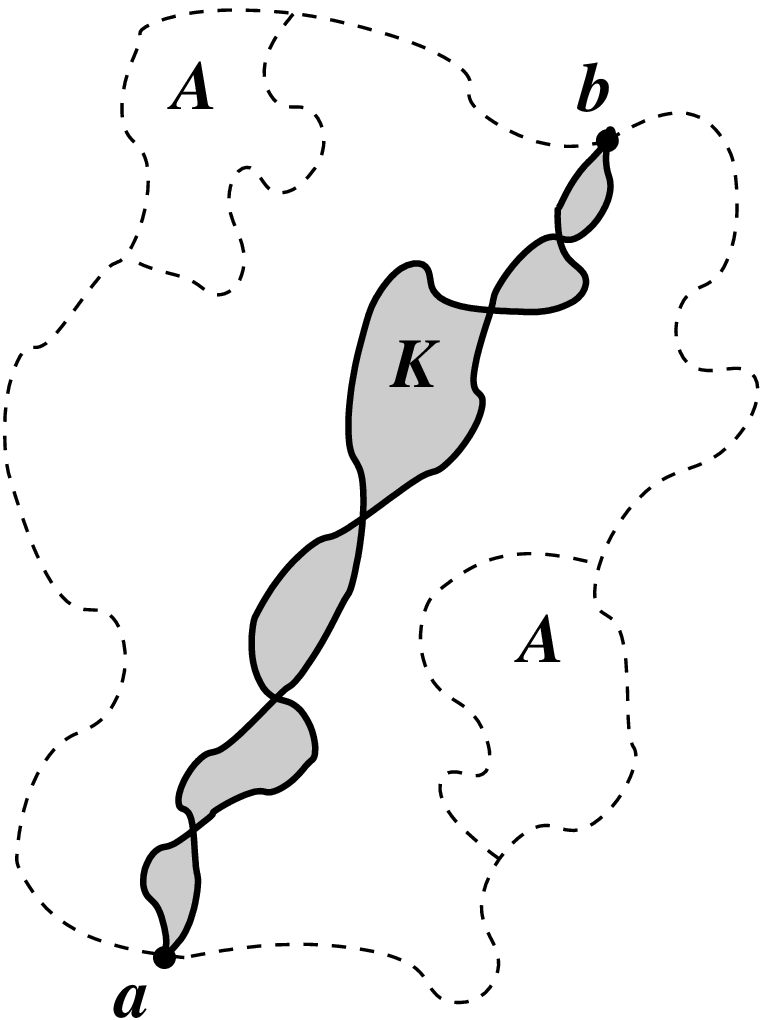}
\caption{Restriction property for sets. A compact set $K$ in $D$ is restricted not to enter the set $A$. The portions of the boundary where the sets $A$ can be attached are shown by dashed lines. They correspond to {\it absorbing} boundaries in the physical models we consider. Left: one-sided restriction. Right: two-sided restriction.}
\label{restriction-sets}
\end{figure}

It is known \cite{LSW-conformal-restriction} that the only ensemble of simple curves that satisfies the two-sided conformal restriction property is the SLE$_{8/3}$. However, we can consider more general sets $K \subset D$ that ``touch'' the boundary of $D$ only at the two fixed points $a$ and $b$ (see the right panel in Fig. \ref{restriction-sets}). \cite{sets-K} In this case we get a one-parameter family of two-sided restriction measures (that is, statistical ensembles of such sets, or clusters, $K$) characterized by the restriction exponent $h$. If the sets $K$ are allowed to ``touch'' the boundary only along, say, the clockwise arc from $a$ to $b$, then we get more general one-sided restriction measures (see the left panel in Fig. \ref{restriction-sets}). All restriction measures are fully classified, and, moreover, there is an explicit construction of all of them.\cite{LSW-conformal-restriction}

The restriction property defined by the condition of avoidance of sets $A$ attached to the boundary immediately implies the following. Consider the right panel in Fig. \ref{restriction-sets}. It is clear that a set $K$ intersects $A$ if and only if its boundary (shown by thick curves) intersects $A$. Thus, the restriction property does not care about the internal structure of the set $K$, and it is sufficient to ``fill it in'' and consider the boundaries of the filled-in sets. This means that two different ensembles of sets that only differ by their internal structure, but have the same fillings and boundaries, lead to the same restriction measure. An example of this is provided by ensembles of Brownian excursions and percolation hulls conditioned to avoid the boundary (see Sec. \ref{restriction in models} for details).

We note in passing that for a certain range of the restriction exponent $h$ ($h < 35/24$), samples of two-sided conformal restriction measures (the filled-in sets $K$) have so-called {\it cut points}.\cite{Werner-restriction-review} These are points with the property that if one of the them is removed, the filled-in set $K$ becomes disconnected. These points are shown on the right panel in Fig. \ref{restriction-sets} as intersections of the ``left'' and ``right'' boundaries of $K$. These points are similar to the so-called ``cutting bonds'', \cite{Coniglio} which are important components in the structure of percolation clusters. In fact, in the mapping to percolation for the SQH transition, the cut points are exactly the cutting bonds of the critical percolation clusters.

For a one-sided restriction measure  (see the left panel in Fig. \ref{restriction-sets}), its sample may touch the portion of the boundary where we are not attaching sets $A$. Then all statistical information related to the restriction property is encoded in the ``left'' filling of the set $K$ or, equivalently, in its ``right'' boundary. In either case, as we have mentioned in the Introduction, the boundaries of restriction measures are variants of SLE$_{8/3}$ known as SLE$(8/3,\rho)$. We shall give more details on SLE$(\kappa,\rho)$ in Sec. \ref{restriction theory}.

As we have already mentioned, it is more natural to think of un-normalized restriction measures. In this case the total weight of a restriction measure can be thought of as a partition function $Z_D(a,b)$, which is the sum of weights of all sets $K$ in the ensemble. We point out that $Z_D(a,b)$ is somewhat arbitrary, since it depends on an arbitrary normalization. This is especially subtle when we imagine obtaining $Z_D(a,b)$ from a partition function in a discrete microscopic model (as in the examples in Sec. \ref{restriction in models} below). Such a derivation will typically involve an infinite normalization in the continuum limit. However, once a particular normalization is chosen for each microscopic model of interest, the partition functions $Z_D(a,b)$ become well defined in the continuum, and contain meaningful information through their dependence on the domain $D$, the marked points $a$ and $b$ on the boundary $\partial D$ where the random sets $K$ intersect the boundary, and on the type of the restriction measure that we consider. [See Sec. \ref{restricion theory: The basic theorem of conformal restriction} for a more in depth discussion of this point, which is based on the definition of the partition function $Z_D(a,b)$ in terms of a physical quantity, namely, the disorder-averaged PCC $g(a,b)$ [see Eq. (\ref{g=Z})], and the notion of current conservation.]

\subsection{Critical curves at $c=0$ and disordered systems}
\label{subsec:critical-curves}

\begin{figure}[t]
\includegraphics[width=0.9\columnwidth]{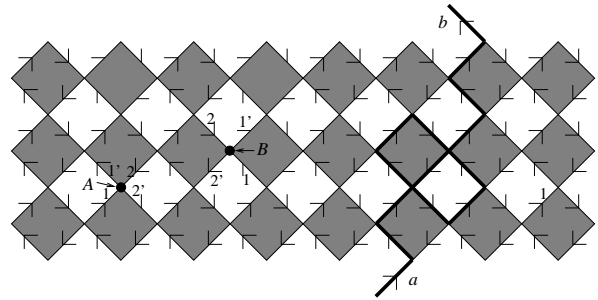}
\caption{
The Chalker-Coddington network model. The fluxes propagate on the links in the directions shown by the arrows.
The bold line connecting the links $a$ and $b$ on the boundary represents a Feynman path or a picture (see main text) contributing to the point contact conductance $g(a,b)$ between these two contacts.}
\label{chalkerbig}
\end{figure}

Consider now a finite 2D disordered conductor occupying a domain $D$. We can place small contacts at points $a$ and $b$ on the boundary of $D$ and measure the boundary PCC $g(a,b)$ between the contacts. In this paper we will focus specifically on systems described by network models of  Chalker-Coddington (CC) type (see Fig. \ref{chalkerbig}). Then, in general, a diagrammatic approach can be developed for computing the disorder-averaged conductance $\langle g(a,b) \rangle$ (more details will be presented below for specific models). In particular, ``Feynman'' paths drawn on a network for a system defined in the domain $D$ determine contributions to $\langle g(a,b) \rangle$. All these paths begin at the point $a$ and end at the point $b$ and are connected, which is a crucial feature of a disordered system. Indeed, for a system with quenched disorder we must average not the partition function, but the free energy. While the partition function generates all paths, the free energy generates connected paths only.

Let us examine under what conditions the Feynman paths satisfy restriction. In order to do so we consider two sets of Feynman paths:

(1) The Feynman paths for $\langle g(a,b) \rangle$ for the system defined in the domain $D \!\setminus\! A$.

(2)The Feynman paths for $\langle g(a,b) \rangle$ in $D$ which do not enter $A$.

It is easily seen that paths from the two sets will have the same weight after disorder averaging if the rules for generating the paths do not depend explicitly on the domain in which they are defined. The only dependence on the domain is that the paths are drawn in that domain.  In other words, the crucial condition for a set of curves to satisfy restriction is that the weights of the curves are {\it intrinsic} --- namely, the weight of a curve  may be determined by examining its shape, without reference to the shape of the system (domain) it is in. A simple example of an intrinsic weight is the probability of a random walk on a square lattice, which is $4^{-N}$, where $N$ is the total number of steps in the walk. This weight is intrinsic since it depends only on the length of the walk, but not on the domain in which the walk happens.

\begin{figure}[t]
\centering
\includegraphics[width=0.9\columnwidth]{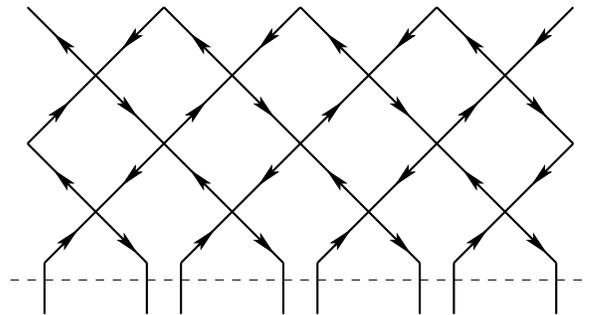}
\caption{CC network with absorbing boundary.}
\label{A boundary}
\end{figure}

An important caveat has to be added to this statement: special boundary conditions must be chosen in order to allow us to identify the weights of the paths described in items 1 and 2 above. These boundary conditions may be described as the ``absorbing boundary conditions'' and often come up naturally in the study of disordered systems: they describe, as already mentioned, the presence of ideal leads attached to the boundary. Indeed, a given path may approach the boundary, and then a certain weight will be associated with the path turning back into the bulk or escaping the system through the boundary. For network models, these weights are determined by parameters ascribed to a particular node that is on the boundary of the network. For the weights to be intrinsic, they must not depend on whether a particular node lies on the boundary or in the bulk of the system. Therefore, a boundary node should be such as to allow a path going through that node to escape the system. A boundary with such nodes is called {\it absorbing}, and in physical terms it is realized by attaching ideal leads to the disordered system. A microscopic picture of the absorbing boundary for the CC network is shown in Fig. \ref{A boundary}.
The role and the importance of the absorbing boundary conditions will be described in more detail in Sec. \ref{restriction in models}, where we shall also describe in more detail the assignment of classical statistical weights to sets of Feynman paths, after disorder averaging.

As in the formal definition of the restriction property above, it is sufficient to consider not the Feynman paths themselves with all their internal structure (multiple loops and crossings), but their fillings and boundaries. This will be implicitly assumed in the following. In particular, in the case of the CC model considered in Sec. \ref{restrictionINChalker} we will introduce the ``pictures'' that emerge as important geometric objects determining contributions to conductances. They will derive from disorder-averaged pairs of Feynman paths on the links of the network model, and will have loops. However, as for any sets satisfying restriction, it will be sufficient to consider the filled pictures (i.e., the geometrical objects that result when all internal `holes' of a picture are filled) and the boundaries of these filled pictures,
especially in the continuum limit. In the presentation given below, we sometimes will refer to both filled and unfilled objects as ``pictures'' to simplify the discussion. However, when we need to distinguish a picture and its filling, we will make the distinction explicit.

In order to establish conformal restriction we must {\it assume} that an alternative way of obtaining the paths in the first set (in item 1 above) is to conformally map the paths from domain $D$ onto domain $D \!\setminus\! A$. But this assumption is the standard assumption of conformal invariance of critical systems in two dimensions. So we expect conformal restriction to hold for the set of disorder-averaged Feynman paths for a disordered system at criticality.

In order to connect this conformal restriction property to probability theory we must also show that the weights obtained for configurations of paths after disorder averaging are positive, such that they can be considered as {\it classical} statistical weights. This will hold in the systems of interest to us, as discussed in Sec. \ref{restriction in models}. We expect that a similar formulation, utilizing Feynman paths, is possible for a large class of network models describing other disordered systems.

In all cases that we consider, the relevant classical geometric objects describing disorder-averaged PCCs at critical points become samples of conformal restriction measures in the continuum limit. This immediately leads to the following consequences. First, this means that the disorder-averaged PCCs are equal (up to some normalization factor) to the partition functions that we have introduced above:
\begin{align}
\label{g=Z}
\langle g(a,b) \rangle = Z_D(a,b).
\end{align}
The normalization factor that is involved in this relation reflects the freedom of normalization of the partition function $Z_D(a,b)$ that we have mentioned above at the end of Sec. \ref{subsec:conf-restriction}. Once this normalization is fixed for a given system, the meaningful dependence on $D, a, b$ is the same for the PCC and the partition function.

The relation (\ref{g=Z}) alone has very strong implications. We will see in Sec. \ref{restriction theory} that the partition functions $Z_D(a,b)$ transform under conformal maps as two-point functions of (Virasoro) primary operators located at positions $a$ and $b$. This turns out to imply that the current insertions at the absorbing boundary (for a two-sided restriction) or at a juxtaposition of the absorbing and a reflecting boundary (for a one-sided restriction) are primary CFT operators, and one can use tools from CFT to study their correlation functions.

\begin{figure}[t]
\centering
\includegraphics[width=0.9\columnwidth]{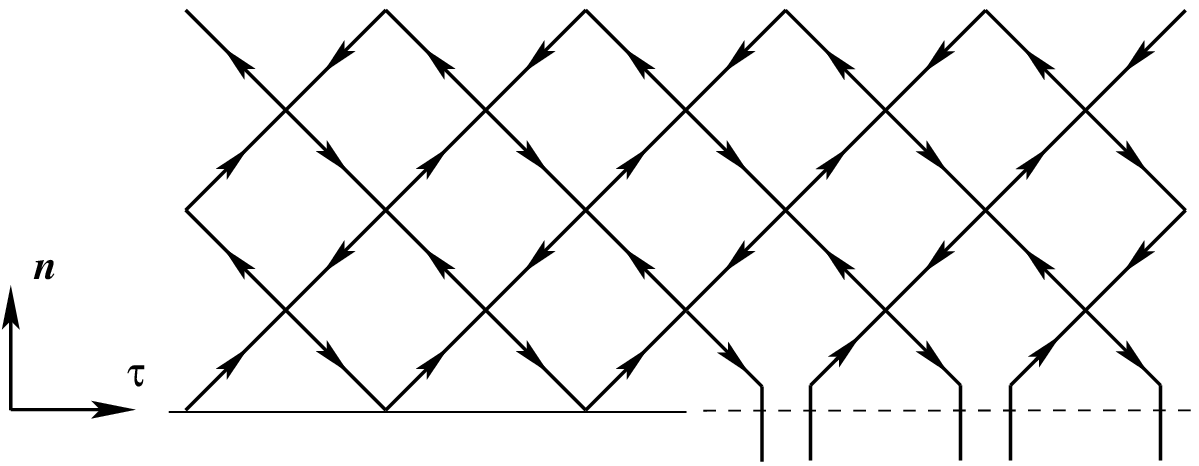}
\vskip 3mm
\includegraphics[width=0.9\columnwidth]{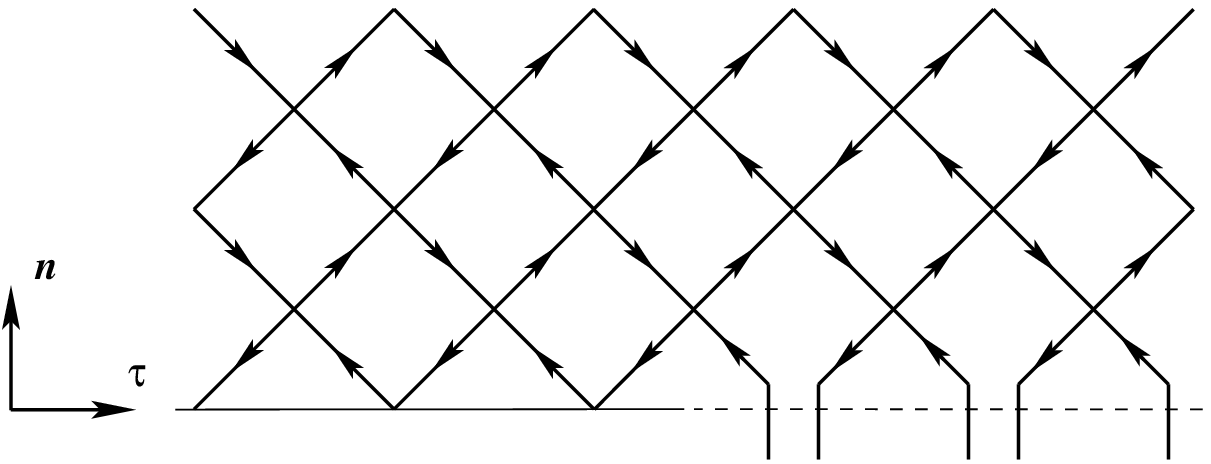}
\caption{
CC networks with two types of boundary condition changes. Top (bottom): ``right'' (``left'') boundary juxtaposed with the absorbing boundary. } \label{R-A and L-A boundaries}
\end{figure}

Second, the boundaries of the relevant classical objects (i.e.,, of the pictures) are described by SLE$(8/3,\rho)$. Based on the specific physical situations  that we consider, the parameter $\rho$ in this description can take three possible values that we will call $\rho_A$, $\rho_{RA}$, and $\rho_{LA}$. The first of these, $\rho_A$, corresponds to the two-sided restriction measure, which is relevant for a point contact placed at the absorbing boundary. The other two values correspond to one-sided restriction measures that appear when we place a point contact at a juxtaposition of the absorbing boundary with one of two possible reflecting boundaries (as, e.g., depicted at point point $b$ in Fig.  \ref{GrayFeymanns}). These two possible reflecting boundaries appear due to the fact that we consider network models with a directionality (an ``arrow'') on the links (designed to capture the physics of conductors with broken time-reversal invariance). Thus, we can have ``right'' and ``left'' reflecting boundaries that would, away from the critical point, support ``edge states'' propagating towards the point contact or away from it, correspondingly (see Fig. \ref{R-A and L-A boundaries}). More precisely, to distinguish the two types of reflecting boundaries, we introduce the following notation. Let $\hat{\bm{n}}$ be the inward normal unit vector, $\hat{\bm{z}}$ be the unit vector along the $z$ axis normal to the plane of the network, and $\hat{\bm{\tau}} = \hat{\bm{n}} \times \hat{\bm{z}}$ a unit vector tangential to the boundary. The triple $\hat{\bm{\tau}}, \hat{\bm{n}}, \hat{\bm{z}}$ is a right-hand triad. The vector $\hat{\bm{\tau}}$ can be in the direction of the current flow along the boundary, or can be opposite to it, and this is the distinction between the two types of reflecting boundaries. We will call a reflecting boundary ``right'' if the direction of the current flow at the boundary is along $\hat{\bm{\tau}}$. Similarly, on a ``left'' boundary, the current flows opposite to $\hat{\bm{\tau}}$. If we introduce $x$ and $y$ coordinates in Fig. \ref{R-A and L-A boundaries} in the usual way, then $\hat{\bm{\tau}}, \hat{\bm{n}}$ will be the unit vectors in the $x$ and $y$ directions, respectively. The reflecting lower boundaries in the top (bottom) panel of Fig. \ref{R-A and L-A boundaries} are examples of ``right'' (``left'') boundaries.

We will argue in Sec. \ref{sec:calculation1} that in all models considered in this paper one obtains the value $\rho_A = 2/3$, which turns out to give a scaling dimension $h_A = 1$ for the conserved current operator. At the same time, the values of $\rho_{RA}$ and $\rho_{LA}$ ($h_{RA}$ and $h_{LA}$) are known analytically only for the SQH critical point, as well as for the classical limit of the CC model. For the IQH transition, these values are know from numerical simulations of the CC model.\cite{Obuse2009} 

\section{Restriction in specific models of disordered electronic systems}
\label{restriction in models}

In this section, we present a detailed analysis of boundary PCC's in four models of disordered electronic systems: the Chalker-Coddington (CC) model for the IQH transition, the SU$(2)$ network model for the SQH transition, the classical limit of the CC model for diffusion in strong magnetic fields, and a weakly coupled nonlinear sigma model for a metal in class D.

\subsection{Feynman paths and restriction in the  Chalker-Coddington model}
\label{restrictionINChalker}

In Ref. \onlinecite{Chalker88}, Chalker and Coddington proposed the following network model to describe the IQH plateau transition. The network consists of links and nodes as shown in Fig. \ref{chalkerbig}. The links carry complex fluxes $z_i$, and the nodes represent (unitary) scattering matrices ${\cal S}$ connecting incoming $(z_1, z_2)$ and outgoing $(z_{1'}, z_{2'})$ fluxes:
\begin{align}
\label{smat}
\Big(\begin{array}{c}
z_{1'} \\ z_{2'}
\end{array} \Big)
= {\cal S}
\Big(\begin{array}{c} z_1 \\ z_2 \end{array} \Big)
= \Big(\begin{array}{cc}
\alpha & \beta \\
\gamma & \delta
\end{array} \Big)
\Big(\begin{array}{c} z_1 \\ z_2 \end{array} \Big).
\end{align}
For the time being, the scattering amplitudes $\al, \ldots, \de$ are assumed to be complex numbers constrained only by the unitarity of $\cal S$,  different for different nodes, which allows us to formulate our model for disordered samples with any realization of disorder. A particular distribution for the scattering amplitudes will be specified later.

There are two types of nodes in the network, the $A$ and the $B$ nodes, which live on one ($A$) or the other ($B$) sublattices of nodes, as indicated on Fig. \ref{chalkerbig}. Unitary scattering matrices always admit the so-called polar decomposition, which for the sublattice $S$ ($A$ or $B$) is written as follows:
\begin{align}
\label{Sdecomposition}
{\cal S}_S = \left( \!\! \begin{array}{cc} e^{i \phi_{1}} & \!\!\! 0 \\
0 & \!\!\! e^{i \phi_{2}} \end{array} \!\!\! \right)
\left(\!\!\! \begin{array}{cc} \sqrt{1 - t_S^2} & \!\!\! t_S \\
- t_S & \!\!\! \sqrt{1 - t_S^2} \end{array} \! \right)
\left( \!\! \begin{array}{cc} e^{i \phi_{3}} & \!\!\! 0 \\
0 & \!\!\! e^{i \phi_{4}} \end{array} \!\!\! \right).
\end{align}
Such parametrization is redundant, but when the scattering matrices are multiplied together, the elements of the diagonal unitary matrices are combined in such a way that the resulting phase factors are associated with links rather than  with nodes. In the CC model these link phases are assumed to be independent random numbers uniformly distributed between $0$ and $2\pi$. Note also that the only negative entry in the nodal scattering matrix [the middle factor in Eq. (\ref{Sdecomposition})] corresponds to the scattering from the lower incoming channel (labeled 1) to the lower outgoing channel (labeled $2'$) in the usual pictorial representation of the CC network (see Fig. \ref{chalkerbig}).

Parameters $t_A, t_B$ have a simple probabilistic meaning: $t_A^2$ is the probability to turn right upon reaching an $A$ node, and $t_B^2$ is the probability to turn left upon reaching a $B$ node. The model is isotropic when the possible values of the nodal parameters $t_S$ are related by
\begin{align}
t_A^2 + t_B^2 =1.
\label{isotropy}
\end{align}
When this equality is satisfied, the probabilities for turning left (or right) at a node are the same for the two sublattices of nodes. Then, depending on whether $t_A < t_B$, at large scales the system flows either to the insulating state with zero two-probe conductance $g$, where all the states are localized, or to the quantum Hall state, where only the bulk states are localized, but there are edge states giving a quantized value of the conductance. The transition between these regimes happens (by symmetry) when $t_A = t_B = 2^{-1/2}$. This determines the critical point in the isotropic CC model.

Let us label the links of the network by integers $j$. We define an (open) Feynman path $f$ to be an ordered sequence $j_1 = a, j_2, \ldots, j_{N(f)+1} = b$ of oriented links on the network that form a continuous path from link $a$ to link $b$, where $a$ and $b$ are distinct. Here $N(f)+1$ is the total number of links in the path $f$. Then $N(f)$ is the number of turns along the path $f$, which is the same as the number of times the path goes through a node. A given link $j$ can be traversed a multiple number of times $n_j(f)$ in a given path $f$, except for the first and the last links, where $n_a(f) = n_b(f) = 1$.  [The numbering is such that $j_k \neq a,b$ for $k = 2, 3,\ldots,N(f)$.]

For the CC model away from the critical point we need to separately keep track of the number of left ($L$) and right ($R$) turns on each sublattice of nodes. Denoting these numbers for a given path $f$ by $N_{A,R}(f)$, $N_{A,L}(f)$, $N_{B,R}(f)$, and $N_{B,L}(f)$, we obviously have $N_{A,R}(f) + N_{A,L}(f) + N_{B,R}(f) + N_{B,L}(f) = N(f)$.

If we ``forget'' the order of the links traversed by a path $f$, but retain the multiplicity $n_j(f)$ of each link, we get what we will call a ``picture''. More generally, a picture $p$ is a map from a subset of all links to the set of positive integers $p: j \to n_j$. In other words, a picture $p$ can be represented by positive integers $n_j$ associated with some links on the network. It is actually more convenient to associate $n_j = 0$ with the links that do not belong to a picture. This convention will allow us to write unrestricted summations over the links of the network.

\begin{figure}[t]
\centering
\includegraphics[width=0.9\columnwidth]{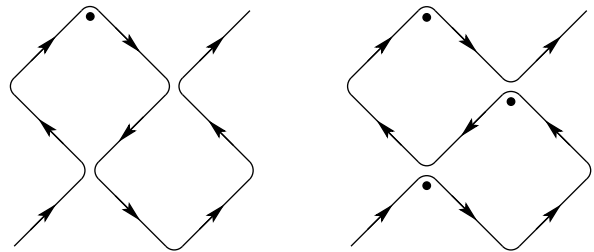}
\caption{ Two paths that correspond to the same picture. Dots mark the turns that contribute with the minus sign, so that for the path on the left $N_-(f) = 1$, and for the one on the right $N_-(f)=3$.}
\label{figure eight}
\end{figure}

It is clear that every path $f$ gives rise to a picture $p(f)$. However, two or more paths that traverse the same set of links in different order, will correspond to the same picture. The simplest example is given by the figure ``eight'' shown in Fig. \ref{figure eight}. Moreover, there are pictures that do not come from any legitimate path. For example, if the sum of the integers $n_j$ on the two incoming links is not equal to the sum of the integers on the two outgoing links at a given node, the picture with such integers cannot come from a legitimate path. We denote by $F(p)$ the set of all paths that give rise to a given picture $p$. Thus, for all $f \in F(p)$ we have $p(f) = p$, and for some pictures $p$ the set $F(p)$ is empty. In fact, there is a precise relation between pictures and Feynman paths outlined in the Appendix \ref{Appendix A}, where, in particular, we show how may distinct paths correspond to a given picture.

In what follows we will only encounter pictures that come from Feynman paths. Note that for all $f \in F(p)$ the number of turns $N(f)$ is the same:
\begin{align}
N(f) &= \sum_{j \in f} n_j - 1 = N(p), & f &\in F(p).
\end{align}
Thus, this number [as well as the individual link numbers $n_j(f) = n_j(p)$] characterizes a picture rather than a single path $f$, and we will emphasize this by denoting this number by $N(p)$ whenever appropriate.

Now we consider the quantum mechanical amplitude $A_f(a, b)$ for a path $f$ that goes from a link $a$ to a link $b$ on the network. The amplitude is given by the product of the phase factors $e^{i n_j \phi_j}$ for each link $j \in f$, and matrix elements of the scattering matrices of the nodes encountered by the path $f$. Let $N_-(f)$ be the number of turns that contribute a negative factor $-t_S$ to the amplitude. Then we have
\begin{align}
A_f(a,b) &= (-1)^{N_-(f)} t_A^{N_{A,R}(f)} (1 - t_A^2)^{N_{A,L}(f)/2} \nonumber \\
& \quad \times t_B^{N_{B,L}(f)} (1 - t_B^2)^{N_{B,R}(f)/2} e^{i \sum_j  n_j(f) \phi_j}.
\end{align}

The total amplitude $A(a,b)$ for getting from link $a$ to link $b$ is given by the sum of $A_f(a, b)$ over all the paths that go from $a$ to $b$ (we denote this set by ${\cal F}_{ab}$):
\begin{align}
A(a,b) = \sum_{f \in {\cal F}_{ab}} A_f(a, b).
\end{align}
We can rewrite this sum by breaking it into the sum over all pictures that come from any of the paths in ${\cal F}_{ab}$ [we denote this set of pictures by ${P}({\cal F}_{ab})$], and the subsequent sum over all the paths giving rise to a specific picture:
\begin{align}
A(a,b) &= \sum_{p \in P({\cal F}_{ab})} e^{i \sum_j  n_j(p) \phi_j} S(p),
\label{G-sum-over-pictures}
\end{align}
where
\begin{align}
S(p) &= \sum_{f \in F(p)} (-1)^{N_-(f)} t_A^{N_{A,R}(f)} (1 - t_A^2)^{N_{A,L}(f)/2} \nonumber \\
& \quad \times t_B^{N_{B,L}(f)} (1 - t_B^2)^{N_{B,R}(f)/2}.
\label{S(p)-general}
\end{align}
For the isotropic model (for which $t_A=t_B$), this simplifies to
\begin{align}
S(p) &= \sum_{f \in F(p)} (-1)^{N_-(f)} t_A^{N_{R}(f)} (1 -
t_A^2)^{N_{L}(f)/2}, \label{S(p)-isotropic}
\end{align}
where $N_R(f)$ and $N_L(f)$ are the total numbers of right and left turns along the path $f$. At the critical point of the isotropic model, this further simplifies to
\begin{align}
S(p) &= 2^{-N(p)/2} \sum_{f \in F(p)} (-1)^{N_-(f)}.
\label{S(p)-critical}
\end{align}

A physically relevant observable is the point contact conductance (PCC) $g(a,b)$ between the links $a$ and $b$ given by $g(a,b) = |A(a,b)|^2$. It is worth pointing out that the amplitude $A(a,b)$ determining the PCC is different from the Green's function $G(a,b)$ (the propagator) between the two points. The PCC in the CC model is defined\cite{Janssen99} by cutting the two links $a$ and $b$ of the network and using the resulting open half-links as sources and drains for the current. Thus, the PCC, as any other conductance, is a property of an open system, while the Green's function is a property of a closed system. The difference is also manifest in the graphical representation of the two quantities: while the PCC gets contributions only from open Feynman paths that go through the initial and the final links only once, the Green's function would include all paths between the links.

The conductance $g(a,b)$ is a random quantity that depends on all the phases $\phi_j$. In the following, we will only be concerned with the disorder averages of PCC's over the distribution of the phases. We will denote such averages by angular brackets. Using the representation (\ref{G-sum-over-pictures}) of the propagator as a sum over pictures, we can write
\begin{align}
g(a,b) &= \sum_{p_1, p_2 \in P({\cal F}_{ab})} e^{i \sum_j  [n_j(p_1) - n_j(p_2)] \phi_j} S(p_1) S(p_2).
\end{align}
Averaging\cite{footnotePhaseAveraging} this expression over the random phases $\phi_j$ forces the numbers on each link  to be the same for the pictures $p_1$ and $p_2$.  This can be written as
\begin{align}
\Big\langle e^{i \sum_j  [n_j(p_1) - n_j(p_2)] \phi_j} \Big\rangle = \prod_j \delta_{n_j(p_1), n_j(p_2)} = \delta_{p_1, p_2},
\label{phase-average}
\end{align}
which implies that different pictures do not interfere when we compute their contributions to $\langle g(a,b) \rangle$. Therefore, we obtain the following expression for the PCC
\begin{align}
\langle g(a,b) \rangle = \sum_{p \in P({\cal F}_{ab})} S^2(p) = \sum_{p \in P({\cal F}_{ab})} W(p).
\label{conductance-CC}
\end{align}
The significance of this formula is that the disorder-averaged PCC $\langle g(a,b) \rangle$ is represented as a sum of positive quantities $W(p) = S^2(p)$ which can be interpreted as classical positive probability weights associated with pictures $p$.

We note here in passing that the quantity $S(p)$ is the sum of the amplitudes for the paths $f \in F(p)$ in the CC model where all the link phases are set to zero. It is known that this (non-random) model without phases on the links belongs to class D in the AZ classification, and is equivalent to the non-random 2D (doubled) Ising model, \cite{Nishimori} equivalent to free Dirac fermions. Therefore, the sum $A_0(a,b)= \sum_{p \in p({\cal F}_{ab})} S(p)$ can possibly be computed explicitly by diagonalizing the transfer matrix for this non-random network model. The sum is real, and its square $A_0^2(a,b) = g_0(a,b)$ (the point contact conductance of the CC model without link phases) is straightforward to compute:
\begin{align}
g_0(a,b) = \Big( \sum_p S(p) \Big)^2
= \sum_p S^2(p) + \sum_{p_1 \neq p_2} S(p_1) S(p_2),
\end{align}
which differs from the average conductance of the actual CC model [Eq. (\ref{conductance-CC})], where the second summand is absent.

The pictures arising from Feynman paths of the Chalker-Coddington model, as defined above, can be seen to satisfy restriction. Consider the average $\langle g(a,b) \rangle$, where $a$ and $b$ are on the boundary of the sample; the sample is defined in the domain $D$ as in Fig. \ref{restriction-sets}. Assume that the boundary conditions are absorbing on that part of the boundary which goes counter-clockwise from $a$ to $b$ (see the left panel in Figs. \ref{restriction-sets} and \ref{A boundary}). The choice of absorbing boundary conditions is crucial. Electrons approaching a node on the absorbing boundary can continue their path to the outside of the boundary thus ``leaking out''. As a consequence, the scattering matrix on the boundary remains to be given by the middle factor in (\ref{Sdecomposition}), just as in the bulk. The fact that the scattering matrix is the same both at the boundary and in the bulk, and the fact that the statistical weights are intrinsic to the pictures, together ensure the restriction property. Indeed, let $A$ be a set appropriate for the definition of one-sided restriction,  as depicted in the left panel in Fig. \ref{restriction-sets}. Then the pictures contributing to $\langle g(a,b) \rangle$ for the system occupying the domain $D \!\setminus\! A$, are the same as the pictures contributing to $\langle g(a,b) \rangle$  for the system occupying the domain $D$ which do not enter $A$, with the same weights.

We note that the restriction property is satisfied in the CC model at the discrete level [i.e.,, at the level of the (CC) lattice model], and even away from criticality. At the critical point in the continuum we can, in addition, assume conformal invariance. Consequently, we obtain one-sided conformal restriction for (the continuum limit of) the pictures. This property immediately implies the following important result: The contacts where we inject and extract currents in the CFT description in the continuum become insertions of (Virasoro) {\it primary} operators with certain dimensions $h_{LA}$ or $h_{RA}$ (depending on the type or the reflecting boundary) next to the contact. This also means that the right boundary of a picture is described by \SLEr{8/3}{\rho_{LA}} or \SLEr{8/3}{\rho_{RA}} for some values of $\rho_{LA}, \rho_{RA}$ related to $h_{LA}, h_{RA}$ by Eq. (\ref{hfunctionofrho}).

Note that in order to establish one-sided conformal restriction, apart from the assumption of conformal invariance, little had to be known about the actual weights of the Feynman paths, or the quantum nature of this problem. However, a few properties were essential:

(i) The weights of pictures are intrinsic, they do not depend on the shape of the domain, and are determined by the same rules on the boundary and in the bulk of the system,

(ii) The weight of each picture is a positive quantity,

(iii) The pictures are connected: loops, or ``vacuum to vacuum'' diagrams, are absent. This is related to the vanishing of the central charge.

\noindent We will see shortly that the same properties hold for paths in other network models.

The above arguments also hold when the whole boundary of the system is absorbing. In this case, the pictures are seen to satisfy two-sided restriction, and in the continuum they are created by insertions of certain primary boundary operators of dimension $h_A$. Their boundaries (both left and right) are then described by \SLEr{8/3}{\rho_A}. We will argue in Sec. \ref{sec:calculation1} that $\rho_A = 2/3$ and $h_A = 1$. As we have mentioned above, the parameter $\rho$ for one-sided restriction can take two possible values $\rho_{RA}$ and $\rho_{LA}$, depending on the two types of reflecting boundary conditions on that part of the boundary which goes clockwise from $a$ to $b$ (see Fig. \ref{R-A and L-A boundaries} above). In the CC model, we can not at present analytically determine the values of $\rho_{RA}$ and $\rho_{LA}$ from microscopic considerations. They can be found numerically, \cite{Obuse2009} and then various critical exponents and correlation functions will be determined by these values and the theory of conformal restriction. At the same time, in the other three models described below, these values are known exactly, and the theory that we present for those models is complete.

We have already mentioned that the restriction property can be completely formulated in terms of the boundaries of filled-in pictures. Therefore, one can further rearrange the sum over pictures in Eq. (\ref{conductance-CC}) by grouping together pictures which have the same fillings. Labeling such fillings by $K$ [similar to the notation for samples of restriction measures used above (see Fig. \ref{restriction-sets})] and denoting the set of pictures with the same filling $K$ as ${\cal P}(K)$, we can write
\begin{align}
\<g(a,b)\> &= \sum_K W(K), & W(K) &= \sum_{p\in {\cal P}(K)} W(p).
\end{align}
Notice that while the sum over pictures in Eq. (\ref{conductance-CC}) is infinite even for a finite network, the corresponding sum over fillings  is in this case finite.

Let us mention here that at the microscopic level the cut points discussed in Sec. \ref{subsec:conf-restriction} correspond to particular links in a filling $K$. Namely, these are links $j$ that have multiplicity one (i.e., $n_j=1$) in {\it each} picture $p \in {\cal P}(K)$. It is easy to see that a filling can be broken into ``irreducible'' parts by removing the ``cut links'', and that the probability weight of a filling is given by the product of the weights of its irreducible components.

We finish this section with a comment about a recent paper by Ikhlef, Fendley, and Cardy (Ref. \onlinecite{Ikhlef11}). In this paper, the authors consider a certain truncation of the CC model and its integrable deformations. We notice here that this truncation has a natural interpretation in our language of pictures. Namely, it is equivalent to keeping only those pictures where all link multiplicities $n_j(p) \leqslant 1$. In this case, it is actually easy to show that the sign factors $(-1)^{N_-(f)}$ are the same for all the paths $f \in F(p)$ contributing to a given picture $p$. [The numbers $N_-(f)$ do depend on a particular path, but their parity is the same for all the paths $f \in F(p)$.] However, the truncated model with $t_A = t_B = 1/\sqrt{2}$ appears to be non-critical. To make it critical, one has to introduce an additional weight $z$ for every visited link, and tune $z$ to a particular value $z_c$. As a consequence, at the critical point of the isotropic truncated model, the weight of a picture is given by
\begin{align}
W(p) = S^2(p) = \Big(\frac{z_c}{2}\Big)^{N(p)}|F(p)|^2,
\end{align}
where in this case $N(p) + 1$ is simply the number of links in the picture, and $|F(p)|$ is given by the first factor in Eq. (\ref{number of paths}) from Appendix \ref{Appendix A}, $|F(p)| = \det L_k$. (In the second factor, the outdegrees of all the vertices are either 1 or 2, and all the edge numbers $m_{ij} = 1$.) One can also consider ``higher'' truncations, where only the pictures with $n_j(p) \leqslant k$ are kept. It is easy to see that all such truncated models, including the one in Ref. \onlinecite{Ikhlef11}, satisfy the restriction property on the lattice, and, therefore, should be described by conformal restriction theory (with exponents depending on the truncation level $k$) at their conformally invariant critical points (possibly obtained again by some fine tuning of the link weights).

\subsection{Network model for the spin quantum Hall effect}
\label{restriction in SQH}

\begin{figure}[t]
\centering
\includegraphics[width=0.9\columnwidth]{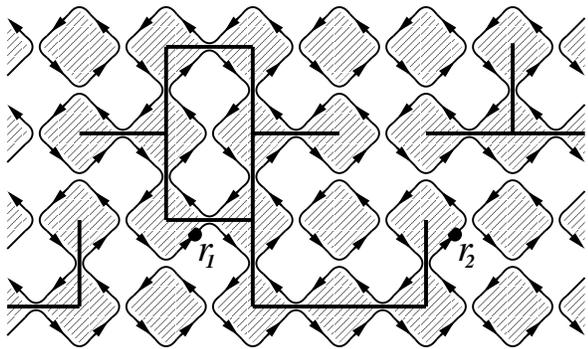}
\caption{The average point contact conductance between the links at $r_1$ and $r_2$ is given by the probability that these points are connected by a percolation hull. }
\label{percolation}
\end{figure}

The spin quantum Hall (SQH) transition was studied, numerically, in Ref. \onlinecite{Kagalovsky99}, and a simple physical picture of the SQH effect was given in Ref. \onlinecite{Senthil1999b}. Average conductances within the network model employed in Ref. \onlinecite{Kagalovsky99} can be obtained exactly through a mapping to bond percolation on a square lattice.\cite{Gruzberg99, Cardy00, Beamond02, Mirlin03, Subramaniam08, Bondesan11} An example of a percolation configuration that appears in the mapping is shown in Fig. \ref{percolation}. Within this approach, average point contact conductances that we focus on in this paper, are explicitly given in terms of probabilities in the percolation problem. These probabilities are intrinsic probabilities of percolation hulls that join the two point contacts. As such, they satisfy restriction property with respect to absorbing boundaries, similar to the weights of the pictures in our treatment of the CC model above.

Specifically, consider the average PCC $\langle g(a,b) \rangle$ between the links $a$ and $b$, at the critical point of the SQH network model. As is shown in Ref. \onlinecite{Subramaniam08}, this is equal to
\begin{align}
\langle g(a,b) \rangle \propto P(a,b) = \sum_{p} P(p;a,b),
\end{align}
where $P(a, b)$ is the probability that the links $a$ and $b$ are connected by a percolation hull. Every hull $p$ that joins these links contributes to the last expression above with its own probability $P(p;a,b)$.

We see that the geometric objects that satisfy restriction with respect to absorbing boundaries for the network model of the SQH transition are percolation hulls. At the critical point and in the continuum limit these hulls become SLE$_6$ lines (this is rigorously known for the site percolation on triangular lattice and its variants,\cite{Smirnov2001, Smirnov2009, Binder2007} and is believed to be true for other percolation models). In relation to conformal restriction, SLE$_6$ was studied in Ref. \onlinecite{LSW-conformal-restriction} where it was shown that chordal SLE$_6$ conditioned not to intersect the real line satisfies two-sided conformal restriction with exponent $h_A=1$. It then follows that the right boundary of such conditioned SLE$_6$ is SLE$(8/3, \rho_A)$ with $\rho_A = 2/3$. Furthermore, SLE$_6$ conditioned not to intersect the positive half-line satisfies the one-sided conformal restriction with the exponent $h = 1/3$. Its right boundary is SLE(8/3,$-2/3$).

As we will see in Sec. \ref{sec:calculation1}, the microscopic picture of the SQH critical point as the critical bond percolation on a square lattice allows us to identify the boundary operators in the corresponding CFT, and, consequently, obtain exact results for various correlation functions of these operators. Physically, these correlation functions are average PCCs in the presence of various complicated boundary conditions. In the case of SQH transition, they are known explicitly, including values of conformal dimensions. In particular, we will show that $h_{LA} = 1/3$ and $h_{RA} = h_A = 1$ ($\rho_{LA} = -2/3$, $\rho_{RA} = \rho_A = 2/3$). However, the main point of this paper is that in all systems that we consider, the same exact results (including non-trivial spatial dependence of PCCs) are valid, except that the values of the two exponents $h_{RA}$ and $h_{LA}$ are not always known exactly.

\subsection{Classical limit of the Chalker-Coddington model}
\label{restriction in classical CC}

The two network models considered so far both had critical points separating insulating states. In this section, we consider a classical variant of the CC network model which leads to a critical (metallic) behavior for any values of parameters. This classical model describes diffusive transport of electrons in high magnetic fields. The model has been studied in detail in Ref. \onlinecite{XRS1997}. Here we briefly summarize results of this analysis.

The classical limit of the (isotropic) CC model is obtained if we neglect quantum interference effects. Thus, we consider a classical particle performing a random walk along the links of the network. Every time the walker approaches a node, it turns right with probability $R = t_A^2$ or left with probability $T = 1 - t_A^2$. Notice that in this limit the model becomes non-random since the (random) phases that were present on the links in the quantum CC model are not considered any more. Observables in this model are not random quantities then, and we do not need to average them. For example, the PCC $g(a,b)$ in this model is simply given by the probability that a random walker starting at the link $a$ reaches link $b$ for the first time without returning to $a$. Thus,
\begin{align}
g(a,b) = \sum_p P(p;a,b),
\end{align}
where $P(p;a,b)$ is the probability for a random walker to follow a particular path $p$ between the links $a$ and $b$. These probabilities are easily seen to satisfy restriction with respect to absorbing boundaries.

\begin{figure}[t]
\centering
\includegraphics[width=0.9\columnwidth]{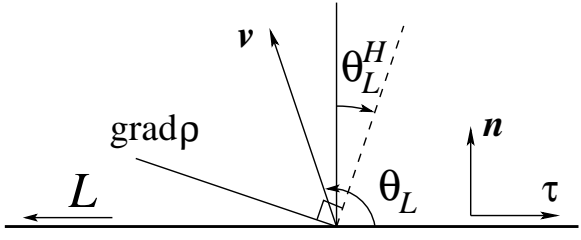}
\caption{Various directions and angles at a ``left'' boundary. Here $T = 1/4$, $R = 3/4$, so that $\gamma = \tan \theta^H_L = -1/3$.}
\label{Hall angle L}
\end{figure}

In an infinite system, the time evolution of probability $\rho(\bm{r},t)$ for a random walker to be at point $\bm{r}$ at time $t$ can be obtained by the Fourier transform. In the long-wave limit, this results in a diffusive spectrum
\begin{align}
-i\omega_k &= D k^2, & D = \frac{a^2}{4\Delta t} \frac{RT}{R^2 + T^2},
\end{align}
where $a$ is the lattice spacing, and $\Delta t$ is the time step. This leads to the diffusive behavior of the coarse-grained probability density $\bar\rho$ in the continuum limit:
\begin{align}
\partial_t \rho(\bm{r},t) = D \nabla^2 \rho(\bm{r},t).
\label{diffusion}
\end{align}
The corresponding probability current (which can be microscopically defined in various equivalent ways) is
\begin{align}
j_x(\bm{r},t) &= -(\sigma_{xx}^0 \partial_x + \sigma_{xy}^0 \partial_y)\rho(\bm{r},t),
\nonumber \\
j_y(\bm{r},t) &= -(\sigma_{yx}^0 \partial_x + \sigma_{yy}^0 \partial_y)\rho(\bm{r},t),
\end{align}
where
\begin{align}
\sigma_{xx}^0 &= \sigma_{yy}^0 = \frac{RT}{R^2 + T^2}, &
\sigma_{xy}^0 = - \sigma_{yx}^0 = -\frac{T^2}{R^2 + T^2}
\label{conductivities-cassical-CC}
\end{align}
are the components of the classical conductivity tensor. We see that the coarse-grained density $\rho$ is playing the role of the electrochemical potential $\phi$, related to the electric field in the system by $\bm{E} = -\nabla \phi$.

In a steady state the diffusion equation (\ref{diffusion}) reduces to the Laplace equation
\begin{align}
\nabla^2 \rho(\bm{r},t) = 0.
\label{Laplace}
\end{align}

\begin{figure}[t]
\centering
\includegraphics[width=0.9\columnwidth]{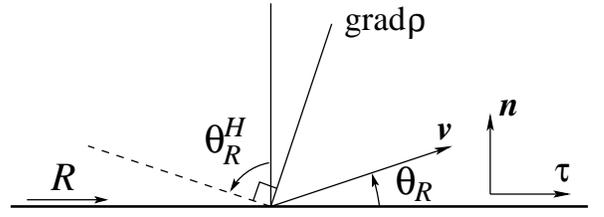}
\caption{Various directions and angles at a ``right'' boundary. Here $T = 1/4$, $R = 3/4$, so that $\tan \theta^H_R = -1/\gamma = 3$.}
\label{Hall angle R}
\end{figure}

As in the cases considered above, for a system with boundaries one can distinguish three types of boundary conditions: absorbing (corresponding to ideal leads) and ``left'' and ``right'' reflecting (hard wall), see their definitions at the end of Sec. \ref{sec:c=0}. At absorbing boundaries the probability density is constant:
\begin{align}
\rho = \rho_0 \qquad \text{on absorbing boundaries}.
\label{absorbingBC}
\end{align}
Non-zero values of the constant $\rho_0$ correspond to a system which is driven (or biased) by a constant influx of particles though an absorbing boundary. In a system that has no applied bias of this sort, $\rho_0 = 0$ reflects the fact that a random walker that hits an absorbing boundary escapes to the lead and never returns to the system.

On the other hand, at a reflecting boundary one must impose the vanishing of the normal component $j_n$ of the current. In Ref. \onlinecite{XRS1997} the authors only considered what we call a ``left'' reflecting boundary. In this case the requirement that $j_n = 0$ leads in the continuum to the following boundary condition:
\begin{align}
(\partial_n - \gamma \partial_\tau) \rho = 0,
\label{reflectingBC-1}
\end{align}
where we have introduced the notation
\begin{align}
\gamma = \frac{\sigma_{xy}^0}{\sigma_{xx}^0} = -\frac{T}{R}.
\end{align}

It is useful to depict this boundary condition by drawing a straight line orthogonal to the direction of the gradient of the density at the boundary. In Fig. \ref{Hall angle L} we show a classical Hall system occupying the upper half plane. We assume that the boundary of the system (along the $x$ axis) is a ``left'' boundary. Then, as follows from Eq. (\ref{reflectingBC-1}), the line orthogonal to $\nabla \rho$ is in the direction $(T, R)$. This direction of the vanishing component of the density gradient is shown as a dashed line, together with the direction of the gradient itself (without an arrow). We define the Hall angle $\theta^H_L$ as the angle that the dashed line makes with the inward normal. Then we have
\begin{align}
\tan \theta^H_L = \gamma.
\end{align}
Notice that the sign in this expression is consistent, since the angle $\theta_H$ is negative in the usual sense (we go clockwise from the normal to the dashed line).

Similarly, at a ``right'' reflecting boundary the vanishing of $j_n$ leads to
\begin{align}
\Big(\partial_n + \frac{1}{\gamma} \partial_\tau \Big) \rho = 0.
\label{reflectingBC-3}
\end{align}
We now depict this boundary condition at a ``right'' boundary of the Hall system occupying the upper half plane (see Fig. \ref{Hall angle R}). The gradient of the density in this case is parallel to the vector $(T, R)$. The direction in which the component of $\nabla \rho$ vanishes is then $(-R,T)$. We show both these directions in Fig. \ref{Hall angle R}. The Hall angle defined as before, now satisfies
\begin{align}
\tan \theta^H_R = \frac{R}{T} = -\frac{1}{\gamma} = -\cot\theta^H_L.
\end{align}
It follows then that  the Hall angles at the two types of reflecting boundaries are related by
\begin{align}
\theta^H_R - \theta^H_L = \frac{\pi}{2}.
\label{theta-Hall-relation}
\end{align}

\begin{figure}[t]
\centering
\includegraphics[width=0.9\columnwidth]{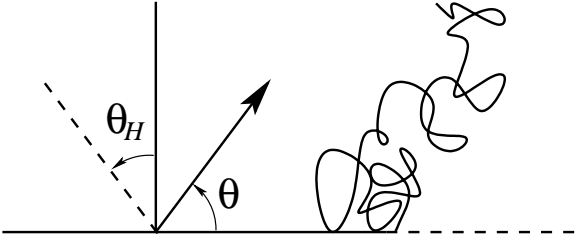}
\caption{The Hall angle $\theta_H$ is measured from the normal to the reflecting boundary (solid line). The angle $\theta$ measured from the positive direction on the real axis is the angle of reflection of a Brownian motion started at the boundary point where the reflecting boundary switches to the absorbing boundary (dashed line).}
\label{Hall angle}
\end{figure}

A steady (time independent) density and current profiles in the continuum limit are governed by the Laplace equation $\nabla^2 \rho = 0$ supplemented by the boundary conditions (\ref{absorbingBC}), (\ref{reflectingBC-1}), and (\ref{reflectingBC-3}) on the tree types of boundaries. There is an alternative (and equivalent) description of the continuum limit in terms of reflected Brownian excursions. These excursions are continuum limits of random walks on the network, which are conditioned not to hit absorbing boundaries, and which are reflected in the direction $e^{i\theta}$ upon hitting a reflecting boundary placed along the horizontal (real) axis (see Fig. \ref{Hall angle}).

This direction can be obtained microscopically from the classical CC model as follows. Imagine a random walker at the ``left'' boundary shown in the bottom on Fig. \ref{R-A and L-A boundaries}. Then in one time step the walker moves by the distance $2a$ ($a$ is the lattice spacing between the middle points of the links of the network) along the boundary to the left with probability $T$ (left turn) or normal to the boundary with probability $R$ (right turn). Then the expected displacement of the walker after one time step from the boundary is $2a(R \hat{\bm{n}} - T \hat{\bm{\tau}})$. This displacement is in the direction of the vector $\bm{v} \propto (-T, R)$ shown in Fig. \ref{Hall angle L}. We denote by $\theta_L$ the angle the vector $\bm{v}$ makes with the tangent vector $\hat{\bm{\tau}}$. Notice that the direction of $\bm{v}$ is related to the direction of the dashed line by a reflection across the normal to the boundary, so we have
\begin{align}
\theta_L = \frac{\pi}{2} - \theta^H_L.
\end{align}
Similarly, on a ``right'' reflecting boundary we have the relation
\begin{align}
\theta_R = \frac{\pi}{2} - \theta^H_R
\end{align}
(see Fig. \ref{Hall angle R}). This is, actually, a general relation between the direction of reflection and the direction along which the gradient vanishes, as can be inferred, for example, from Dubedat's paper on reflected Brownian motions.\cite{Dubedat-ReflBM} As a consequence of Eq. (\ref{theta-Hall-relation}), the angles of reflection at the two types of reflecting boundaries are related by
\begin{align}
\theta_L - \theta_R = \frac{\pi}{2}.
\label{theta-relation}
\end{align}

Brownian excursions reflected at an angle $\theta$ from a part of the boundary are known to satisfy conformal restriction property with the (one-sided) restriction exponent \cite{LSW-conformal-restriction, Werner-restriction-review}
\begin{align}\label{his1minusthetaoverpi}
h = 1 - \frac{\theta}{\pi}.
\end{align}
In terms of the Hall angles $\theta^H_{R,L}$, we get the two possible values of the one-sided restriction exponents corresponding to the two possible types of reflecting boundaries:
\begin{align}
h_{RA} &= \frac{1}{2} + \frac{\theta^H_{R}}{\pi}, &
h_{LA} &= \frac{1}{2} + \frac{\theta^H_{L}}{\pi}.
\end{align}
Then the relation (\ref{theta-Hall-relation}) implies the following:
\begin{align}
h_{RA} - h_{LA} = \frac{1}{2}.
\label{h-relation}
\end{align}
While this relation holds for the classical limit of the CC model (classical diffusion in magnetic field), {\it a priori} it is not valid for other systems we consider in this paper. For example, for the SQH transition, the exponents are known exactly to be $h_{RA} = 1$, $h_{LA} = 1/3$ (see Sec. \ref{sec:calculation1}), so they do not satisfy the relation (\ref{h-relation}).

The diffusive behavior in a magnetic field can also be described by a simple Gaussian theory\cite{XRS1997} of a complex scalar field $z({\bm r})$ with the action
\begin{align}
S_0 &= \frac{\sigma_{xx}^0}{4} \int d^2 r \,  \partial_\mu z  \partial_\mu \bar z
+ \frac{\sigma_{xy}^0}{4} \int d^2 r \, \epsilon_{\mu\nu} \partial_\mu z  \partial_\nu \bar z,
\label{S0}
\end{align}
where $\bar z$ is the complex conjugate of $z$ (and we use the convention where field configurations are weighted by $e^{-S_0}$ in the functional integral). The propagator of the field $z$
\begin{align}
d({\bm r}, {\bm r}') = \frac{\sigma_{xx}^0}{4} \big\langle {\bar z}({\bm r}) z({\bm r}') \big\rangle_0,
\end{align}
where $\langle \ldots \rangle_0$ stands for the average in the field theory with the action $S_0$, satisfies (with respect to the coordinate $\bm r$) the same equations and boundary conditions as the density $\rho$ above. Its relation to transport properties of the diffusive system are described in detail in Ref. \onlinecite{XRS1997}.

\subsection{Metal in class D}
\label{restriction in class D}

At the mean field level, the problem of thermal transport of a disordered superconductor with broken spin rotation and time-reversal symmetries belongs to class D in the AZ classification. \cite{Senthil2000, Read-Green2000, Bocquet2000} A generic model in this class can have an insulating, a thermal quantum Hall, and a metallic state. The metallic state at weak disorder can be described by a nonlinear sigma model. We will be interested in the weak coupling regime of this model where perturbation theory is justified, and the replica and the supersymmetry formulations give the same results. In the compact replica formulation, the sigma model action is\cite{Senthil2000}
\begin{align}
S &= \frac{\sigma_{xx}^0}{8} \int d^2 r \, {\rm tr} \,  \partial_\mu \! Q \, \partial_\mu \! Q
+ \frac{\sigma_{xy}^0}{8} \int d^2 r \,{\rm tr} \, \epsilon_{\mu\nu}Q \, \partial_\mu \! Q \, \partial_\nu \! Q,
\label{S-sigma-model}
\end{align}
where $Q$ is a $2n \times 2n$ matrix from the coset ${\text O}(2n)/\text{U}(n)$, and $\sigma_{xx}^0$ and $\sigma_{xy}^0$ are bare longitudinal and Hall thermal conductivities in a certain normalization. A possible parametrization for the sigma model field $Q$ is
\begin{align}
Q = \begin{pmatrix}
\sqrt{1 - Z Z^\dagger} & Z \\ Z^\dagger & - \sqrt{1 - Z^\dagger Z}
\end{pmatrix},
\label{Q-parametrization-Z}
\end{align}
where $Z$ is a complex antisymmetric matrix.

When the bare $\sigma_{xx}^0$ is large, we can treat the sigma model perturbatively. The Hall conductivity is not renormalized perturbatively. At the same time, at one loop, the diagonal conductivity is renormalized (with increasing system size $L$) as
\begin{align}
\frac{d \sigma_{xx}}{d \ln L} &= \frac{1}{2\pi},
\end{align}
so at sufficiently large scale $L$ the conductivity is logarithmically large $\sigma_{xx}(L) \sim \ln L$. If we consider such a large metallic system in class D, then the leading [in inverse powers of $\sigma_{xx}(L)$] behavior of correlation functions (including transport properties) of the system will be described by the first nontrivial (quadratic) term of the expansion of the action (\ref{S-sigma-model}) in powers of the matrix $Z$ that appears in Eq. (\ref{Q-parametrization-Z}). In terms of the matrix elements $z_{ij}$ of this matrix, this quadratic term is simply
\begin{align}
S_0[Z] &= \frac{\sigma_{xx}(L)}{4} \int d^2 r \,  \partial_\mu z_{ij}  \partial_\mu \bar z_{ij}
\nonumber \\ & \quad
+ \frac{\sigma_{xy}^0}{4} \int d^2 r \, \epsilon_{\mu\nu} \partial_\mu z_{ij}  \partial_\nu \bar z_{ij}.
\label{S0-Z}
\end{align}
This is, essentially, the same action as $S_0$ (\ref{S0}) except that there are $n(n-1)$ copies of the complex field $z$.

We conclude that the leading transport behavior of metal in class D is the same as diffusion in magnetic field described in the previous section, and can be alternatively described by conformal restriction theory. The value of the Hall conductivity $\sigma_{xy}^0$ is more or less arbitrary, and therefore, in this case we again deal with arbitrary values of the Hall angles $\theta_{L,R} \in [-\pi/2, \pi/2]$, and the one-sided restriction exponents $h_{L,R} \in [0,1]$.

We comment here that there are different network models in class D, that have been studied numerically.\cite{Chalker01, Mildenberger07} In one of these models, the so called O(1) model, the random phases on the links can only take values $\pm 1$ independently. The model appears to have only the metallic phase. It is natural to ask whether one can identify proper geometric objects and establish the restriction property directly at the level of the O(1) network model, similar to how it has been done for the CC model above. The same question exists for other network models in class D, including the Cho-Fisher model\cite{Cho-Fisher} and the network model equivalent to the Ising spin glass.\cite{Nishimori} At present this remains an interesting open problem. We note, however, that a straightforward generalization of the averaging over the phases in the CC model [see Eq. (\ref{phase-average})] to the O(1) model leads to objects (analogs of the pictures in the CC model) where the numbers of the advanced and retarded paths on a given link have the same parity. It is clear then that if we truncate this description by retaining only the objects which include links visited by either retarded or advanced paths at most once, we obtain exactly the truncated model of Ikhlef {\it et al}.\cite{Ikhlef11} This illustrates the drastic nature of the truncation procedure which leads to the same model starting from networks in different symmetry classes.

\section{The theory of conformal restriction}
\label{restriction theory}

In this section we review basic properties of conformal restriction measures and and their relationship with  multiple SLE$(8/3, \rho)$.

\subsection{Basic theorem of conformal restriction}
\label{restricion theory: The basic theorem of conformal restriction}

The basic theorem of conformal restriction states that the statistics of a restriction measure is determined by a single real parameter $h$ called the restriction exponent.\cite{LSW-conformal-restriction} This is true for both two-sided and one-sided restriction measures. The proof is based on the idea that the statistics of a restriction measure in a domain $D$ is fully determined by the collection of  probabilities
\begin{align}
P_A \equiv P[K \cap A = \varnothing] = P[K \in D \!\setminus\! A]
\end{align}
that a sample $K$ from this measure avoids an arbitrary subset $A$ attached to the boundary of $D$ (or, alternatively, that $K$ stays in the subdomain $D' = D\!\setminus \!A$), as described in Sec. \ref{subsec:conf-restriction}. One then proceeds to show that $P_A$ is uniquely determined by a single non-negative parameter $h$ that shall be called the restriction exponent.

In fact, for the case where the restriction measure lies in the upper half plane $\mathbb{H}$, is anchored at the origin and aims at infinity, it was proved that $P_A$ is given by the following expression:
\begin{align}
\label{basictheorem}
P_A = |\Phi_A'(0)|^h,
\end{align}
where $\Phi_A$ is the conformal map from $\mathbb{H}\setminus A$ to $\mathbb{H}$, which fixes the origin and infinity $\Phi_A(0)=0$, $\Phi_A(\infty)=\infty$ and has unit derivative at infinity $\Phi_A'(\infty)=1$.

To prove this assertion, one proceeds as follows. Take two sets $A$ and $B$ and consider their union $A\cup B$. The probability $P_{A\cup B}$ that a restriction measure avoids $A \cup B$ can be written as  $P_{A\cup B} = P_A P_{B|A}$, where the second factor on the right-hand side is the conditional probability to avoid $B$ given that $A$ is avoided. Now, the conformal restriction property, assumed to hold, means that this second factor can be written as $P_{B|A} = P_{\Phi_A(B)}$, where we have employed the map $\Phi_A$ to ``remove'' the avoided set $A$. Thus, we get the functional equation
\begin{align}\label{veiledfunctional}
P_{A\cup B} = P_A P_{\Phi_A(B)}.
\end{align}
The rest of the proof consists of solving this equation. In order to appreciate the equation's structure we switch notations, and instead of labeling the probability $P_A$ by the set $A$ we label it by the map $\Phi_A$ and write $P(\Phi_A)$. Now $A \cup B$ is associated with the map $\Phi_{A \cup B} = \Phi_{\Phi_A(B)} \circ \Phi_A $. The functional equation (\ref{veiledfunctional}) takes the form
\begin{align}\label{functional}
P(\Phi_{\Phi_A(B)} \circ \Phi_A)  = P(\Phi_A) P[\Phi_A(B)].
\end{align}
Thus, $P$ maps the composition operation in the space of conformal maps into simple multiplication. It is clear that (\ref{basictheorem}) obeys (\ref{functional}) as the factor $|\Phi_A'(0)|$ is the Jacobian of the transformation $\Phi_A$ at the origin, and as such gets multiplied as successive maps are composed. One can show that, in fact, $|\Phi_A'(0)|$ is the only solution up to the arbitrary parameter $h$. \cite{LSW-conformal-restriction}

Once the basic theorem is established in the form of Eq. (\ref{basictheorem}) for restriction measures in the upper half plane, it can be generalized to arbitrary simply connected domains. \cite{LSW-conformal-restriction} First, we transport the restriction measure with exponent $h$ from the upper half plane to a domain $D$ using a conformal map $f_D$ chosen in such a way that $f_D(0) = a$ and $f_D(\infty) = b$ for two marked points $a$ and $b$ on the boundary $\partial D$. If we choose a set $A$ attached to $\partial D$ such that both points $a$ and $b$ belong to the common boundary of $D$ and its subset $D' = D\!\setminus \! A$, then the analog of Eq. (\ref{basictheorem}) in this general situation is
\begin{align}
P[K \cap A = \varnothing] = P[K \in D \!\setminus\! A] = |\Phi_A'(a)|^h |\Phi_A'(b)|^h,
\label{basictheorem-D}
\end{align}
where $\Phi_A$ is a conformal map from $D'$ to $D$ normalized such that $\Phi_A(a) = a$, $\Phi_A(b) = b$.

One may consider the total weight of all sets $K$ extending from $a$ to $b$ in the domain $D$ and define this object as the partition function $Z_D(a,b)$ in Sec. \ref{subsec:conf-restriction} above. This definition requires some care, as we have only been dealing with probabilities until now, in effect dividing by $Z_D(a,b)$. To make sense of the definition, we assign the total weight one for some given, but arbitrary domain, and then demand that the partition function is consistent with conformal restriction (which is nevertheless a statement about probabilities). It then may be shown that conformal restriction is only consistent with the following transformation law:
\begin{align}
Z_{D_1}(a,b) = |f'(a)|^h |f'(b)|^h Z_{D_2}(f(a),f(b)),
\label{transformation}
\end{align}
where $f(z)$ is a conformal map from $D_1$ to $D_2$ that maps the marked points $a$ and $b$ to $f(a)$ and $f(b)$. This transformation law is that of a two-point correlation function of {\it primary} operators, in terminology of CFT.

All the results in this section so far are valid for both two-sided and one-sided restriction measures. However, to derive Eq. (\ref{transformation}), we first concentrate on the (easier) two-sided case. In this case the relation (\ref{transformation}) is a direct consequence of (\ref{basictheorem-D}) if $D_1 \subseteq D_2$, and $f(a) = a$, $f(b) = b$. Indeed, the ratio of the two partition functions is simply the probability that a sample $K$ of the restriction measure stays in the smaller domain $D_1$, and this probability, by Eq. (\ref{basictheorem-D}), is the product $|f'(a)|^h |f'(b)|^h$. The opposite situation $D_2 \subseteq D_1$ is also a straightforward consequence when we use the fact that for the inverse map $f^{-1}: D_2 \to D_1$ the derivative is $(f^{-1})' = 1/f'$. Thus, the restriction property can be used to relate total weights for both {\it decreasing} and {\it increasing} sequences domains, thereby, extending the result (\ref{transformation}) to arbitrary domains $D_1$ and $D_2$ as long as their boundaries agree at the points $a$ and $b$. The last condition can be relaxed as we can always rotate and rescale the domains to make $a$ coincide with $f(a)$ and $b$ coincide with $f(b)$. These two operations would only use the assumption that partition functions are invariant under rotations and are multiplied by powers of the rescaling factor under scale transformations. This quite natural scaling property is much weaker than the conformal covariance. Nevertheless, making this assumption we see that the transformation law (\ref{transformation}) holds for arbitrary two-sided restriction measures. We shall argue in Sec. \ref{subsec:superuniversal} below that for all the models of our interest we have $h = h_A = 1$ on absorbing boundaries.

In the case of one-sided restriction we can derive Eq. (\ref{transformation}) as follows. Consider a domain $D_1$ with {\it four} marked points $a$, $b$, $c$, and $d$ on the boundary. The points $c$ and $d$ are where the reflecting boundary condition switched to the absorbing one, and the points $a$ and $b$ on the absorbing portion of the boundary are the ``attachment'' points for the sets $K$. Consider all possible subsets $D_1' \subset D_1$ which agree with $D_1$ at the points $a$ and $b$ but do not include the points $c$ and $d$. Assume that the boundary conditions in $D_1$ are absorbing along the whole boundary. We now can define the partition function $Z_{D_1}(a,b,c,d)$ by demanding that it reduces to $Z_{D_1'}(a,b)$ if we restrict the sets $K$ contributing to $Z_{D_1}(a,b,c,d)$  to stay in $D_1'$. Formally, we define
\begin{align}
Z_{D_1}(a,b,c,d) P_{D_1}[K \in D_1'] = Z_{D_1'}(a,b),
\label{Z-four-points}
\end{align}
where $P_{D_1}[K \in D_1']$ is the probability that the set $K$ (drawn in $D_1$ with its partly reflecting boundary conditions) stays in $D_1'$.

Now take a domain $D_2$ and let $f$ be a conformal map from  $D_1$ to $D_2$. Let $D_2' = f(D_1')$. We assume that in $D_2$ there are reflecting boundary conditions between $f(c)$ and $f(d)$ and the rest of the boundary is absorbing. In $D_2'$, the boundary conditions are absorbing everywhere. We notice that
\begin{align}
P_{D_1}[K \in D_1'] = P_{D_2}[K \in D_2']
\label{probabilities}
\end{align}
by the conformal transport of probabilities. Even though the domains in question have marked points $c$ and $d$ (and their images) on their boundaries, the probabilities above are conformally {\it invariant}, since, as we will argue in Sec. \ref{subsec:superuniversal}, the conformal dimensions of the boundary changing operators at $c$ and $d$ are zero. Next, for the domains $D_1'$ and $D_2'$, the transformation property (\ref{transformation}) is valid (both domains have absorbing boundaries):
$Z_{D_1'}(a,b) = |f'(a)|^{h_A} |f'(b)|^{h_A} Z_{D_2'}(f(a),f(b))$. Dividing this by the equal probabilities in Eq. (\ref{probabilities}) gives:
\begin{align}
\frac{Z_{D_1'}(a,b)}{P_{D_1}[K \in D_1']} = |f'(a)|^{h_A} |f'(b)|^{h_A} \frac{Z_{D_2'}(f(a),f(b))}{P_{D_2}[K \in D_2']}.
\end{align}
Using the Eq. (\ref{Z-four-points}) we rewrite this as
\begin{align}
Z_{D_1}(a,b,c,d) &= |f'(a)|^{h_A} |f'(b)|^{h_A} \nonumber \\
&\quad \times Z_{D_2}(f(a),f(b),f(c),f(d)).
\end{align}
Finally, we take the limit $a\to c$ and $b\to d$ with an appropriate rescaling of the partition functions to get a well-defined finite limit. Namely, we define
\begin{align}
Z_D(a,b) = \lim_{a\to c, b\to d} |a-c|^{h'} |b-d|^{h'} Z_D(a,b,c,d),
\end{align}
and similarly for $Z_{D_2}(f(a),f(b))$. We expect that for a certain value $h'$ independent of the domains and the marked points the limit exists and is finite. This procedure is essentially the same as the operator product expansion in field theory. Upon taking the limit $a\to c$ and $b\to d$, the extension of (\ref{transformation}) to one-sided restriction follows.

So far we have defined partition functions in the continuum. Here we want to make contact with microscopic models. To do this, a somewhat stronger assumption of conformal invariance, than the one we have made so far, is needed to obtain (\ref{transformation}). Since the probability that a current inserted through the absorbing boundary will persist a macroscopic distance away is vanishing in the continuum limit, some care must be taken in relating the discrete weights and the partition functions in the continuum. The procedure we describe now essentially uses current conservation, and not much else.

Let us choose some microscopic scale $\delta$ of the order of the lattice spacing. Consider a point $a$ on an absorbing boundary, and a segment of length $l$ containing this point $a$. The length $l$ is assumed to be much larger than $\delta$ but much smaller than any other macroscopic scale (such as the scale on which the boundary curves or the size of the system). Through this segment we inject uniformly the current ${l/\delta}$ (in units of elementary charges per unit time). The number $l/\delta$ is, essentially, the number of point contacts that will each contribute to the total current through the segment. We do the same procedure around another point $b$ on the boundary. Now we define the partition function, or the two-probe conductance, as the total weight of pictures connecting the point contacts near point $a$ with those near point $b$, multiplied by $(l/\delta)^2$. Now we take the continuum limit $\delta \to 0$, and then also take $l \to 0$. The result will be the point contact conductance in the continuum. A crucial point in this construction is that the total current through a segment is naturally assumed to be a conformally invariant quantity.

Note that since $l$ is much larger than the lattice spacing, the definition is lattice independent, and the particular assumption of conformal invariance made here seems no stronger than the one we have made thus far (which concerned only probabilities, not partition functions).  We nevertheless cannot make the latter statement rigorous, and (\ref{transformation}) remains an assumption as it pertains to total weights defined from a microscopic model. We also note that we have defined here the operator which inserts current through an absorbing boundary, but similar procedures may be employed to handle other boundary conditions.

\subsection{Description of critical curves through SLE}

\begin{figure}[t]
\centering
\includegraphics[width=0.9\columnwidth]{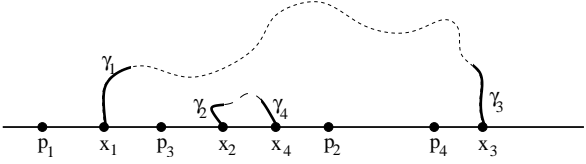}
\caption{Two SLE's interacting with four points $p_i$ in the upper half plane $D=\mathbb{H}$.} \label{TwoSLE}
\end{figure}

The above description of a restriction measure by the probabilities of avoidance of certain sets given by (\ref{basictheorem}) is somewhat implicit. To obtain a more explicit description of restriction measures, which will be also useful for computations, we can concentrate on the boundaries of a restriction sample (a cluster). These cluster boundaries are known to be variants of \SLE{8/3} distorted in a particular way, and are described by a variant of the Schramm-Loewner equation. In particular, the ensemble of the \SLE{8/3} curves satisfies the two-sided conformal restriction with the restriction exponent $h = 5/8$.\cite{LSW-conformal-restriction} In this section, we describe the SLE method for critical curves.

Suppose we have some statistical system defined in a domain $D$ in which $2n+m$, $x_i$, $i=1,\dots,2n$ and $p_j$, $j=1,\dots,m$ special points are marked on the boundary. Figure \ref{TwoSLE} illustrates this situation for a system in the upper half plane, so that the marked points lie on the real axis. The partition function of such a system is denoted by $Z_D(x_1, \dots, x_{2n}; p_1, \dots, p_m)$. The points $x_i$ are beginnings and ends of curves that are crated in the statistical system. Namely, we assume that there exist $n$ curves in the upper half plane, the $i$'th curve connecting $x_i$ with $x_{n+i}$ on the real axis. $p_j$ denote points where boundary conditions are changed or operators are inserted, etc. The shape of the curves depends on the positions of the $m$ marked points $p_j$ in some way. We shall consider only the case where the operators inserted at $p_j$ are primary fields of weights $h_j$. It turns out that the points $x_i$ also correspond to insertions of primary fields, and that all these have the same weight $h_\kappa = (6 - \kappa)/(2\kappa)$. Then the partition function $Z_D$ transforms under conformal maps of the domain $D$ and the marked points on its boundary as a correlation function of primary operators in CFT:
\begin{align}
& Z_D(x_1, \dots, x_{2n}; p_1,\dots, p_m) =
\prod_{i=1}^{2n} |g'(x_i)|^{h_\kappa} \prod_{j=1}^{m} |g'(p_j)|^{h_j} \nonumber \\
& \times Z_{\tilde D} \big[g(x_1), \dots, g(x_{2n}); g(p_1), \dots, g(p_m)\big],
\label{partitionConformalInv}
\end{align}
where $g$ is a conformal map from $D$ to $\tilde D$.

An important consequence of Eq. (\ref{partitionConformalInv}) is obtained if we consider both domains $D$ and $\tilde D$ to be the upper half plane $\mathbb H$. In this case the allowed conformal transformations are M\"obius transformations with real coefficients. These transformations form the SL$(2,\mathbb{R})$ group and are generated by
\begin{align}\label{SL2RLs}
L_{-1} &=  \sum_{i=1}^{2n} \partial_{x_i} + \sum_{j=1}^m \partial_{p_j}, \nonumber \\
L_{0} &=  \sum_{i=1}^{2n} \big(x_i \partial_{x_i} + h_i \big)
+ \sum_{j=1}^m \big( p_j \partial_{p_j} + h_\kappa \big),  \nonumber \\
L_{+1} &= \sum_{i=1}^{2n} \big( x_i^2 \partial_{x_i} + 2h_i x_i \big)
+ \sum_{j=1}^m \big( p_j^2 \partial_{p_j} + 2h_\kappa p_j \big).
\end{align}
Then the infinitesimal form of the transformation law (\ref{partitionConformalInv}) for the M\"obius transformations gives the global conformal invariance of the partition function $Z_{\mathbb H}$ in the upper half plane:
\begin{align}
\label{SL2R}
L_{\pm1,0} Z_{\mathbb H} = 0.
\end{align}

Let us now mark $2n$ disjoint segments on the curves, $\gamma_i$, $i=1,\dots,2n$, so that the $i$'th segment connects the point $x_i$ to a point $z_i$ on the $i$-th curve for $i=1,\dots,n$ or the $i-n$-th curve for $i=n+1,\dots,2n$ (see figure (\ref{TwoSLE})). Let us also denote the contribution to the partition function from configurations of the statistical system in which the segments take a particular shape $\gamma_i$ as $Z(\gamma_1, \dots, \gamma_{2n}; p_1, \dots , p_m)$. With the shape of the curves $\gamma_i$ fixed, the quantity $Z(\gamma_1, \dots, \gamma_{2n}; p_1, \dots , p_m)$ is the partition function in the {\it slit domain} ${\mathbb H}\setminus \bigcup_i^{2n} \gamma_i$, that is, the upper half plane with the segments $\gamma_i$ removed.  The full partition function is obtained by summing these partial contributions over the shapes of all segments $\gamma_i$. Somewhat abusing notation for the sum over all possible shapes of curves $\gamma_i$ we write
\begin{align}
& Z_{\mathbb H}(x_1, \dots, x_{2n}; p_1, \dots, p_m) \nonumber \\
& \quad = \sum_{\{\gamma_i\}} Z(\gamma_1, \dots, \gamma_{2n}; p_1, \dots , p_m).
\label{partitionExpectation}
\end{align}

The points $z_i$ sit at the tips of the segments. We can enlarge or reduce the size of the segments by moving the tips $z_i$ along the curves. We shall parametrize this motion along the $i$-th curve by a ``time'' parameter $t_i$. To describe the shape of the segments at arbitrary times $t_i$ SLE makes use of a conformal map from the slit domain $\mathbb{H} \setminus \bigcup_{i=1}^{2n} \gamma_i$ back to the upper half plane. This conformal map $g_{\boldsymbol{t}}(z)$ depends on all the times $t_i$ which together form the vector $\boldsymbol{t}$. This time dependence is given by a stochastic differential equation which determines $g_{\boldsymbol{t}}(z)$ as a function of $t_i$:
\begin{align}\label{multipleSLE}
d_{t_i} g_{\boldsymbol{t}}(z) = \frac{2 dt_i}{g_{\boldsymbol{t}}(z) -
\xi_i(\boldsymbol{t})}.
\end{align}
Here $\xi_i(\boldsymbol{t}) = g_{\boldsymbol{t}}(z_i)$ are the images of the tips of the segments $\gamma_i$ under the map $g_{\boldsymbol{t}}(z)$. Each $\xi_i$ is located on the real axis and obeys the same equation (\ref{multipleSLE}) with respect to $t_j$ for $j \neq i$. However for $z=z_i$ Eq. (\ref{multipleSLE}) is ill-defined and must be replaced by:
\begin{align}
\label{SLEforcing}
& d_{t_i} \xi_i = \kappa \, \partial_{\xi_i} \!
\log Z_{\mathbb H}\big(\xi_1, \dots, \xi_{2n}; g_{\boldsymbol{t}}(p_1), \dots, g_{\boldsymbol{t}}(p_m)\big) dt_i \nonumber \\
& \quad + \sqrt{\kappa} dB_i, 
\end{align}
where $B_i(t)$ are independent standard Brownian motions. When all the times are increased by small increments $dt_i$, the evolution of the conformal map $g_{\boldsymbol{t}}(z)$ can be described by the following system of equations:
\begin{align}
d g_{\boldsymbol{t}}(z) &= \sum_{i=1}^{2n} \frac{2dt_i}{g_{\boldsymbol{t}}(z) - \xi_i(\boldsymbol{t})}, \label{multipleSLE-all-times} \\
d \xi_i &= \kappa \, \partial_{\xi_i} \! \log
Z_{\mathbb H}\big(\xi_1, \dots, \xi_{2n}; g_{\boldsymbol{t}}(p_1), \dots, g_{\boldsymbol{t}}(p_m)\big) dt_i \nonumber \\
& \quad + \sqrt{\kappa} dB_i + \sum_{j \neq i} \frac{2 dt_j}{\xi_i - \xi_j}.
\label{SLEforcing-all-times}
\end{align}
These equations are essentially the same as Eqs. (2) and (3) in Ref. \onlinecite{BBK2005}. Note that Eq. (\ref{multipleSLE-all-times}) fixes the asymptotic behavior of the map $g_{\boldsymbol{t}}(z) \sim z + \big(\sum_i t_i\big)/z$ as $z\to\infty$. The $2n$ curves whose evolution is described by Eqs. (\ref{multipleSLE-all-times}) and (\ref{SLEforcing-all-times}) are often called SLE traces or simply traces. Since the equations are stochastic differential equations, they, in fact, define a probability measure on the traces, or an ensemble.

So far we did not specify the partition function $Z_{\mathbb H}$ that appears in the general SLE equation (\ref{SLEforcing-all-times}). In the next subsection we consider special cases where $Z_{\mathbb H}$ will be uniquely determined by the global conformal invariance (\ref{SL2R}). For the special value $\kappa = 8/3$, this will give a description of general one-sided restriction measures.

\subsection{Description of restriction measures using SLE$(8/3,\rho)$}
\label{SLEtoDescribeRestriction}

One can obtain a restriction measure by taking $m=0$, $n=1$, and $x_2\to\infty$. By translational invariance ($L_{-1} Z_{\mathbb H} = 0$), we immediately obtain that $Z_{\mathbb H}(x_1,\infty)$ is a constant and thus the forcing term in (\ref{SLEforcing}) vanishes. We shall also drop $t_2$, never considering the evolution with respect to this time.  The equations (\ref{multipleSLE}) and (\ref{SLEforcing}) reduce to:
\begin{align}
d g_t (z) &= \frac{2 dt}{ g_t(z) - \xi(t)}, & d \xi  &= \sqrt{\kappa} d B_t.
\end{align}
Different values of $\kappa$ lead to different ensembles of curves. The only value of $\kappa$ for which the restriction property holds is $\kappa=8/3$.\cite{LSW-conformal-restriction} The restriction exponent in this case turns out to be $h=5/8$. This corresponds to the insertion at $x_1$ of a primary operator of weight $h=5/8$.

Thus, the SLE process without forcing thus can only produce a restriction measure of a single exponent $5/8$. It is possible to modify the SLE$_{8/3}$ process somewhat and obtain more general one-sided restriction measures with other restriction exponents. We shall still be interested in the case where a single curve emanates from $x_1$ and ends at infinity. In order to be able to generalize the usual SLE procedure which gives rise to the exponent $5/8$, one has to employ a point splitting procedure. Namely, we shall take the point $x_1$ and replace it by two points $x_1$ and $X_1 < x_1$ [this choice will lead to a restriction measure in the upper half plane whose right boundary will be the SLE(8/3, $\rho$) curve]. The SLE trace shall emanate from $x_1$, and $X_1$ will be a marked point (previously this was denoted by $p_1$). At the end of the procedure the two points are re-fused, and we have a trace emanating from a single point $x_1$ which is also a marked point. Note, however, that independently of whether the points $x_1$ and $X_1$ are fused or not, for any non-zero time $t = t_1$ the partition function appearing in (\ref{SLEforcing}) will depend on three distinct points $p_1 = g_t(X_1)$, $\xi = g_t(z_1)$, and $\infty$. $X_1$ will correspond as usual to an insertion of a primary operator. The global conformal invariance determines in this case partition function to have a simple power law dependence whose exponent is denoted by $3\rho/8$:
\begin{align}
Z_{\mathbb H}(\xi,\infty, p_1) = \frac{1}{(\xi- p_1)^{3\rho/8}}.
\end{align}
With this form of the partition function the general equations (\ref{multipleSLE-all-times}) and (\ref{SLEforcing-all-times}) reduce to
\begin{align}
d g_t (z) &= \frac{2 dt}{g_t(z) - \xi(t)}, &
d \xi  &= \frac{\rho dt}{\xi - g_t(X_1)} + \sqrt{\kappa} d B_t.
\end{align}

The value of $\rho$ determines the restriction exponent:
\begin{align}
h = \frac{(3\rho+10)(2+\rho)}{32}.
\end{align}
Moreover, it determines the conformal weight of the operator at $X_1$, which is given by
\begin{align}
\label{conformalweightofrho}
h' = \frac{\rho(4+3\rho)}{32}.
\end{align}
The process one obtains in this way is termed \SLEr{8/3}{\rho}.

The described procedure shows that boundaries of one-sided restriction measures can be created by fusing the operator creating an \SLE{8/3} curve with a primary operator. Since a restriction measure is fully specified by its exponent, by appropriately choosing the parameter $\rho > -2$ in the above procedure, we can obtain restriction measures with any given exponent $h > 0$. We shall assume in the following that this also holds locally, independently of other curves or marked points. This assumption does not have a rigorous mathematical proof. It rests on the physical assumption that any additional operators or curves can only change the large scale properties of the given curve rather than its local structure.

\subsection{Martingale conditions on partition functions}

When considering more general situations where restriction holds, we shall have to consider more complicated partition functions than those appearing in the previous section \ref{SLEtoDescribeRestriction}. To compute those partition functions it is not sufficient to use the conformal covariance (\ref{partitionConformalInv}) and the global SL$(2,\mathbb{R})$ invariance (\ref{SL2R}). In addition to these conditions we must make use of the idea of the partial summation over segments of curves expressed by  Eq. (\ref{partitionExpectation}), to obtain further conditions on $Z_{\mathbb H}$.

We will derive these well known conditions in this section. In order to set notation, we first consider the situation where $m=0$ and the forcing term in (\ref{SLEforcing}) vanishes, namely,
\begin{align}
\label{noforcing}
d_{t_i}\xi_i =\sqrt{\kappa} dB_i.
\end{align}
In this situation, $2n$ independent curves start at the points $x_i$ and go to infinity. Due to independence of the Brownian motions in (\ref{noforcing}), the statistical weight of the SLE traces in this case is given by
\begin{align}
\prod_{i=1}^{2n} Z(\gamma_i,\infty).
\label{product-measure}
\end{align}
This product of independent SLE measures can be used to define expectation values by
\begin{align}\label{expectationNoInteraction}
\mathbb{E} \big[ f(\gamma_1, \dots, \gamma_{2n} ) \big] &\equiv \frac{1}{{\cal N}_{2n}}
\sum_{\{\gamma_i\}} \Big[ f(\gamma_1, \dots, \gamma_{2n}) \prod_{i=1}^{2n} Z(\gamma_i,\infty) \Big], \nonumber \\
{\cal N}_{2n} &= \sum_{\{\gamma_i\}} \prod_{i=1}^{2n} Z(\gamma_i,\infty),
\end{align}
where the sum over $\gamma_i$ is used in the same sense as in Eq. (\ref{partitionExpectation}) and denotes summing over all possible shapes of the curves. We now consider the general case with a non-trivial forcing term in Eq. (\ref{SLEforcing}). The SLE measure in this case is different from the product (\ref{product-measure}). However, it is possible to describe a general ensemble of SLE curves using the product measure (\ref{product-measure}) produced by independent Brownian motions (\ref{noforcing}). Heuristically we can think of the measure (\ref{product-measure}) produced by (\ref{noforcing}) serving to scan a large class of curves. Very loosely speaking, we can consider curves sampled from the non-interacting measure (\ref{product-measure}) and {\it the same} curves from the general SLE ensemble, and compare their weights. Alternatively, we can think of creating the probability measure in (\ref{SLEforcing}) by first producing the curves using (\ref{noforcing}) and then re-weighting them. A priori it is not at all clear that this should be possible, since the weights of the curves sampled from the two measures can be incomparable (for example, one weight can be zero, while the other non-zero). However, it happens to be possible in the case of the two SLE measures with the same value of $\kappa$: one with the independent forcing (\ref{noforcing}) and the other with a non-trivial forcing (\ref{SLEforcing}).\cite{Werner-Girsanov} More formally, for an expectation value in a general SLE ensemble we have:
\begin{widetext}
\begin{align}
&\sum_{\{\gamma_i\}} Z(\gamma_1, \dots, \gamma_{2n}; p_1, \dots, p_m) f(\gamma_1, \dots, \gamma_{2n})
=  \sum_{\{\gamma_i\}} \frac{Z(\gamma_1, \dots, \gamma_{2n}; p_1, \dots, p_m)
f(\gamma_1, \dots, \gamma_{2n})}{\prod_{i=1}^{2n} Z(\gamma_i,\infty)} \prod_{i=1}^{2n} Z(\gamma_i,\infty)   \nonumber  \\
& \quad = {\cal N}_{2n} \, \mathbb{E} \bigg[\frac{Z(\gamma_1, \dots, \gamma_{2n}; p_1, \dots, p_m)
f(\gamma_1, \dots, \gamma_{2n})}{\prod_{i=1}^{2n} Z(\gamma_i,\infty) }\bigg],
\label{GirsanovForPhysicists}
\end{align}
where ${\cal N}_{2n}$ and $\mathbb{E}$ are the ones defined in Eq. (\ref{expectationNoInteraction}), that is,  $\mathbb{E}$ denotes the expectation value with respect to the measure of the independent SLE's. This method of re-weighting the measure, when done with appropriate mathematical rigor is known under the name of Girsanov's transformation. The condition that the weights of the curves sampled from two measures are comparable, translates in more rigorous terms to the condition of absolute continuity of one measure with respect to the other. (For Girsanov's theorem and other information on stochastic analysis see, for example, Refs. \onlinecite{Oksendal} and \onlinecite{Klebaner}.)

We now make us of the re-weighting procedure to obtain further conditions on $Z_{\mathbb H}$. We freeze all the times $t_i$ except $t_1$ and rename it $t_1=t$. Consider the partition function $Z_{\mathbb H}(x_1, \dots, x_{2n}; p_1, \dots, p_m)$. By definition it is independent of $t$:
\begin{align}
\partial_t Z_{\mathbb H}(p_1, \dots, p_m; x_1, \dots, x_{2n}) = 0.
\label{t-independence}
\end{align}
When $t$ is varied, only the segment $\gamma_1$ is produced, and the partial summation (\ref{partitionExpectation}) can be used in the form
\begin{align}
& Z_{\mathbb H}(x_1, \dots, x_{2n}; p_1, \dots, p_m)
= \sum_{\gamma_1} Z(\gamma_1, x_2 \dots, x_{2n}; p_1, \dots , p_m).
\end{align}
Now we can use the re-weighting procedure (\ref{GirsanovForPhysicists}) with $f=1$:
\begin{align}
Z_{\mathbb H}(x_1, \dots, x_{2n}; p_1, \dots, p_m)
&= {\cal N}_1 \, \mathbb{E} \bigg[\frac{Z(\gamma_1, x_2, \dots, x_{2n}; p_1, \dots, p_m)}{Z(\gamma_1,\infty)}\bigg],
&& {\cal N}_1 = \sum_{\gamma_1} Z(\gamma_1,\infty).
\end{align}
Next we transform both the numerator and the denominator in the last expression using the SLE map from ${\mathbb H}\setminus \gamma_1$ to $\mathbb H$ and recalling the covariance property (\ref{partitionConformalInv}):
\begin{align}
Z_{\mathbb H}(x_1, \dots, x_{2n}; p_1, \dots, p_m) &= {\cal N}_1 \,
\mathbb{E}\bigg[ \prod_{i=2}^{2n} |g_t'(x_i)|^{h_\kappa} \prod_{j=1}^{m} |g_t'(p_j)|^{h_j}
Z_{\mathbb H} \big[\xi_1(t), g_t(x_2), \dots, g_t(x_{2n}); g_t(p_1), \dots, g_t(p_m)\big] \bigg].
\end{align}
Notice that the singular derivative $g_t'(x_1)$ has canceled between the numerator and the denominator. The last equation is valid at any time $t$. We now substitute this into Eq. (\ref{t-independence}) (taking into account that the normalization ${\cal N}_1$ is $t$-independent):
\begin{align}
\partial_t \mathbb{E}\bigg[ \prod_{i=2}^{2n} |g_t'(x_i)|^{h_\kappa} \prod_{j=1}^{m} |g_t'(p_j)|^{h_j}
Z_{\mathbb H} \big[\xi_1(t), g_t(x_2), \dots, g_t(x_{2n}); g_t(p_1), \dots, g_t(p_m)\big] \bigg] = 0.
\label{martingale}
\end{align}
Conditions of this type often appear in stochastic analysis when one studies special types of stochastic processes called martingales.\cite{Oksendal, Klebaner} Roughly speaking,  a martingale is a stochastic process  $M(t)$ whose expectation value is constant in time: $\partial_t \mathbb{E}[M(t)] = 0$. For this reason we call consequences of Eq. (\ref{martingale}) derived below martingale conditions on partition functions.

Now we set $t=0$ in (\ref{martingale}). It is then straightforward to use the stochastic equations (\ref{multipleSLE}) and (\ref{noforcing}) and Ito's formula\cite{Oksendal, Klebaner} to transform this equation into a Fokker-Planck equation:
\begin{align}
\bigg[\frac{\kappa}{2} \partial_{x_1}^2
- 2 \sum_{i=2}^{2n} \bigg(\frac{h_\kappa}{(x_i - x_1)^2} - \frac{1}{x_i - x_1} \partial_{x_i}\bigg)
- 2 \sum_{j=1}^m \bigg( \frac{h_j}{(p_j - x_1)^2} - \frac{1}{p_j - x_1} \partial_{p_j} \bigg)  \bigg]
Z_{\mathbb H}(x_1, \dots, x_{2n}; p_1, \dots, p_m) = 0.
\end{align}
This equation was derived using evolution with respect to $t_1$ only. Similar equations result when we use other times $t_i$, so in the end we get $2n$ martingale conditions
\begin{align}
\bigg[\frac{\kappa}{2} \partial_{x_i}^2
- 2 \sum_{j \neq i}^{2n} \bigg(\frac{h_\kappa}{(x_j - x_i)^2} - \frac{1}{x_j - x_i} \partial_{x_j} \bigg)
- 2\sum_{k=1}^m \bigg( \frac{h_k}{(p_k - x_i)^2} - \frac{1}{p_k - x_i} \partial_{p_k} \bigg) \bigg]
Z_{\mathbb H}(x_1, \dots, x_{2n}; p_1, \dots, p_m) = 0,
\label{FP-equations}
\end{align}
which the partition function $Z_{\mathbb H}$ must satisfy in addition to the conditions of conformal invariance (\ref{SL2R}).

The Fokker-Planck equations can be written in a compact way using tyhe following differential operators:
\begin{align}
\L_{-m}(x_i) = \sum_{j \neq i}^{2n} \bigg(\frac{(m-1) h_\kappa}{(x_j - x_i)^m} - \frac{1}{(x_j - x_i)^{m-1}} \partial_{x_j} \bigg)
+ \sum_{k=1}^m \bigg( \frac{(m-1) h_k}{(p_k - x_i)^m} - \frac{1}{(p_k - x_i)^{m-1}} \partial_{p_k} \bigg).
\end{align}
\end{widetext}
Notice that each operator insertion except $x_i$ contributes a term to ${\cal L}_{-m}(x_i)$. The translational invariance condition $L_{-1} Z_{\mathbb H} = 0$ can be written as
\begin{align}
\partial_{x_i} Z_{\mathbb H} = {\cal L}_{-1}(x_i) Z_{\mathbb H}.
\end{align}
Thus, trading the derivative $\partial_{x_i}$ for ${\cal L}_{-1}(x_i)$, the Fokker-Planck equations (\ref{FP-equations}) can be written as
\begin{align}
\Big( \frac{\kappa}{2} {\cal L}_{-1}^2(x_i) - 2 {\cal L}_{-2}(x_i) \Big) Z_{\mathbb H}(x_1, \dots, x_{2n}; p_1, \dots, p_m) = 0.
\label{extraconditions}
\end{align}

\section{Conformal restriction and CFT in the Coulomb gas formalism}
\label{sec:CFT}

In the following Sec. \ref{sec:calculation1} we will make use of conformal restriction to obtain certain information on the transport behavior of the system. The development can be cast solely in the language of conformal restriction, however, we shall make use of ideas which are already well developed in the language of CFT, so we prefer to mix the two approaches in the presentation. In this section, we set up notations related to CFT in the so-called Coulomb gas formalism.\cite{Dotsenko-Fateev} To find more details the reader may consult the Refs. \onlinecite{YellowBook} and \onlinecite{BB-review, IAG-review, RBGW07}.

As we have already mentioned, from a field theory perspective restriction models are described by conformal field theory (CFT) with vanishing central charge, $c=0$. In the Coulomb gas formalism, correlations function of CFT are computed by a certain ansatz. The CFT correlation function is replaced by a correlation function for the Gaussian free field, with a possible introduction of certain additional non-local operators, called screening charges. We shall describe the procedure briefly below.

Consider a Gaussian free field, namely a fluctuating bosonic field with the action:
\begin{align}\label{GFFAction}
S = \frac{1}{8\pi} \int \! d^2 r \big(\nabla \varphi \big)^2.
\end{align}
This action describes a CFT with $c=1$. We can modify the central charge to $c<1$ by introducing the so called background charge $-2\alpha_0$, where $\alpha_0$ is related to the central charge as follows:
\begin{align}
c = 1 - 24 \alpha_0^2.
\label{c}
\end{align}
Next, consider vertex operators, namely, the exponentials of the free field
\begin{align}\label{vertexOperators}
V_\alpha(z,\bar{z}) = e^{ i \sqrt{2} \alpha \varphi(z,\bar{z})},
\end{align}
where the parameter $\alpha$ is called the Coulomb charge of the operator.
In the CFT with the background charge the vertex operator $V_\alpha$ has the conformal weight
\begin{align}
h(\alpha) = \alpha (\alpha - 2\alpha_0).
\label{h-alpha}
\end{align}
The last ingredient we need are the screening operators defined as:
\begin{align}\label{ScreeningOperators}
Q_{\pm}  = \int \! dz d\bar{z} \, V_{\alpha_{\pm}}(z,\bar{z}),
\end{align}
where $\alpha_{\pm}$ are the positive and negative solutions of  $h(\alpha_{\pm})=1$:
\begin{align}
\alpha_\pm = \alpha_0 \pm \sqrt{\alpha_0^2 + 1}.
\end{align}

With these ingredients the recipe to compute a CFT correlation function of primary operators $O_{h_i}(z_i,\bar{z})$ using Coulomb gas is (roughly) given by:
\begin{align}\label{ansatz}
&\< O_{h_1}(z_1,\bar{z}_1)  O_{h_2}(z_2,\bar{z}_2) \dots  O_{h_N}(z_N,\bar{z}_N)  \>_{\mathrm{CFT}} =  \\
&\< V_{\alpha_1} (z_1, \bar{z}_1) V_{\alpha_2} (z_2, \bar{z}_2) \dots V_{\alpha_N} (z_N, \bar{z}_N) Q_+^{m_+} Q_-^{m_-} \>_{\mathrm{GFF}} \nonumber.
\end{align}
The subscript GFF indicates that the correlation function is computed with a weight $e^{-S}$ with action (\ref{GFFAction}), while the CFT subscript indicates a correlation function in a conformal field theory. The charges $\alpha_i$ are chosen such that $h(\alpha_i)= h_i$. The number of insertions, $m_{\pm}$, of the screening operators is arbitrary, but a constraint which reads
\begin{align}
m_+ \alpha_+ + m_- \alpha_- + \sum_i \alpha_i = 2 \alpha_0,
\end{align}
must be satisfied. This recipe can be motivated in different ways, but as far as we know cannot be justified fully and rigorously.

A special place is given to operators whose charges $\alpha_{m,n}$ are given by
\begin{align}
\label{KacCharges}
\alpha_{r,s}  = \frac{1}{2}(1-r)\alpha_+ + \frac{1}{2}(1-s)\alpha_-.
\end{align}
The primary operators corresponding to these charges are denoted by $\psi_{r,s}$. $r$ and $s$ are the numbers of screening operator insertions $Q_+$ and $Q_-$ necessary to compute two-point correlation functions of $\psi_{r,s}$. The operators, $\psi_{r,s}$ have a special role in the representation theory of conformal symmetry (the Virasoro algebra), particularly when $r$ and $s$ are positive integers. In this case correlation functions with insertions of $\psi_{r,s}$ satisfy differential equations of order $rs$. The relation (\ref{KacCharges}) is called the `Kac table', originally meant to be used only when $r$ are $s$ are positive integers, but often is extended to include all integer and even half integer indices. We shall often make use of this (extended) Kac table parametrization.

We may use (\ref{KacCharges}) as a convenient parametrization of the primary operators appearing in the theory. Every primary operator has a conformal weight $h$. With this conformal weight, a conformal charge $\alpha$ may be associated, which satisfies $h(\alpha) = h$. In fact, for a given value of $\alpha_0$, the Eq. (\ref{h-alpha}) for a conformal weight has two solutions for the Coulomb charge:
\begin{align}
\alpha = \alpha_0 \pm \sqrt{\alpha_0^2 + h}.
\label{alpha-h}
\end{align}
These charges can always be written as $\alpha = \alpha_{r,s}$ for some choice of numbers  $r$ and $s$, not necessarily positive integers.

The parametrization (\ref{KacCharges}) has several advantages. First, the conformal charge $\alpha$ naturally appears in formulas, and turns out to be a particularly convenient parametrization. Second, the parametrization is widely used in CFT literature. And third, if $\alpha$ happens to be given by $\alpha_{r,s}$ for some positive integers $r$ and $s$, there is some chance that the operator will have special properties, closely related to the representation theory of CFT, so keeping track of $r$ and $s$ is generally a good idea.

Notice that given a central charge $c$, Eq. (\ref{c}) has two solutions for the possible background charge: $\alpha_0 = \pm \sqrt{(1-c)/24}$. In the SLE language, these correspond to two dual values of $\kappa$ describing distinct ``phases'' of SLE curves: ``dilute'' ($\kappa < 4$) and ``dense'' ($\kappa > 4$).\cite{Dupa-review} A convenient choice of parametrization\cite{RBGW07} is such where
\begin{align}
2\alpha_0(\kappa) &= \frac{\sqrt\kappa}{2} - \frac{2}{\sqrt\kappa}, \\
\alpha_+(\kappa) &= \frac{\sqrt\kappa}{2}, & \alpha_-(\kappa) &= - \frac{2}{\sqrt\kappa}.
\end{align}
This value of $\alpha_0$ is positive in the dense phase and negative in the dilute phase. In terms of $\kappa$, the general Kac table charges and weights become
\begin{align}
\alpha_{r,s}(\kappa) & = \frac{4s - \kappa r + \kappa - 4}{4\sqrt{\kappa}}, \\
h_{r,s}(\kappa) &= \frac{(\kappa r - 4s)^2 - (\kappa - 4)^2}{16\kappa}.
\end{align}

The models we are interested in are related to conformal restriction, and the central charge for them is $c=0$. The two values of $\kappa$ that correspond to it are $\kappa = 8/3$, related to self-avoiding walks, and $\kappa = 6$ related to percolation. Since the boundaries of restriction measures are always simple (``dilute'') curves SLE($8/3,\rho$), in order to describe them we choose $\kappa = 8/3$ and
\begin{align}
\alpha_0\big(\tfrac{8}{3}\big) &= - \frac{1}{2\sqrt{6}}, &
\alpha_+\big(\tfrac{8}{3}\big) &= \sqrt{\frac{2}{3}}, & \alpha_-\big(\tfrac{8}{3}\big) &= -\sqrt{\frac{3}{2}}.
\end{align}
With this choice, the charges and weights of operators appearing in the Kac table are
\begin{align}
\alpha_{r,s}\big(\tfrac{8}{3}\big) &= \frac{3s - 2r - 1}{2\sqrt{6}}, \\
h_{r,s}\big(\tfrac{8}{3}\big) &= \frac{(2r - 3s)^2 - 1}{24}.
\label{h-8/3}
\end{align}
On the other hand, to describe percolation hulls that in the continuum are \SLE{6} curves, we need to choose $\kappa = 6$ and
\begin{align}
\alpha_0(6) &= \frac{1}{2\sqrt{6}}, &
\alpha_+(6) &= \sqrt{\frac{3}{2}}, & \alpha_-(6) &= -\sqrt{\frac{2}{3}}.
\end{align}
In this case the general Kac charges and weights are
\begin{align}
\alpha_{r,s}(6) &= \frac{2s - 3r + 1}{2\sqrt{6}}, \\
h_{r,s}(6) & = \frac{(3r - 2s)^2 - 1}{24}.
\label{h-6}
\end{align}

The conformal covariance properties of the partition functions given in Eqs. (\ref{partitionConformalInv}) and (\ref{SL2R}) imply that these are CFT correlation functions, where the points $p_i$ and $x_i$ (see Fig. \ref{TwoSLE}) correspond to the insertion of primary operators of conformal weights $h_\kappa$ for the SLE traces beginning at $x_i$ and $h_i$ for the points $p_i$. Moreover, the martingale conditions in the form of the second order differential equations (\ref{extraconditions}) imply that the operators inserted at points $x_i$ are degenerate at level two, to use the CFT language. Using the Kac table parametrization, these can be identified with $\psi_{1,2}$, as was demonstrated by Bauer and Bernard\cite{Bauer-Bernard-2003} for an arbitrary value of $\kappa$. Here we provide an interpretation of this operator in the CFT/Coulomb gas language for $c=0$. In this case, the operator has conformal weight $5/8$ and the corresponding Coulomb charge $\alpha = \sqrt{3/8}$ can be written as $\alpha_{1,2}(8/3)$. Thus, the operator creating an \SLE{8/3} trace is $\psi_{1,2}$.

The CFT interpretation of \SLEr{\kappa}{\rho}, which is anchored at a point $x$ on the boundary, is the insertion of the operator $\psi_{1,2}$ at $x$ and another operator $O_{h'}$ of conformal dimension $h'$ at $x-0^+$. The relation between $h'$ and $\rho$ is given by Eq. (\ref{conformalweightofrho}). Comparing this equation with (\ref{h-alpha}) for $\alpha_0(8/3) = -1/(2\sqrt{6})$ ($c=0$), we see that
\begin{align}
\alpha' = \sqrt{\frac{3}{32}} \, \rho
\end{align}
is a Coulomb charge representing $O_{h'}$. We have two operators  close together at $x$, $\psi_{1,2}$, and $O_{h'}$. These in fact fuse together to produce another primary operator $O_h(x)$, whose conformal weight depends on $\rho$ as in Eq. (\ref{hfunctionofrho}). A conformal charge consistent with $h$ appearing in (\ref{hfunctionofrho}) is given by
\begin{align}
\alpha =  \alpha_{1,2}(8/3) + \alpha'.
\end{align}
This last formula can be read as follows: an operator of charge $\alpha_{1,2}(8/3)$ (which creates the SLE trace) is fused with an operator of charge $\alpha'$ (this is the operator $O_{h'}$) to produce an operator $O_h$ whose charge is given by the simple sum of the two operators which were fused to produce it. This is in fact no surprise, as charges  add up in the  Gaussian free field formulation of CFT. $O_h$ may then be called the {\it simple} fusion product of $O_{h'}$ and $\psi_{1,2}$.

We can also consider fusions of multiple $\psi_{1,2}$. These may lead to SLE lines that form small loops touching the boundary, but the simple fusion where the charges add leads to multiple SLE lines starting at the same point and going to infinity of other distant points.\cite{BB-review, RBGW07, BBK2005} These are created by insertions of the so-called $n$-leg operators $\psi_{1,n+1}$. For $\kappa = 8/3$ these operators create $n$ SAWs and have dimensions
\begin{align}
h_{1,n+1}\big(\tfrac{8}{3}\big) &= \frac{n(3n+2)}{8}.
\end{align}
For $\kappa = 6$ they create $n$ percolation hulls, and their dimensions are
\begin{align}
h_{1,n+1}(6) &= \frac{n(n-1)}{6}.
\end{align}

Multi-leg operators can also be defined in the bulk. It is known\cite{BB-review, RBGW07} that the bulk $n$-leg operator is $\psi_{0,n/2}$ in the extended Kac table. The conformal weights of these operators for the two values of $\kappa$ relevant for $c=0$ are
\begin{align}
h_{0,n/2}\big(\tfrac{8}{3}\big) &= \frac{9n^2 - 4}{96}, &
h_{0,n/2}(6) &= \frac{n^2 - 1}{24}.
\label{n-leg-bulk}
\end{align}

\section{Point contact conductances in the simplest settings}
\label{sec:calculation1}

In this section we will use the relation (\ref{g=Z}) between PCCs at critical points of the disordered systems considered above and conformal restriction measures to compute PCCs in the simplest settings. These are the settings where the PCC in question is either itself a two- or three-point function of primary boundary operators, or related to such functions in a simple way. All such two- and three-point functions are essentially fixed by global conformal invariance (see, for example, Ref. \onlinecite{YellowBook}). For convenience, we reproduce the standard argument in the language of conformal restriction, and constraints imposed on partition functions.

\begin{figure}[t]
\centering
\includegraphics[width=0.47\columnwidth]{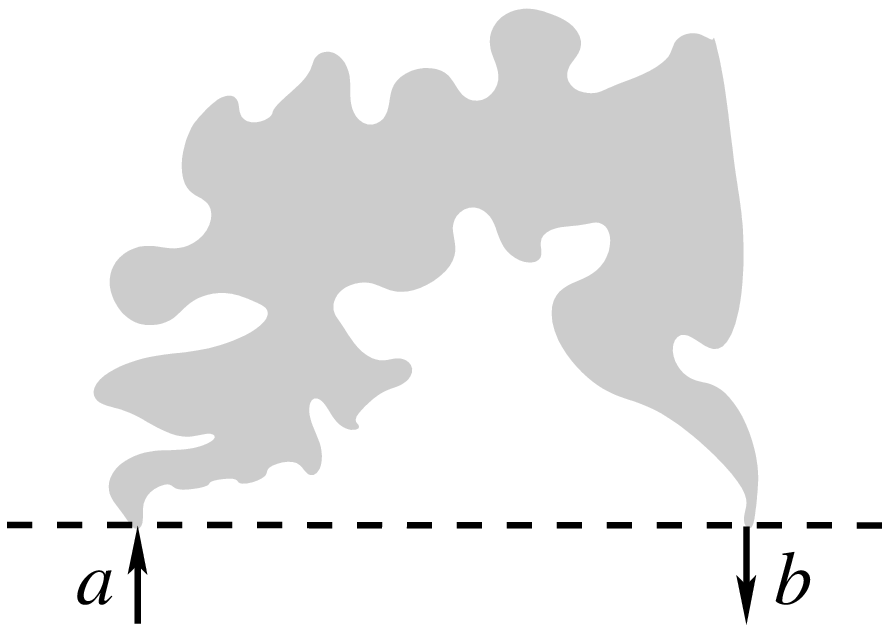}
\hfill
\includegraphics[width=0.47\columnwidth]{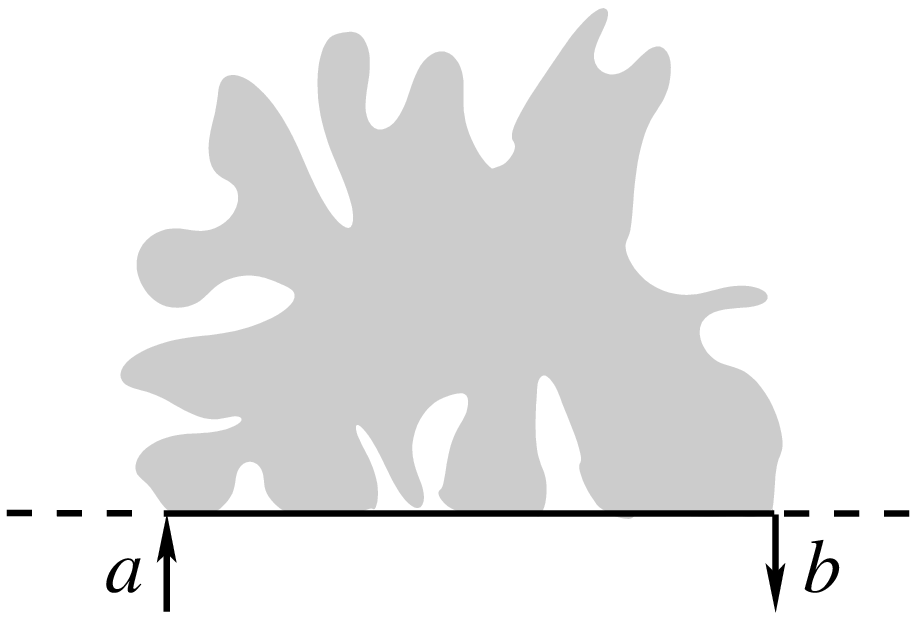}
\caption{
Two-point point contact conductance. Left: contacts are placed on the absorbing boundary. Right: contacts are placed at juxtapositions of the absorbing and a reflecting boundary.}
\label{2point-Fig}
\end{figure}

The simplest quantity one can compute is the average two-point conductance $\langle g(a,b) \rangle$ between points $a$ and $b$ on the straight absorbing boundary of a critical system occupying the upper half plane. We have already seen that this conductance is the same as the partition function $Z_{\mathbb{H}}(a,b)$. In CFT language, these  are primary (boundary) operators with scaling dimensions $h_A$. This immediately implies that the same quantity is given by a correlation functions of currents, $\langle g(a,b) \rangle = \<j_a j_b\>$, where the operator $j_a$ injects current into the system through a link at the point $a$ and $j_b$ extracts the current through a link at the point $b$ (see the left panel in Fig. \ref{2point-Fig}).
\begin{align}
\label{2-point-g-absorbing}
\langle g(a,b) \rangle = \frac{C}{|a-b|^{2h_A}}.
\end{align}

We will argue below that for all systems that we consider $h_A = 1$ is the weight of the conserved current operator. At the same time, the constant $C$, while universal upon fixing the normalization of the current operators in the continuum, does depend on the particular critical point. This constant is related to the critical longitudinal conductivity $\sigma_{xx}$, as can be seen from the following consideration. We can integrate the expression (\ref{2-point-g-absorbing}) over $a \in [1, R]$ and $b \in [-R, -1]$ to obtain a two-probe conductance
\begin{align}
G &= \int_1^R da \int_{-R}^{-1} db \frac{C}{|a-b|^{2}} = C \ln \frac{(R + 1)^2}{4 R}.
\end{align}
By conformal invariance, this is the same as the two-probe conductance of a rectangle of length $L$ and width $W$ that can be obtained from the upper half plane by a conformal map by an elliptic integral. In the limit $R \gg 1$ we get $W \gg L$, and the conformal map reduces essentially to the logarithmic map $w = \ln z$ to an infinite strip of width $\pi$, with metallic contacts of width $W = \ln R$ placed along the boundaries at Im$\,w = 0$ and Im$\,w = \pi$, at the distance $L = \pi$ apart. In this limit the conductance $G$ becomes
\begin{align}
G \approx C \ln R = \sigma_{xx} \frac{W}{L},
\end{align}
which is the Ohm's law with
\begin{align}
\sigma_{xx} = \pi C.
\end{align}
Thus, as we have claimed, the universal constant $C$ in the PCC (\ref{2-point-g-absorbing}) is related to the critical conductivity $\sigma_{xx}$ which depends on a particular critical system. For the SHQ transition, for example, this conductivity is know exactly to be equal to $\sigma_{xx} = \sqrt{3}/2$ (in natural units). \cite{Cardy00} For the IQH critical point this conductivity is not known analytically, and for the classical limit of the CC model it can take any value in the range $[0, 1/2]$ [see Eq. (\ref{conductivities-cassical-CC})].

Now let us illustrate how the result (\ref{2-point-g-absorbing}) can be understood in terms of pictures as samples of a restriction measure, and their SLE boundaries. The pictures that contribute to $g(a,b)$ include two lines, the inner and outer boundaries of the picture, both lines start at $a$ and end at $b$ (see the left panel of Fig.  \ref{2point-Fig}). To define a partition function for the process described by Eqs. (\ref{multipleSLE}) and (\ref{SLEforcing}) that creates those lines, we must split point $a$ and point $b$ each into three points $a^-$, $a^0$ and $a^+$, $b^-$, $b^0$ and $b^+$. $a^-$ is now the origin of the outer line, $a^+$ is the origin of the inner line and at $a^0$ we place an operator with a proper weight such that after fusing $a^-$, $a^0$  and $a^+$, we get an operator of a weight corresponding to $j$. The same is done at $b$. We end up with a partition function $Z_{\mathbb H}(a^0,b^0;a^-,b^-,a^+,b^+)$. The fused partition function, $Z_{\mathbb H}(a,b)$, obtained when all the points associated with $a$ are fused and all the points associated with $b$ are fused, gives $\langle g(a,b) \rangle = Z_{\mathbb H}(a,b)$. The fused partition function also satisfies (\ref{SL2R}), and these three conditions are sufficient to specify it completely up to a constant factor, and we get back to Eq. (\ref{2-point-g-absorbing}). The same arguments apply to the average PCC between two contacts placed at juxtapositions of the absorbing boundary and a reflecting boundary (see the right panel in Fig. \ref{2point-Fig}). Depending on whether the reflecting segment of the boundary (illustrated by the solid line) is ``left'' or ``right'' (see Sec. \ref{subsec:critical-curves} for definitions), we get
\begin{align}
\label{2-point-g-mixed}
\langle g(a,b) \rangle &= \frac{C}{|a-b|^{2h_{LA}}} & \text{or} && \frac{C}{|a-b|^{2h_{RA}}}.
\end{align}

\begin{figure}[t]
\centering
\includegraphics[width=0.45\columnwidth]{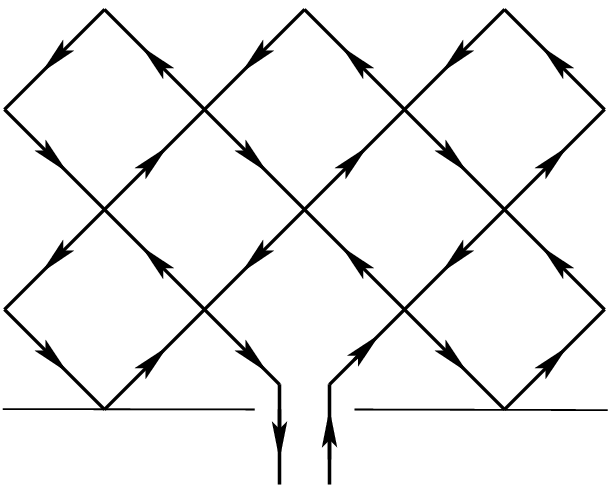}
\hfill
\includegraphics[width=0.45\columnwidth]{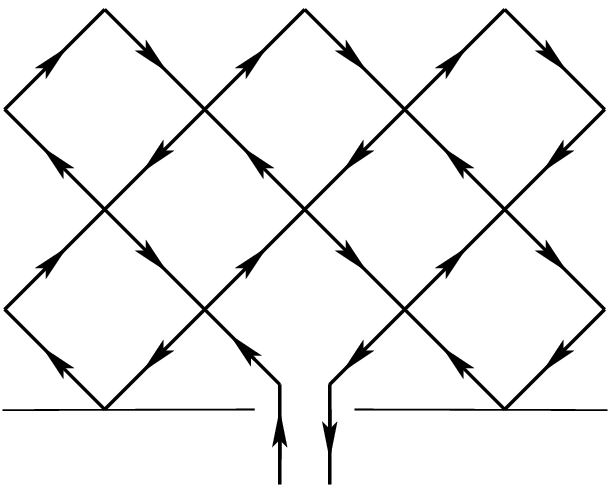}\\
\vskip 2mm
\includegraphics[width=0.45\columnwidth]{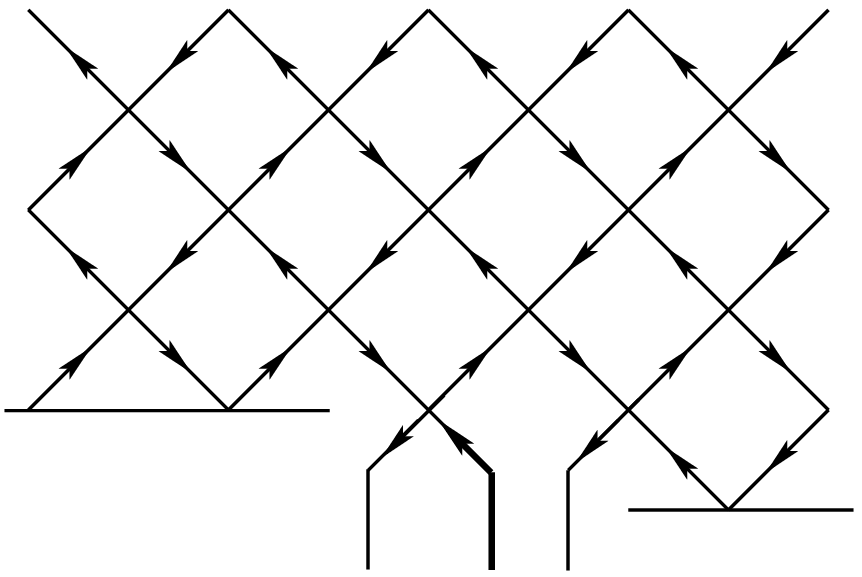}
\hfill
\includegraphics[width=0.45\columnwidth]{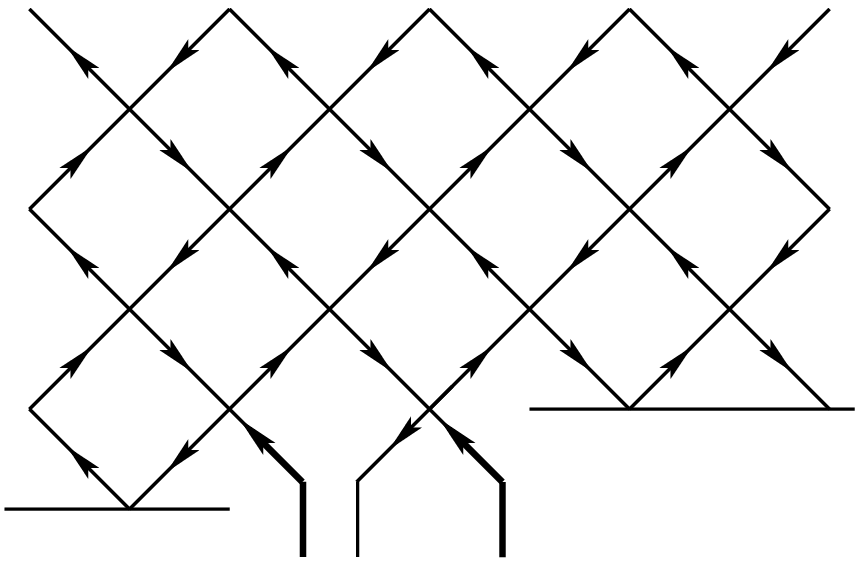}
\caption{Top: Point contacts placed at reflecting boundaries through small openings. The dimensions of current insertions at these contacts are denoted by $h_R$ and $h_L$. Bottom: Point contacts placed at small openings between reflecting boundaries of opposite chirality. The corresponding dimensions are $h_{RL}$ and $h_{LR}$.}
\label{fig:PC-reflecting}
\end{figure}

We can also consider point contacts placed at small openings in reflecting boundaries (see the top panel in Fig. \ref{fig:PC-reflecting}). We denote the dimensions of current insertions through such contacts by $h_R$ and $h_L$. The two situations shown on the top in Fig. \ref{fig:PC-reflecting} are related by reflection across the vertical line. Such reflection should not change the dimensions of the current insertions, so we expect these dimensions to be equal:
\begin{align}
h_R = h_L.
\label{hR=hL}
\end{align}
Then the two point PCCs between such contacts is
\begin{align}
\label{2-point-g-reflecting}
\langle g(a,b) \rangle &= \frac{C}{|a-b|^{2h_R}}.
\end{align}
Point contacts can be also placed at small openings between reflecting boundaries with the opposite chirality (see the bottom part of Fig. \ref{fig:PC-reflecting}). Injecting current through the middle (incoming) link on the left figure inserts an operator with dimension $h_{RL}$. In this situation there is no symmetry relating $h_{RL}$ and $h_{LR}$, so we expect these dimensions to be different in general.

We can consider slightly more complicated setups without any additional input. Indeed, Eqs. (\ref{SL2R}) contain three independent conditions which determine all three point functions up to a constant. The three point function of primary operators of dimensions $h_a, h_b, h_c$ inserted at points $a,b,c$ has the form:
\begin{align}
\frac{C}{|a-b|^{h_a+h_b-h_c} |a-c|^{h_a + h_c - h_b} |b-c|^{h_b+h_c-h_a}}.
\end{align}

\begin{figure}[t]
\centering
\includegraphics[width=0.47\columnwidth]{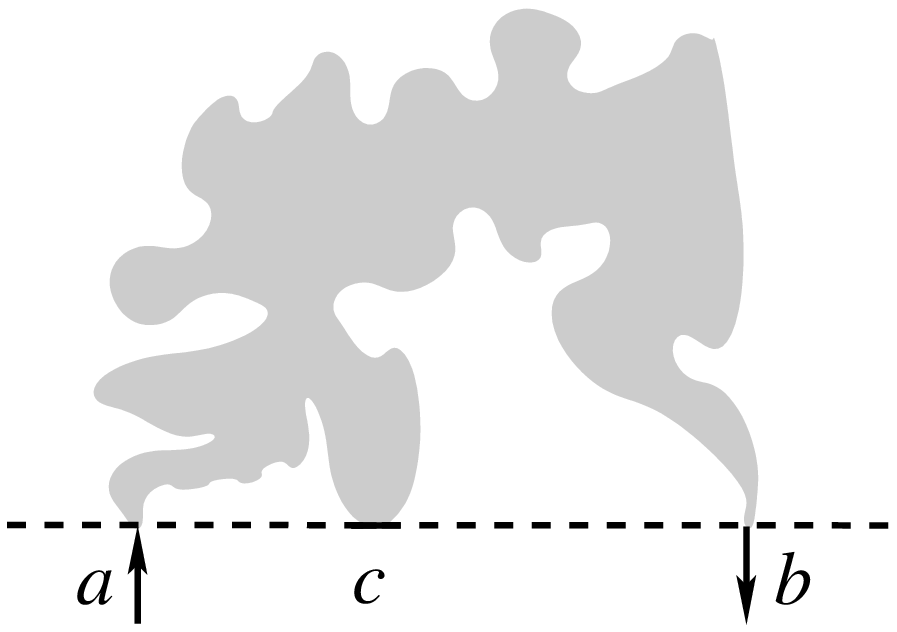}
\hfill
\includegraphics[width=0.47\columnwidth]{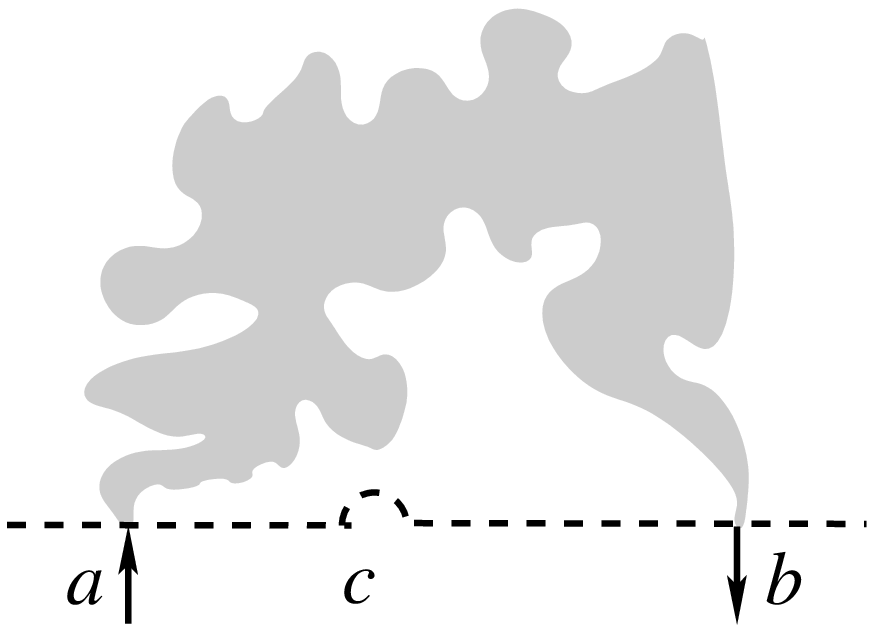}
\caption{A three point function. Left: the change $\delta \langle g(a,b) \rangle$ in the two-point conductance is positive when a small segment of reflecting boundary of length $\epsilon$ is inserted near the point $c$. Right: $\delta \langle g(a,b) \rangle$ is negative when a small bump of radius $\epsilon$ on the absorbing boundary is inserted near the point $c$.}
\label{fig:g-change}
\end{figure}

As a physical example we can consider the following setup shown in the left panel of Fig. \ref{fig:g-change}. It shows a critical system whose boundary is mostly absorbing but with a small insertion (of length $\epsilon$) of a reflecting boundary near the point $c$. This insertion makes possible for a picture contributing to $\langle g(a,b) \rangle$ to touch this small reflecting boundary segment, thereby increasing the overall conductance. The difference between the conductance in this case and the one without the insertion is then represented by pictures that necessarily touch the reflecting segment. This corresponds to a three point function of two current insertion operators (of dimension $h_A$) and one operator that ``forces'' the picture to ``touch'' the point $c$. Denoting the dimension of this operator by $h_T$, we have
\begin{align}
\delta \langle g(a,b) \rangle = \frac{C \epsilon^{h_T}}{|a-b|^{2h_A - h_T} |a-c|^{h_T} |b-c|^{h_T}}.
\label{delta-g}
\end{align}

Similarly, we can consider the situation where the absorbing boundary is slightly moved into the system, forming a small semicircular bump of radius $\epsilon$ centered at $c$. This is shown in the right panel of Fig. \ref{fig:g-change}. In this case the conductance $\langle g(a,b) \rangle$ is reduced by essentially the same amount [Eq. (\ref{delta-g})] since now the pictures that were going through the semicircle (the same as the picture on the left panel of Fig. \ref{fig:g-change}) do not contribute anymore.

\begin{figure}[t]
\centering
\includegraphics[width=0.8\columnwidth]{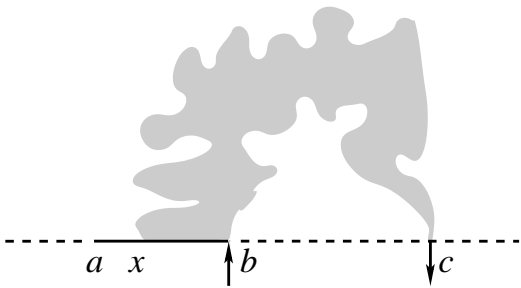}
\caption{Feynman picture drawn in gray for current inserted at an interface between reflecting and absorbing boundaries (the reflecting boundary, represented by a heavy line, occupies the segment $[a,b]$, the dotted line represents absorbing boundaries). The point $x$ denotes the location at which the diagram lifts-off the real axis.}
\label{GrayFeymanns}
\end{figure}

Another situation, shown in Fig. (\ref{GrayFeymanns}), leads to a three-point function. Here a reflecting boundary extends from $a$ to $b$, and the rest of the boundary is absorbing. The current is injected into the sample at the point $b$ and extracted through a point $c$ on the absorbing boundary. The weight of the operator at $c$ is  $h_A$.  The weight $h_b$ of the operator at $b$ is either $h_{LA}$ or $h_{RA}$ , and depends on the particular situation. Point $a$ is a point at which the boundary conditions change, and {\it a priori} we do not know if it may be described by a primary operator. We do know however that for every picture (the gray area on Fig. \ref{GrayFeymanns}) there is a point $x$ between $a$ and $b$ at which the picture will lift off the real axis, never to visit the real axis again before exiting at $c$. For all pictures with a given point $x$, the boundary to the left of $x$ serves as the absorbing boundary (microscopically, a shift of one lattice spacing into the bulk is necessary). The fact that the boundary conditions to the left of $x$ are effectively absorbing allows us to identify $x$ as a point anchoring a one sided restriction measure. Indeed, the left boundary of the picture is anchored at $x$ and obeys restriction with respects to sets placed on the absorbing boundary to the left of $x$. Denoting the dimension of the operator at the point $x$ by $h_l$, the contribution $\langle g(b,c;a,x) \rangle$ of the pictures with a given $x$ to the conductance $\langle g(b,c;a) \rangle$ is given by the three-point function
\begin{align}
\frac{C}{|x-b|^{h_l + h_b - h_A}|x-c|^{h_l + h_A - h_b}|b-c|^{h_b + h_A - h_l}}
\end{align}
Summing (integrating) over all lift-off points $x$ we get
\begin{align}
\langle g(b,c;a) \rangle = \int_a^b \! dx \, \langle g(b,c;a,x) \rangle.
\label{g-integral}
\end{align}
While for arbitrary weights $h_l, h_A, h_b$ this integral can be expressed in terms of a hypergeometric function, the actual final expression [see Eq. (\ref{g-reflecting-segment})] is quite simple due to the special values $h_A = h_l = 1$.

In the next subsections will will discuss the weights $h_A, h_T, h_l$, and $h_{L,R}$ in the models that we consider in this paper. We will argue that the weights $h_A = 1$, $h_T = 2$, and $h_l = 1$ are {\it superuniversal} and do not depend on a particular model and its symmetry class. We will exhibit other operators with superuniversal weights, and will give explicit constructions of these for specific systems of our interest. Other weights ($h_{L,R}$) will also be discussed for each system separately.

\subsection{Superuniversal weights}
\label{subsec:superuniversal}

In this section we show that the weights of some operators that have appeared in the previous sections are superuniversal, that is, they do not depend on the symmetry class of a particular disordered system, but only on the conformal restriction property.

The first example is the operator that injects current through an absorbing boundary. Its dimension $h_A = 1$ is superuniversal, which is related to the fact that we are describing a critical conductor with a finite average conductivity and various conductances. Indeed, suppose two leads are connected to the conductor: one lead connects to a segment $[a,b]$ of the boundary, while the other connects to the segment $[c,d]$. The conductance between the leads is given by the Kubo (Landauer) formula
\begin{align}
\< g \>  = \sum_{i\in [a,b]} \sum_{j\in [c,d]} \<j_i j_j\>,
\end{align}
where $i \in [a,b]$, for example, denotes that site $i$ belongs to segment $[a,b]$ . In the continuum limit this becomes
\begin{align}
\< g \>  = \epsilon^{2 h_A-2} \int_{[a,b]} \int_{[c,d]} \! dx dy \, \<j(x) j(y)\>,
\end{align}
where $h_A$ is the conformal weight of $j$, and $\epsilon$ is the lattice spacing. For this expression to remain finite (neither zero, nor infinity) as expected for a critical conductor, the weight $h_A$ must be equal to 1. If $h_A > 1$ we will have an insulator as $\epsilon \to 0$, while if $h_A < 1$ we will have a superconductor.

Another way to arrive at the same conclusion is to consider the total current through a segment $I = \int_a^b dx j(x)$. This total current should be a conformally-invariant object of dimension 0. But if we perform a conformal transformation that takes $x$ to $f(x)$, the current transforms to
\begin{align}
I = \int_{f(a)}^{f(b)} \!  df \, |f'(x)|^{h_A - 1} j(f).
\end{align}
The conformal invariance of $I$ implies $h_A = 1$.

The second example of a superuniversal operator appears in the situations shown in Fig. \ref{fig:g-change}. In these cases we consider diagrams in which current is forced to pass through a point $x$ on the boundary of the system. The boundary is absorbing around this point. A diagram forced to touch the point $x$ looks locally as a sample of a two-sided restriction measure, and the weight of this measure can be easily seen to be $2$. Indeed, to select only diagrams that pass through the point $x$ we introduce a small semicircle $A$  of radius $\epsilon$ centered at $x$ The difference between the overall weight of all diagrams and the weight of diagrams that avoid $A$ is proportional to the weight of the pictures that pass through $x$ (as $\epsilon$ tends to zero). The conformal restriction property allows us to compute the weight of the diagrams that avoid $A$ by effecting a conformal transformation that removes $A$ [see Eq. (\ref{transformation})]. Such transformation $f(z)$ that also fixes the points $a$ and $b$ can be easily written explicitly (see Appendix \ref{conformal-map}). Then, using Eq. (\ref{transformation}) we can find the change in the PCC after the deformation of the boundary to the first non-vanishing order in $\epsilon$ as
\begin{align}
\delta \langle g(a,b) \rangle &= \big(1 - |f'(a)|^{h_A} |f'(b)|^{h_A} \big) \langle g(a,b) \rangle \nonumber \\
&\approx \frac{2 h_A |b-a|^2}{|c-a|^2 |c-b|^2} \epsilon^2 \langle g(a,b) \rangle \nonumber \\
&= \frac{C \epsilon^2}{|a-b|^{2h_A - 2} |a-c|^{2} |b-c|^{2}} \nonumber \\
&= \frac{C \epsilon^2}{|a-c|^{2} |b-c|^{2}}.
\end{align}
This is exactly the equation (\ref{delta-g}) with $h_T = 2$, a universal value independent, in particular, of $h_A=1$.

The dimension $h_T = 2$ is consistent with the fact that in the CFT language the removal of the semicircle $A$ is effected by the insertion of the stress energy tensor $T$. The stress energy tensor $T$ has always dimension $2$, and for $c=0$ is a primary operator.

The third superuniversal weight is that of lift-off points: $h_l = 1$ (see Fig. \ref{GrayFeymanns}). This can be seen in the following way: the lift-off point must always be integrated over to obtain a physically measurable average conductance. The conductance is  conformal invariant. For example, to compute the overall contribution of diagrams in Fig. (\ref{GrayFeymanns}) we must take $ \<  \int_{a}^b dx \,  O_{h_l}(x) \dots \>$, where $O_{h_l}(x)$ is the operator which creates a lift-off at point $x$, and the ellipsis denotes other operators which must be inserted (in the example shown in Fig. \ref{GrayFeymanns}, the other operators are inserted at the points $b$ and $c$). Upon a M\"obius conformal transformation $f(z)$ that maps the upper half plane to the upper half plane, the integral $\int_{a}^b dx \,  O_{h_l}(x)$ transforms to
\begin{align}
\int_{f(a)}^{f(b)} \! df \, |f'(x)|^{h_l-1} O_{h_l}(f).
\end{align}
Conformal invariance of the PCC that is obtained from this expression implies that $h_l=1$.

Since the operator $O_{h_l}$ is of dimension one, its integral is also a primary operator, but of dimension\cite{footnoteNotationBC} $h_{BC} = 0$. This means that in Fig. \ref{GrayFeymanns}, instead of summing over $x$, we may simply place an operator of dimension zero at $a$. Thus, the operator that changes the boundary condition from absorbing to reflecting is a primary of dimension zero. Note that this result is completely general. This has been noticed and used before by Cardy\cite{Cardy1992} in the context of percolation to compute the crossing probability in a rectangle. But, for other systems with central charge $c=0$, to the best of our knowledge, the value $h_{BC} = 0$ was not reported before. This shows that the integral expressing the average conductance in Eq. (\ref{g-integral}) is simply given by the three point function of operators with weights $h_{BC}=0$, $h_b$, and $h_A = 1$:
\begin{align}
\langle g(b,c;a) \rangle = \frac{C}{|a-b|^{h_b - 1} |b-c|^{h_b+1}|a-c|^{1-h_b}}.
\label{g-reflecting-segment}
\end{align}

\begin{figure}[t]
\centering
\includegraphics[width=\columnwidth]{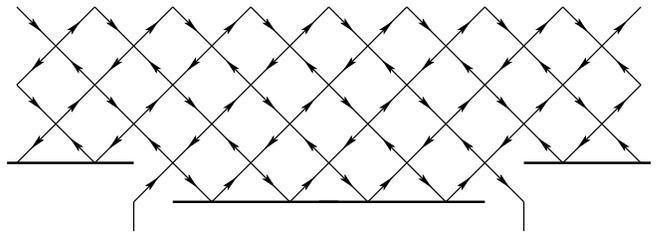}
\caption{Contacts placed at points where the two reflecting boundary conditions, left and right, are switched. The average PCC between these contacts does not depend on the distance between them due to the current conservation.}
\label{fig:dimension-zero}
\end{figure}

One more superuniversal weight $h_0$ is a special case of $h_{RL}$ (or $h_{LR}$) which is obtained when current is injected through the smallest possible opening between reflecting boundaries with the opposite chirality (see Fig. \ref{fig:dimension-zero}). In this situation the current cannot exit through the same opening, and the average PCC between two such contacts does not depend on the distance between them due to current conservation. Therefore, the dimension of such current insertion is zero: $h_0 = 0$.

We note that most of the results for superuniversal weights are consequences of conservation laws, and can be derived from Ward identities, such as those associated with the current operator or the stress energy tensor. However, some of the results for the superuniversal weights are stronger. For example, as we shall see in the next section, for critical percolation the operator that creates two percolation hulls at the absorbing boundary, namely $\psi_{1,5}$, has weight two. This is the superuniversal weight $h_T = 2$ of the stress energy tensor. Both operators locally correspond to the case that a unit flux of charge impinges on the boundary and then escapes to infinity. Since they are locally equivalent, they must have the same weight.
However, $\psi_{1,5}$ is not globally equivalent to the stress energy tensor. The difference between the operators becomes apparent at the macroscopic level: while the stress energy tensor only forces a hull to touch down on the absorbing boundary, solutions to the null vector equation associated with a $\psi_{1,5}$ operator may be selected in such a way as to obtain a certain topology of current flow through boundary points. Such global distinction is important, for example, in the proof of Watts' formula.\cite{Watts-formula}

\subsection{Weights of operators for the SQH effect}

We now turn to the Kac table classification of the operators appearing in the SQH effect problem. These are directly related to hulls of percolation clusters.\cite{Gruzberg99} The operator creating $n$ percolation hulls (multiple \SLE{6}) at a boundary (also called the boundary $n$-leg operator) is known\cite{Saleur-Bauer-1989} to be $\psi_{1,n+1}$ in the theory with $\kappa = 6$. According to Eq. (\ref{h-6}), its weight is
\begin{align}
h_{1,n+1}(6) &= \frac{n(n-1)}{6}.
\end{align}

Consider now the operator that injects current through a single link (a point contact) at the origin on the absorbing boundary. This creates a percolation hull that is not allowed to hit the boundary until it goes out through the other point contact. Consider the positive real axis. By known\cite{Aizenman:Duplantier:Aharony} percolation arguments this absorbing boundary can be replaced by a reflecting boundary plus insertion of an additional hull emanating from a point immediately to the right of the origin. This extra hull ``screens'' the current carrying hull from approaching the boundary. The same can be done for the negative real axis. This absorbing boundary can also be equivalently described by a reflecting boundary plus insertion of a hull, this time immediately to the left of the origin. The overall outcome is that we can think of current inserted on the absorbing boundary as three hulls inserted on a reflecting boundary. This is the three-leg operator $\psi_{1,4}$ with dimension $h_A = h_{1,4}(6) = 1$, the superuniversal value. This value is consistent with what is known rigorously\cite{LSW-conformal-restriction} about the \SLE{6} conditioned not to touch the boundary.

\begin{figure}[t]
\centering
\includegraphics[width=0.47\columnwidth]{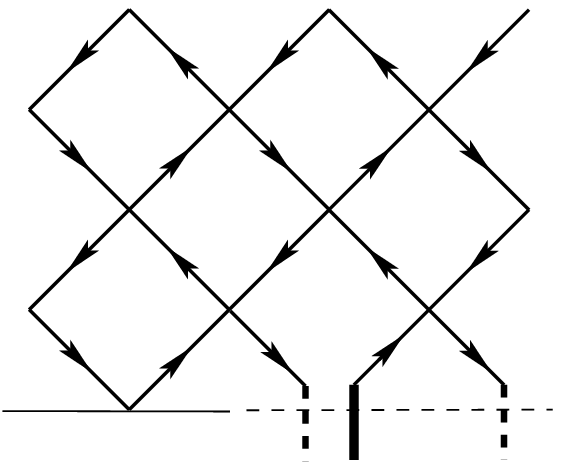}
\hfill
\includegraphics[width=0.47\columnwidth]{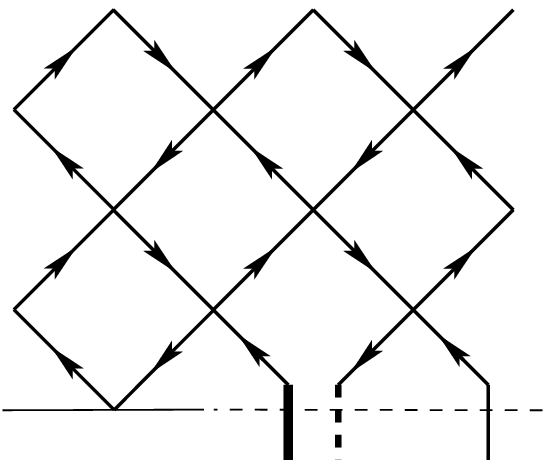}
\caption{Point contacts at the juxtapositions of the absorbing boundary with the right and left boundaries. The current is injected through the fat links. The dashed links represent the origins of ``screening'' percolation hulls, see the main text.}
\label{R-A and L-A contacts}
\end{figure}

Similarly, we can consider the operators that inject current at points of juxtaposition of a reflecting and the absorbing boundaries. These are shown in Fig.~\ref{R-A and L-A contacts} and have weights $h_{RA}$ and $h_{LA}$. Let the current be injected at the origin through links that are shown by fat lines. The absorbing boundary along the positive real axis can be replaced by a reflecting boundary plus a ``screening'' hull inserted immediately to the right of the origin (its origin is marked by a dashed link immediately to the right of the current insertion). The reflecting boundary conditions on the negative real axis are in fact different for $h_{RA}$ and $h_{LA}$. For $h_{LA}$ the negative real axis is simply the reflecting boundary in the percolation language. Then it is clear that the weight $h_{LA}$ is that of the two-leg operator in percolation, that is, $h_{LA} = h_{1,3}(6) = 1/3$. At the same time, for $h_{RA}$ the negative real axis is described by a reflecting boundary plus an insertion of an additional screening hull immediately to the left of the origin. This makes $h_{RA}$ the weight of the three-leg operator just like $h_A$: $h_{RA} = h_{1,4}(6) = 1$.

The dimensions for operator insertions at reflecting boundaries are easily seen to correspond to the two-hull operators: $h_R = h_L = h_{1,3}(6) = 1/3$. They are equal to each other, in accord with Eq. (\ref{hR=hL}). Similarly, it is easy to see that $h_{RL} = h_{1,4}(6) = 1$ and $h_{LR} = h_{1,3}(6) = 1/3$.

Other superuniversal weights are also easily confirmed for the SQH transition. Indeed, forcing a hull to go through a point on the absorbing boundary is equivalent to having two percolation hulls starting at this point, plus two screening hulls to screen the absorbing boundary. Thus, the relevant operator is the four-leg operator with $h_T = h_{1,5}(6) = 2$. For a lift-off point we have the two halves of the current-carrying hull that comes to this point and leaves it, plus one screening hull that ensures that the current does not touch the boundary again. This is the three-leg operator, and $h_l = h_{1,4}(6) = 1$. The switch between the absorbing and a reflecting boundary can be described by a single screening hull, and the corresponding weight is $h_{BC} = h_{1,2}(6) = 0$. The same one-leg operator injects current through a switch between the left and the right reflecting boundaries (see Fig. \ref{fig:dimension-zero}), and $h_0 = h_{1,2}(6) = 0$.

We comment here that the degeneracies between certain weights ($h_A = h_{RA} = h_{RL}$ and $h_{LA} = h_R = h_L = h_{LR}$) are a consequence of the {\it locality} property of percolation: even one ``absorbing'' link next to a point contact creates a screening hull.

\subsection{Weights of operators for classical CC}

Weights of all boundary operators that we have introduced can be found exactly for the classical CC model (diffusion in a magnetic field). In particular, this confirms the superuniversal weights. First consider injecting a current through the absorbing boundary. The dimension of such current insertion is $h_A = 1$. Indeed, since there are no reflecting boundaries around, the problem is equivalent to that of diffusion. It is easy to see, using Green's functions, that the fraction of diffusing particles that will reach a height $y$ from the real axis if they are released from the point $i\epsilon$ is $\epsilon/y$, the rest will be trapped by the absorbing boundary.  Since releasing the diffusing particles is equivalent to the insertion of current through the absorbing boundary (we assume that $\epsilon$ is the lower cutoff scale), and the $\epsilon$ dependence marks the weight of the operator, we indeed see that the weight for current insertion through the absorbing boundary is $h_A = 1$. The rigorous formulation of this result in terms of Brownian excursions and its proof are due to Virag.\cite{Virag2003}

Now let us consider lift-off points. A touch-off point is a point at which current arrives at a point on a reflecting boundary, immediately leaves it, and is conditioned never to touch the boundary again, say, to the left of the point at which it arrived. In the classical CC, due to lack of interference, one may separate the past of the diffusing particle from its future, relative to the moment it reached the lift-off point. The past and the future are actually two independent restriction measures, whose weights simply add up\cite{LSW-conformal-restriction} due to lack of interference. The weight of the future measure is $1 - \theta/\pi$ as in Eq. (\ref{his1minusthetaoverpi}). For the past measure we must apply time reversal in order to be able to apply Eq. (\ref{his1minusthetaoverpi}); this takes $\theta \to \pi - \theta$, which means that for this measure, $h = \theta/\pi$. The sum of the two weights is the weight of the touch of point $h_l=1$.

The simple decoupling (independence) of the past from the future also occurs if we force the current to pass through a boundary point. Here both the past and the future are current insertion (extraction) operators having the same weight $h_A = 1$. Summing up the weights of the two measures, we obtain that the operator in question has weight $h_T = 2$.

A general way of finding weights of current insertions at a boundary is to solve a boundary value problem for the electric potential $\phi$ that gives the conductance of a rectangular sample (Hall bar) of length $L$ and width $W$. For long samples with $L/W \gg 1$ the conductance is exponentially small:
\begin{align}
g \sim e^{- \pi h L/W},
\end{align}
where $h$ is the scaling dimension of the most relevant operator contributing to the conductance.
Details of this calculation are given in the Appendix \ref{sec:weights-calssical-CC}. Results are presented in Table \ref{DimensionsTable}, where we summarize the dimensions of various operators for the systems of our interest.

\begin{table}[t]
\begin{tabular}{|c|c|c|c|}
\hline  & IQH transition & SQH transition & Classical CC model \\
\hline $h_A$ & 1 & $h_{1,4}(6) = 1$ & 1  \\
\hline $h_T$ & 2 & $h_{1,5}(6) = 2$ & 2 \\
\hline $h_l$ & 1 & $h_{1,4}(6) = 1$ & 1 \\
\hline $h_{BC}$ & 0 & $h_{1,2}(6) = 0$ & 0 \\
\hline $h_{RA}$ & 0.8 & $h_{1,4}(6) = 1$ & $1/2 + \theta^H_R/\pi$  \\
\hline $h_{LA}$ & 0.32 & $h_{1,3}(6) = 1/3$ & $1/2 + \theta^H_L/\pi$  \\
\hline $h_R$ & --- & $h_{1,3}(6) = 1/3$ & 0  \\
\hline $h_L$ & --- & $h_{1,3}(6) = 1/3$ & 0  \\
\hline $h_{RL}$ & --- & $h_{1,4}(6) = 1$ & 1/2 \\
\hline $h_{LR}$ & --- & $h_{1,3}(6) = 1/3$ & 0  \\
\hline $h_0$ & 0 & $h_{1,2}(6) = 0$ & 0  \\
\hline
\end{tabular}
\caption{Dimensions of various operators. The first four lines represent the superuniversal weights. The values of $h_{RA}$ and $h_{LA}$ for the IQH are obtained from numerical simulations of PCC in the CC model in Ref. \onlinecite{Obuse2009}. All other dimensions are exact. The angles $\theta^H_{R,L}$ are the Hall angles at the two types of reflecting boundaries. The symbols ``---'' mean that the corresponding exponents are not known to us.}
\label{DimensionsTable}
\end{table}

\section{Conclusions and outlook}
\label{sec:conclusions}

We have revisited the problem of the plateau transition in the integer quantum Hall (IQH) effect and related Anderson localization-delocalization transitions in two spatial dimensions. Specifically, we have considered the Chalker-Coddington network model and related models. In all cases we have focused on the so-called boundary point-contact conductances (PCCs) at critical points, and their behavior in the presence of various boundaries (absorbing and reflecting). While most of our results are general and apply to all problems we consider, let us concentrate here on the most interesting case, the IQH problem.

There are two key observations that allow us to analyze the problem. The first observation is that microscopic expressions for PCCs can be written as a sum of {\it positive} contributions related to certain geometric objects that we call {\it pictures} (see Sec. \ref{restrictionINChalker}). Written as a sum over pictures, a PCC can be interpreted as a partition function of an ensemble of pictures, each picture having a certain statistical weight. The second observation is that these statistical weights are {\it intrinsic} and satisfy the so-called restriction property with respect to {\it absorbing} boundaries. Namely, whenever we deform an absorbing boundary, the pictures that continue to contribute to a PCC are the ones that are present in the new (deformed) system. The pictures that intersect the deformed boundary do not contribute any more, and the PCC is renormalized. When we combine the restriction property with the assumed conformal invariance at the IQH transition point, we can employ the recently developed mathematical theory of conformal restriction measures. This theory is closely related to conformal field theories (CFTs) with zero central charge. As a result, we get several results that were already mentioned in the introduction and discussed at length in the main part of the paper. Let us briefly repeat them here.

First, PCCs in various geometries can be studied as correlation functions of (Virasoro) {\it primary} CFT operators. This statement alone allows to calculate exact forms for PCCs that reduce to three-point functions. Secondly, we predict the values of conformal dimensions of some of the primary operators that appear in the theory based on very general arguments. Finally, the connection with conformal restriction is established for other disordered systems, including the spin quantum Hall transition, the classical limit of the Chalker-Coddington model (diffusion in a magnetic field), and the metal in class D. For these systems many more dimensions of primary operators can be obtained exactly.

The relation between the Chalker-Coddington model and conformal restriction that we have discovered allows for an approach to the study of the critical properties of the IQH transition that is alternative to the ones used before. A full understanding of the transition by these methods will require much more work. We plan to extend this paper in several directions.

We hope to be able to consider PCCs in more complicated geometries, where the necessary CFT correlation functions will be four-point or higher functions. In these cases, knowing the conformal dimensions of the primary operators involved will not be sufficient. In addition, we would have to understand whether there are degenerate operators\cite{BPZ} related to the existence of null vectors in the corresponding representations of the Virasoro algebra. Potentially, this can be established by analyzing fusions of the operators that we have already identified.

In this paper we have considered only boundary PCCs. The necessary mathematical theory of sets $K$ that ``touch'' the boundary of a domain at two points is called the ``chordal restriction theory''. In principle one can define conductances between point contacts in the bulk of a network (obtained by cutting some of the internal links),\cite{Janssen99} or between a boundary and a bulk contact. Upon disorder averaging these conductances should also satisfy a (suitably modified) restriction property. The corresponding mathematical theory of the ``bulk'' restriction or the ``radial'' restriction has not been worked out, and we plan to develop it, and its relations to bulk CFT operators.

For the bulk theory the cut points that we have mentioned in Secs. \ref{subsec:conf-restriction} and \ref{restrictionINChalker} will likely be important objects. It is known from the conformal restriction theory\cite{Werner-restriction-review} that two-sided restriction measures have cut points for any restriction exponent $h$ in the range $5/8 \leqslant h < 35/24$. The cut points form a fractal set of Hausdorff dimension
\begin{align}
d_{\text{cut}}(h) &=  2 - \frac{\big(\sqrt{24 h +1} - 1\big)^2 - 1}{12} \nonumber \\ &
= 2 - 2h + \frac{2\sqrt{24 h +1} - 1}{12}.
\end{align}
As we have argued, in our problems two-sided restriction measures correspond to PCCs between contacts placed on the absorbing boundary, in which case the restriction exponent is the dimension of the conserved current operator: $h_A = 1$. Then the dimension of the set of cut points is $d_{\text{cut}}(1) = 3/4$. In the case of the SQH transition (percolation), this dimension can be written as
\begin{align}
d_{\text{cut}}(1) = 2 - 2 h_{0,2}(6),
\end{align}
where $h_{0,2}(6) = 5/8$ is the dimension of the {\it bulk\/} four-leg operator [see Eq. (\ref{n-leg-bulk}) for $n=4$] well known to be related to the critical exponent of the correlation length for percolation
\begin{align}
\nu_{\text{perc}} = (2 - 2 h_{0,2}(6))^{-1} = d_{\text{cut}}(1)^{-1} = 4/3.
\end{align}
While we know that the localization length exponent for the SQH transition is exactly this $\nu_{\text{SQH}} = 4/3$, it is very different for the IQH transition. The relation of cut points to $\nu_{\text{IQH}}$ (if any) is not clear to us at the moment, but it is tantalizing to speculate that one can get more understanding by focusing on a decomposition of fillings of pictures into irreducible components.

Another possible extension of our results is in the direction of studying the restriction property away from critical points. While the conformal invariance is lost away from critical points, the restriction property for pictures survives, and one can attempt to create a theory of ``massive'' restriction measures. Similar attempts to develop a theory of ``off-critical'' or ``massive'' variants of SLE exist in the literature.\cite{Makarov-Smirnov, Bauer-Bernard-Cantini} Also, one-sided conformal restriction measures can be built from Brownian motions with oblique reflection at boundaries, and one can try to extend this construction by introducing a finite ``killing rate'' for the Brownian particles. In the field theory language this corresponds to adding a mass term to the action (\ref{S0}).

Finally, we hope that the general idea of using notions and methods of stochastic conformal geometry (conformal restriction and SLE) can be fruitful in the study of other critical disordered systems.

\section{Acknowledgements}

We are grateful to D. Bernard, J. Cardy, P. Di Francesco, J. Dubedat, G. Lawler, A. Mirlin, I. Pak, N. Read, S. Smirnov, and P. Wiegmann for useful comments and discussions. EB was supported by a grant from the Israel Science Foundation, grant no. 852/11 and additional support was given by the Binational Science Foundation, grant no. 2010345. IAG was supported by NSF Grants Nos. DMR-0448820, DMR-0213745 and DMR-1105509. This work has been supported, in part, by NSF grant DMR-0706140 (A.W.W.L.).

\appendix

\section{Directed graphs, pictures, and Feynman paths}
\label{Appendix A}

In this appendix we present a relation between pictures and Feynman paths on the CC network. The connection goes through the notion of a directed graph, or {\it digraph} (see Refs. \onlinecite{Tutte, Aigner, Stanley, West}). Namely, for each picture $p$ we construct a digraph with vertices being the network nodes visited by the picture by replacing each link of the picture that is traversed $n_j$ times with $n_j$ directed edges. (All graphs constructed in this way are {\it loopless}: there is no edge in them that connects a vertex to itself.) Then, each Feynman path $f \in F(p)$ corresponds to an Eulerian trail (that is, a sequence of directed edges that visits every edge exactly once) on this digraph. The correspondence is one-to-many, since all permutations among $n_j$ edges of the digraph connecting a pair of vertices correspond to the same Feynman path.

There are several theorems in the theory of directed graphs that are relevant for our discussion and allow us to characterize the pictures that come from Feynman paths, and also count the number of Feynman paths $|F(p)|$ for a given picture $p$.

First, we introduce a few definitions. Let $D$ be a digraph with vertex set $V = \{v_1,..., v_m\}$ and edge set $E = \{e_1,..., e_n\}$. A {\it trail} in $D$ is a sequence $e_1,e_2,\ldots,e_r$ of distinct edges such that the final vertex of $e_i$ is the initial vertex of $e_{i+1}$ for all $1 < i < r - 1$. If, in addition, the final vertex of $e_r$ is the initial vertex of $e_1$, then the trail is called a {\it tour} or {\it cycle}. A trail (tour) is {\it Eulerian} if it visits every edge of $D$ exactly once. A digraph that has no isolated vertices and contains an Eulerian tour is called an Eulerian digraph. The {\it outdegree} of a vertex $v$, denoted outdeg$(v)$, is the number of edges of the digraph with initial vertex $v$. Similarly the {\it indegree} of $v$, denoted indeg$(v)$, is the number of edges of the digraph with final vertex $v$. A digraph is {\it balanced} if indeg$(v)$ = outdeg$(v)$ for all vertices $v$.

\begin{figure}[t]
\centering
\includegraphics[width=0.55\columnwidth]{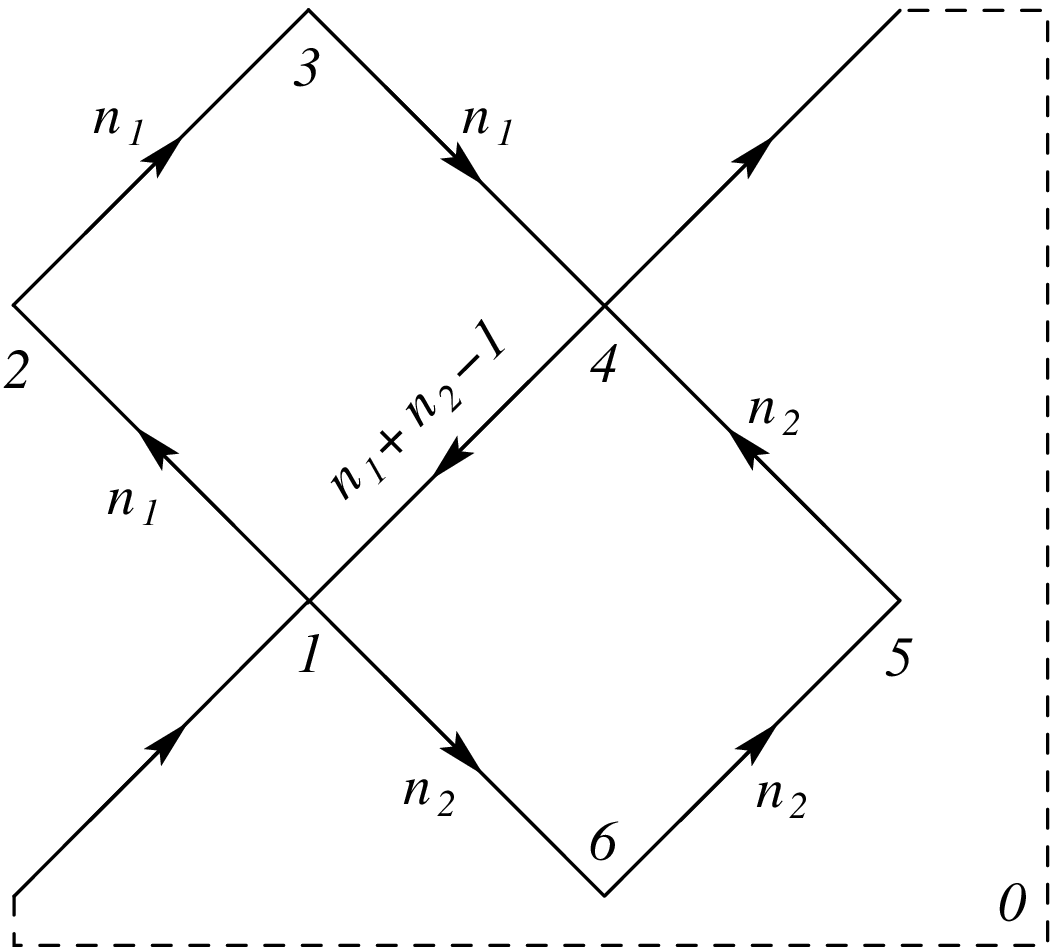}
\hfill
\includegraphics[width=0.35\columnwidth]{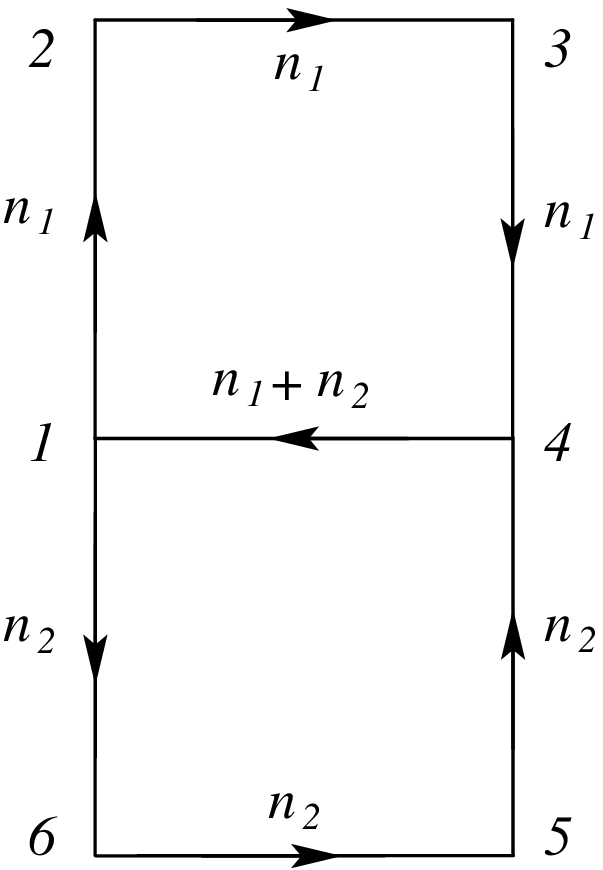}
\caption{
Left: a balanced picture (see the main text). The dashed line joins the initial and final vertices of every Feynman path corresponding to this picture into a single vertex labeled by $0$ here. Removing this vertex leads to the balanced picture shown on the right.}
\label{picture eight}
\end{figure}

The first theorem that we need is the following: a digraph without isolated vertices is Eulerian if and only if it is connected and balanced. This immediately gives the characterization of pictures that come from Feynman paths: for every vertex of such a picture the sum of numbers $n_j$ on the incoming links must be equal to the sum of numbers $n_j$ on the outgoing links. An example of such a balanced picture is shown in Fig. \ref{picture eight}. Actually, there is a little caveat that we need to mention. As drawn, this picture has the initial and final vertices that are {\it not} balanced. To eliminate this problem, we connect these two vertices into a single vertex $v_0$ (labeled by 0 in Fig. \ref{picture eight}) with $\text{outdeg}(v_0) = \text{indeg}(v_0) = 1$. The resulting picture {\it is} balanced. Moreover, after this is done, the number of Feynman paths corresponding to the original picture is equal to the number of Eulerian tours on the digraph corresponding to the modified picture divided by the multiplicity factors $n_j!$ for each link of the balanced picture.

Next we describe how we can count the number of Feynman paths $|F(p)|$ that correspond to a balanced picture $p$. We need two more theorems. One of them is the so-called BEST theorem that relates the number of Eulerian tours on a digraph $D$ to the number of spanning (directed) trees on $D$. Here is a precise
formulation. Let $D$ be a connected balanced digraph with vertex set $V$. Fix an edge $e$ of $D$, and let $v$ be the initial vertex of $e$. Let $T(D, v)$ denote the number of oriented (spanning) subtrees of $D$ with root $v$, and let $E(D,e)$ denote the number of Eulerian tours of $D$ starting with the edge $e$. Then
\begin{align}
E(D, e) = T(D, v) \prod_{u \in V} \big(\text{outdeg}(u) - 1\big)!.
\label{BEST theorem}
\end{align}

The other theorem, the so-called matrix-tree theorem, gives the number of spanning trees $T(D, v)$ with a given initial vertex $v$ in terms of the minor of the {\it Laplacian matrix} of the digraph. Let us denote the number of edges going from vertex $v_i$ to vertex $v_j$ by $m_{ij}$. The Laplacian matrix
$L = L(D)$ of a directed graph $D$ with vertex set $V = \{v_1,\ldots,v_m\}$ is the $m \times m$ matrix
\begin{align}
L_{ij} &=
\begin{cases}
-m_{ij} & \text{if } i \neq j, \\
\text{outdeg}(v_i) &\text{if } i = j.
\end{cases}
\end{align}
The matrix-tree theorem states the following: Let $D$ be a digraph with vertex set $V = \{v_1,..., v_m\}$, and let $1 \leqslant k \leqslant m$. Let $L$ be the Laplacian matrix of $D$, and define $L_k$ to be $L$ with the $k$-th row and column deleted. Then
\begin{align}
T(D,v) = \det L_k.
\label{matrix-tree theorem}
\end{align}
The result is independent of $k$.

\begin{figure}[t]
\centering
\includegraphics[width=0.8\columnwidth]{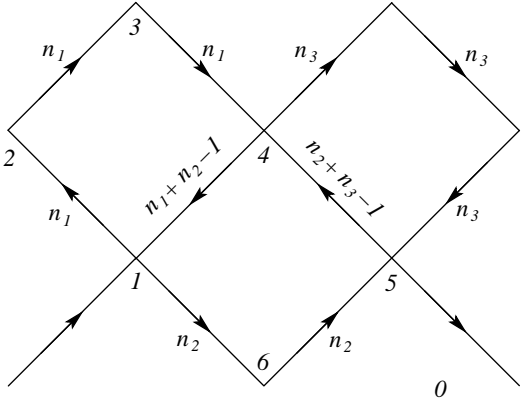}
\caption{A more complicated balanced picture.}
\label{picture heart}
\end{figure}

Combining Eqs. (\ref{BEST theorem}) and (\ref{matrix-tree theorem}) with the known degeneracy of the Eulerian tours that give the same Feynman path, we finally obtain the following general formula:
\begin{align}
|F(p)| & = \det L_k \frac{\prod_{u \in V} \big(\text{outdeg}(u) - 1\big)!}
{\prod_{i,j} m_{ij}!}.
\label{number of paths}
\end{align}
To illustrate this formula, consider the balanced picture shown on the left in Fig. \ref{picture eight}. Instead of counting the number of Eulerian trails that start at the beginning of every Feynman path, we can join the initial and final vertices of these paths into a single vertex (labeled by $0$ on the left in the figure), and count the number of Eulerian tours on the corresponding digraph. It is clear from the above discussion that the extra vertex $0$ does not enter into the calculation of $E(D,e)$ (even though the denominator in the formula (\ref{number of paths}) for $|F(p)|$ should still contain the edge multiplicities $m_{ij}$ from the original picture). Therefore, we remove it and obtain the balanced picture shown on the right in Fig. \ref{picture eight}. Labeling the remaining vertices as shown, we obtain the following Laplacian matrix:
\begin{align}
L = \begin{pmatrix}
n_1 + n_2 & -n_1 & 0 & 0 & 0 & -n_2 \\
0 & n_1 & -n_1 & 0 & 0 & 0 \\
0 & 0 & n_1 & -n_1 & 0 & 0 \\
-n_1 - n_2 & 0 & 0 & n_1 + n_2 & 0 & 0 \\
0 & 0 & 0 & -n_2 & n_2 & 0 \\
0 & 0 & 0 & 0 & -n_2 & n_2
\end{pmatrix}.
\end{align}
Deleting the first row and the first column, we get
\begin{align}
\det L_1 = (n_1 + n_2)n_1^2 n_2^2.
\end{align}
The formula (\ref{number of paths}) now gives
\begin{align}
|F(p)| &= (n_1 + n_2)n_1^2 n_2^2
\nonumber \\ & \quad \times
\frac{\big[(n_1 + n_2 - 1)! (n_1 -
1)! (n_2 - 1)! \big]^2}
{(n_1 + n_2 - 1)! \big[n_1! \, n_2!\big]^3} \nonumber \\
& = \frac{(n_1 + n_2)!}{n_1! \, n_2!}.
\end{align}
In this particular case $|F(p)|$ has a combinatorial interpretation as the number of distinct orderings of going around the top and the bottom plaquettes on the right picture in Fig. \ref{picture eight}. However, in the more complicated cases there is no such simple interpretation, while the general
formula (\ref{number of paths}) is still straightforward to use. For example, for the picture shown in Fig. \ref{picture heart} we have
\begin{align}
|F(p)| &= \frac{(n_1 + n_2 + n_3 - 1)!}{n_1! \, (n_2 - 1)! \, n_3!}
+ \frac{(n_1 + n_2 + n_3 - 2)!}{(n_1 - 1)! \, n_2! \, (n_3 - 1)!}.
\end{align}

We note here that the weighting factors for Feynman paths $f$ that enter the definition of the quantity $S(p)$ in Eqs. (\ref{S(p)-general})--(\ref{S(p)-critical}) are not determined by the graph-theoretic data for the corresponding digraph (even in the simplest case of the critical point in the isotropic
system). Therefore, it seems that an explicit calculation of $S(p)$ for a given picture $p$ is a much more challenging problem than that for $|F(p)|$.

\section{A conformal map}
\label{conformal-map}

Here we construct the conformal map  $f: {\mathbb H}\setminus A \to {\mathbb H}$, where $A$ is the semicircle of radius $\epsilon$ centered at $c \in {\mathbb R}$, that preserves two points $a < c$ and $b > c$ (see Fig. \ref{fig:g-change}). The semicircle's diameter along the real axis goes from $c_- \equiv c-\epsilon$ to $c_+ \equiv c+\epsilon$. We are interested in the limit where $\epsilon$ is much smaller than other distances in the problem: $\epsilon \ll c-a, b-c$, and will expand results to the first non-trivial order in $\epsilon$.

Let the original domain ${\mathbb H}\setminus A$ be in the complex $z$ plane. We construct the map $w(z)$ in stages. First, we perform a M\"obius transformation $s(z)$ that maps the point $a$ to 0 and the point $b$ to $\infty$:
\begin{align}
s(z) = \frac{z-a}{b-z}.
\end{align}
The images of various points under this map are
\begin{align}
s(a) &= 0, \qquad s(b) = \infty, \nonumber \\
s(c_\pm) &= \frac{c_\pm - a}{b - c_\pm} \approx \frac{c-a}{b-c}\Big[1 \pm \epsilon \Big(\frac{1}{c-a} - \frac{1}{b-c}\Big) \Big].
\end{align}

Since $s(z)$ is a M\"obius transformation, the semicircle $A$ maps to another semicircle in the $s$ plane. The center of this semicircle is
\begin{align}
s_0 &= \frac{1}{2} \big[s(c_+) + s(c_-)\big] = \frac{(c-a)(b-c) + \epsilon^2}{(b-c)^2 - \epsilon^2}
\approx \frac{c-a}{b-c},
\end{align}
and its radius is
\begin{align}
r_0 &= \frac{1}{2} \big[s(c_+) - s(c_-)\big] = \frac{\epsilon (b-a)}{(b-c)^2 - \epsilon^2}
\approx \epsilon \frac{b-a}{(b-c)^2}.
\end{align}

Next we shift everything by $s_0$ and rescale by $r_0$:
\begin{align}
t(s) = \frac{s-s_0}{r_0} = \frac{z(1 + s_0) - a - b s_0}{r_0(b-z)}.
\end{align}
This transformation preserves the infinity, but maps $0$ to
\begin{align}
t_0 = - \frac{s_0}{r_0} = - \frac{(c-a)(b-c) + \epsilon^2}{\epsilon (b-a)}
\approx - \frac{1}{\epsilon} \frac{(c-a)(b-c)}{(b-a)},
\end{align}
and the semicircle in the $s$ plane to the semicircle of unit radius centered at the origin in the $t$ plane.

Now we can perform the Zhukovsky transformation
\begin{align}
u(t) &= \frac{1}{2} \Big(t + \frac{1}{t} \Big) \nonumber \\
&= \frac{1}{2} \Big(\frac{z(1 + s_0) - a - b s_0}{r_0(b-z)} + \frac{r_0(b-z)}{z(1 + s_0) - a - b s_0} \Big),
\end{align}
which removes the semicircle in the $t$ plane, preserves the infinity, and maps $t_0$ to
\begin{align}
u_0 = \frac{1}{2} \Big(t_0 + \frac{1}{t_0} \Big) \approx \frac{t_0}{2}
= - \frac{1}{2\epsilon} \frac{(c-a)(b-c)}{(b-a)}.
\end{align}

One more M\"obius transformation maps $\infty$ back to $b$, and $u_0$ to $a$:
\begin{align}
w(u) = \frac{b(u - u_0) + a}{u - u_0 + 1}.
\end{align}

Finally, the transformation $w(z)$ that we want is obtained by composing all the above maps:
\begin{align}
f &= w \circ u \circ t \circ s.
\end{align}
Under these maps the points $a$ and $b$ are successively mapped as
\begin{align*}
a \stackrel{s}{\to} 0 \stackrel{t}{\to} t_0 \stackrel{u}{\to} u_0 \stackrel{w}{\to} a, &&
b \stackrel{s}{\to} \infty \stackrel{t}{\to} \infty \stackrel{u}{\to} \infty \stackrel{w}{\to} b.
\end{align*}

Now we can wind the derivative of the map $f(z)$ as
\begin{align}
f'(z) &= w'(u) \cdot u'(t) \cdot t'(s) \cdot s'(z)
\nonumber \\
&= \frac{1}{2r_0} \Big( 1 - \frac{1}{t_0^2}\Big) \bigg[\frac{b-a}{(b-z)(u - u_0 +1)}\bigg]^2.
\end{align}
From this expression we immediately see that
\begin{align}
f'(a) = \frac{1}{2r_0} \Big( 1 - \frac{1}{t_0^2}\Big).
\end{align}
To evaluate $f'(b)$ we need first to find
\begin{align}
&(b-z)u = \frac{z(1 + s_0) - a - b s_0}{2 r_0} + \frac{r_0(b-z)^2}{z(1 + s_0) - a - b s_0},
\nonumber \\
&(b-z)u\big|_{z = b} = \frac{b-a}{2 r_0}.
\end{align}
Then
\begin{align}
f'(b) = 2r_0 \Big( 1 - \frac{1}{t_0^2}\Big).
\end{align}

The basic transformation formula for conformal restriction measures, Eq. (\ref{transformation}), now gives
\begin{align}
Z_{{\mathbb H}\setminus A}(a,b) &= |f'(a)|^{h_A}|f'(b)|^{h_A} Z_{\mathbb H}(a,b) \nonumber \\
&= \Big( 1 - \frac{1}{t_0^2}\Big)^{2h_A} Z_{\mathbb H}(a,b).
\end{align}
Relating this to transport properties, we find the change in the average point contact conductance between $a$ and $b$ upon deforming the real axis by the bump $A$:
\begin{align}
\delta \langle g(a,b) \rangle &= \big(1 - |f'(a)|^{h_A} |f'(b)|^{h_A} \big) \langle g(a,b) \rangle
\nonumber \\ &
= \bigg[1 - \Big( 1 - \frac{1}{t_0^2}\Big)^{2h_A} \bigg] \langle g(a,b) \rangle
\nonumber \\ &
\approx \frac{2 h_A}{t_0^2} \langle g(a,b) \rangle
\approx \frac{2 h_A |b-a|^2}{|c-a|^2 |c-b|^2} \epsilon^2 \langle g(a,b) \rangle
\nonumber \\ &
= \frac{C \epsilon^2}{|a-b|^{2h_A - 2} |a-c|^{2} |b-c|^{2}}.
\end{align}

\section{Weights of current insertions in the classical CC model}
\label{sec:weights-calssical-CC}

\begin{figure}[t]
\centering
\includegraphics[width=0.6\columnwidth]{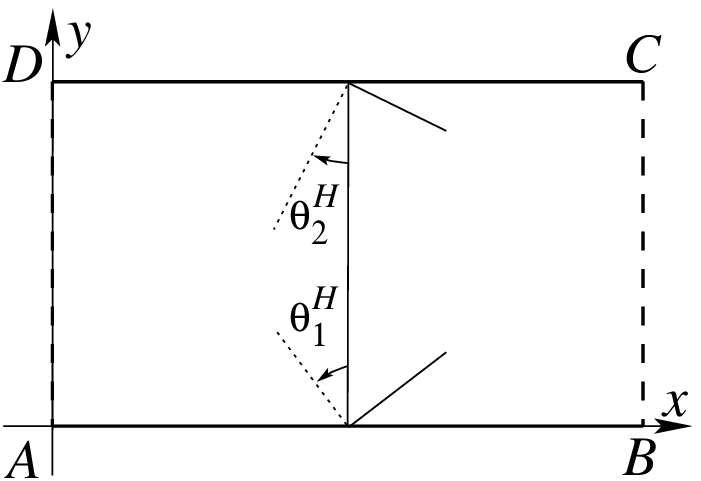}
\hfill
\includegraphics[width=0.3\columnwidth]{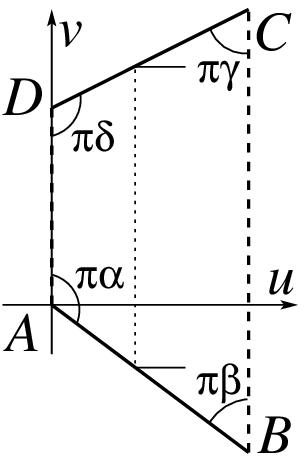}
\caption{Left: Hall bar in the complex $z$ plane. The vertical (dashed) portions of the boundary are attached to ideal leads. The horizontal portions of the boundary are reflecting with possibly different Hall angles $\theta^H_1$ and $\theta^H_2$. Right: the same Hall bar in the complex $w$ plane. In this figure $\theta^H_1 = \theta^H_R$ and $\theta^H_2 = \theta^H_L$. The dotted lines are the directions in which the components of $\nabla \phi$ vanish.}
\label{fig:Hall-bar}
\end{figure}

Let us place the Hall bar in the complex $z = x+iy$ plane so that its corners $A, B, C$, and $D$ (going counterclockwise around the sample) are at the points $0, L, L + iW, iW$ (see Fig. \ref{fig:Hall-bar}). In the bulk of the sample the potential $\phi(x,y)$ satisfies the Laplace equation
\begin{align}
\nabla^2 \phi = 0.
\label{Laplace}
\end{align}
We assume  that two ideal contacts are attached to the sides of length $W$, and that these leads are kept at constant potentials:
\begin{align}
\phi|_{AD} &= 0, && \phi|_{BC} = V.
\label{BC-1}
\end{align}
Boundary conditions at the horizontal sides $AB$ and $CD$ of the Hall bar have to be chosen according to the particular boundary operator we are interested in. To find, say, $h_A$, we require
\begin{align}
\phi|_{AB} &= 0, & \phi|_{CD} = 0. \label{h_A}
\end{align}
Similarly, for $h_{RA}$ and $h_{LA}$ we choose
\begin{align}
\phi|_{AB} &= 0, & (\partial_y - \tan \theta^H \partial_x)\phi|_{CD} = 0, \label{h_x}
\end{align}
where the Hall angle $\theta^H$ can take two possible values. For the dimensions $h_R$, $h_L$, $h_{RL}$, and $h_{LR}$ we choose
\begin{align}
(\partial_y - \tan \theta^H_1 \partial_x)\phi|_{AB} &= 0, & (\partial_y - \tan \theta^H_2 \partial_x)\phi|_{CD} = 0. \label{h_xx}
\end{align}
It is this case that is shown in Fig. \ref{fig:Hall-bar}.

In the case of the dimensions $h_A$ and $h_{LA}, h_{RA}$ it is easier to solve the necessary boundary value problem for the Laplace equation (\ref{Laplace}) in an infinite strip of width $W$. In this situation the boundary conditions (\ref{BC-1}) for the potential at the ends of the strip are not important, and we can simply look for solutions that decay exponentially in the positive $x$ direction. Thus we assume
\begin{align}
\phi(x,y) = e^{-kx} f(y).
\label{separation}
\end{align}
Once possible values of $k$ are found, they are related to the scaling dimension $h$ of the corresponding boundary operator by
\begin{align}
\pi h = k W,
\end{align}
and the leading scaling dimension is given by the smallest non-negative $k$. Substituting this into the Laplace equation (\ref{Laplace}) gives
\begin{align}
f''(y) + k^2 f(y) &= 0, & f(y) = A \sin ky + B \cos ky.
\label{f(y)}
\end{align}

In the case of $h_A$ the boundary conditions (\ref{h_A}) imply that $B = 0$ and $kW = \pi n$, $n = 1, 2, \ldots$. The smallest exponent then is
\begin{align}
h = h_A = 1,
\label{h_A-result}
\end{align}
as expected. In the case of $h_{LA}, h_{RA}$ the boundary conditions (\ref{h_x}) immediately give $B = 0$ and
\begin{align}
&\cos kW + \tan \theta^H_2 \sin kW = 0, \quad \Rightarrow \quad \cot kW = - \tan\theta^H_2,
\nonumber \\
& k W = \theta^H_2 + \pi/2 + \pi n, \quad n \in {\mathbb Z}.
\end{align}
The smallest positive eigenvalue ($n = 0$) leads to
\begin{align}
h_{LA} &= \frac{1}{2} + \frac{\theta^H_L}{\pi}, & h_{RA} = \frac{1}{2} + \frac{\theta^H_R}{\pi},
\label{h_xA-result}
\end{align}
the restriction exponents for reflected Brownian motions, as expected.

To find the two probe conductance in the case when both the top and the bottom sides are reflecting we need to solve the problem [Eqs. (\ref{Laplace}), (\ref{BC-1}), and (\ref{h_xx})]. An exact solution of this problem is possible via the use of the Schwarz-Christoffel conformal map. \cite{Wick, Lippmann-Kuhrt, Rendell-Girvin} The idea is to conformally map the Hall bar in the $z$ plane to the quadrilateral in the complex $w = u + iv$ plane shown on the right in Fig. \ref{fig:Hall-bar}. The angles at the vertices of the sample in the $w$ plane are $\pi \alpha$, $\pi \beta$, $\pi \gamma$, and $\pi \delta$, where
\begin{align}
\alpha &= \frac{1}{2} + \frac{\theta^H_1}{\pi}, &
\beta &= \frac{1}{2} - \frac{\theta^H_1}{\pi} = 1 - \alpha, \nonumber \\
\gamma &=  \frac{1}{2} + \frac{\theta^H_2}{\pi}, &
\delta &= \frac{1}{2} - \frac{\theta^H_2}{\pi} = 1 - \gamma.
\end{align}
Once we find such a map $w(z)$, then in the $w$ plane the dotted lines denoting the direction in which the gradient of the potential vanishes become vertical. The solution of the boundary value problem [Eqs. (\ref{Laplace}), (\ref{BC-1}), and (\ref{h_xx})] in this plane is simply given by
\begin{align}
\phi(u, v) = u = \re w.
\end{align}
This then gives the total potential drop
\begin{align}
V = \re w(B),
\label{voltage}
\end{align}
and the horizontal electric field $E_u = -\partial_u \phi(u, v) = -1$. This means that the current in the bulk is also uniform in this geometry and has the form
\begin{align}
\begin{pmatrix} j_u \\ j_v \end{pmatrix} = \begin{pmatrix} \sigma_{xx} & \sigma_{xy} \\ -\sigma_{xy} & \sigma_{xx} \end{pmatrix}
\begin{pmatrix} -1 \\ 0 \end{pmatrix} = \begin{pmatrix} -\sigma_{xx} \\ \sigma_{xy} \end{pmatrix}.
\end{align}

Notice that in the case when the Hall angles $\theta^H_1$ and $\theta^H_2$ are not equal, the conductance in one direction is not equal to the conductance in the other direction. Thus, we have to be careful with the calculation. We need to evaluate the current through the right lead, where the potential is higher, since this is where the current enters the system. In the direction away from this lead into the bulk the current starts to accumulate on one of the edges, and the current distribution acquires a delta-function contribution at this edge.

The total current entering through the right lead is
\begin{align}
I = |j_u| |v(C) - v(B)| = \sigma_{xx} \im [w(C) - w(B)].
\label{current}
\end{align}
Combining Eqs. (\ref{voltage}) and (\ref{current}) gives the two-terminal conductance:
\begin{align}
g = \frac{I}{V} = \sigma_{xx} \frac{\im[w(C) - w(B)]}{\re w(B)}.
\label{conductance}
\end{align}
In this form the conductance depends on a particular normalization of the conformal map $w(\zeta)$ which was chosen such that $w(D)$ is purely imaginary. This expression can be rewritten in an equivalent form
\begin{align}
g = \frac{\sigma_{xx}}{\sin(\pi \alpha)} \frac{|w(C) - w(B)|}{|w(B)|},
\label{conductance-2}
\end{align}
which is independent of the normalization of $w(\zeta)$.

While it is straightforward to find an exact Schwarz-Christoffel map $w(z)$, in the case of long systems ($L \gg W$) it is much easier to use an approximate map $\tilde{w}(z)$. In this case, the trapezoids (like the one shown in the right panel in Fig. \ref{fig:Hall-bar}) almost degenerate into triangles. We can then use a map appropriate for mapping an infinite strip to an infinite wedge, that is, an exponential map. We will only need to make sure that the opening angle of the wedge is the same as that of the trapezoid in question (determined by the Hall angles $\theta^H_1$ and $\theta^H_2$).
Thus, we use an approximate formula
\begin{align}
g \sim \frac{\sigma_{xx}}{\sin(\pi \alpha)} \frac{|\tilde{w}(C) - \tilde{w}(B)|}{|\tilde{w}(B)|}.
\end{align}
For a long finite strip, this introduces distortions near the leads, but they happen to be negligible, which is confirmed by the exact solution.

It is convenient to treat the cases of equal and unequal Hall angles separately. When $\theta^H_1 = \theta^H_2$, we need to use a simple rotation as the approximate map:
\begin{align}
\tilde{w}(z) = e^{-i \theta^H_1} z.
\label{approx-map-equal}
\end{align}
For the conductance $g$ this gives $g \sim \sigma_{xx} W/L$ and implies
\begin{align}
h_R = h_L = 0.
\label{h_x-result}
\end{align}

When the Hall angles are not equal, we will use the following (shifted and re-scaled) exponential map:
\begin{align}
\tilde{w}(z) = e^{-i \theta^H_1}(e^{k z} - 1).
\label{approx-map}
\end{align}
This maps the corner $A$ to the origin in the $w$ plane, and also the side $AB$ to a straight segment with the argument $- \theta^H_1$. To make sure that the upper side $DC$ has the correct slope, we need to choose
\begin{align}
kW = \theta^H_1 - \theta^H_2.
\end{align}
The images of the vertices of the strip are
\begin{align}
\tilde{w}(A) &= 0, \quad \tilde{w}(B) = e^{-i \theta^H_1}(e^{(\theta^H_1 - \theta^H_2)L/W} - 1),
\nonumber \\
\tilde{w}(C) &= e^{-i \theta^H_2} e^{(\theta^H_1 - \theta^H_2)L/W} - e^{-i \theta^H_1},
\nonumber \\
\tilde{w}(D) &= e^{-i \theta^H_2} - e^{-i \theta^H_1}.
\end{align}
Using Eq. (\ref{theta-Hall-relation}) the conductance $g$ is found to be
\begin{align}
g \sim \sigma_{xx} \frac{e^{(\theta^H_1 - \theta^H_2)L/W}}{\big|e^{(\theta^H_1 - \theta^H_2)L/W} - 1\big|} = \frac{\sigma_{xx}}{\big|1 - e^{\pm \pi L/2W}\big|}.
\label{g-approx}
\end{align}
When $\theta^H_1 = \theta^H_R$ and $\theta^H_2 = \theta^H_L$ the exponential term in the denominator in (\ref{g-approx}) can be neglected, and we get $g_{LR} \sim \sigma_{xx}$. In the opposite case the exponential term in the denominator in (\ref{g-approx}) dominates, and we get $g_{RL} \sim \sigma_{xx} e^{-\pi L/2W}$. These two imply, in turn,
\begin{align}
h_{LR} &= 0, & h_{RL} &= \frac{1}{2}.
\label{h_xx-result}
\end{align}


\begin{thebibliography}{99}

\bibitem{Anderson58} P. W. Anderson, Phys.\ Rev.\ \textbf{109}, 1492 (1958).

\bibitem{AL50} P. A. Lee and T. V. Ramakrishnan, Rev. Mod. Phys. 57, 287 (1985);
{\it 50 years of Anderson localization}, ed. by E. Abrahams (World Scientific, 2010).

\bibitem{Zirnbauer96} M. R. Zirnbauer, J.\ Math.\ Phys.\ \textbf{37}, 4986 (1996).

\bibitem{Altland97} A. Altland and M. R. Zirnbauer, Phys.\ Rev.\ B \textbf{55}, 1142 (1997).

\bibitem{ReviewSymmetryClasses} For a recent review, see e.g. the Introduction of Ref. \onlinecite{RyuEtAl-NJPhys}.

\bibitem{RyuEtAl-NJPhys} S. Ryu, A. Schnyder, A. Furusaki, A. Ludwig, New J. Phys. 12, 065010 (2010).

\bibitem{Evers08} F.\ Evers and A.\ D.\ Mirlin, Rev.\ Mod.\ Phys.\ \textbf{80}, 1355 (2008).

\bibitem{Amsterdam-group}
A.\ de Visser, L.\ A.\ Ponomarenko, G.\ Galistu, D.\ T.\ N.\ de Lang, A.\ M.\ M.\ Pruisken, U.\ Zeitler, and D.\ Maude, J.\ Phys.: Conf.\ Series {\bf 51}, 379 (2006); D.\ T.\ N.\ de Lang, L.\ A.\ Ponomarenko, A.\ de Visser, and A.\ M.\ M.\ Pruisken, Phys.\ Rev.\ B {\bf 75}, 035313 (2007).

\bibitem{Tsui-group}
W.\ Li, G.\ A.\ Cs\'athy, D.\ C.\ Tsui, L.\ N.\ Pfeiffer, and K.\ W.\ West, Phys.\ Rev.\ Lett.\ {\bf 94}, 206807  (2005); W.\ Li, C.\ L.\ Vicente, J.\ S.\ Xia, W.\ Pan, D.\ C.\ Tsui, L.\ N.\ Pfeiffer, and K.\ W.\ West, Phys.\ Rev.\ Lett.\  {\bf 102}, 216801 (2009); W.\ Li, J.\ S.\ Xia, C.\ Vicente, N.\ S.\ Sullivan, W.\ Pan, D.\ C.\ Tsui, L.\ N.\ Pfeiffer, and K.\ W.\ West, Phys. Rev. B {\bf 81}, 033305 (2010).

\bibitem{Amado10}
M.\ Amado, E.\ Diez, D.\ L\'opez-Romero, F.\ Rossella, J.\ M.\ Caridad, F.\ Dionigi , V.\ Bellani, and D.\ K.\ Maude, New.\ J.\ Phys.\ \textbf{12}, 053004 (2009).

\bibitem{Saeed11} K.\ Saeed, N.\ A.\ Dodoo-Amoo, L.\ H.\ Li, S.\ P.\ Khanna, E.\ H.\ Linfield, A.\ G.\ Davies, and J.\ E.\ Cunningham, Phys.\ Rev.\ B \textbf{84}, 155324 (2011).

\bibitem{Huang12}
J.\ Huang, L.\ N.\ Pfeiffer, and K.\ W.\ West, Phys.\ Rev.\ B \textbf{85}, 041304(R) (2012).

\bibitem{Shen12}
T.\ Shen, A.\ T.\ Neal, M.\ L.\ Bolen, J.\ J.\ Gu, L.\ W.\ Engel, M.\ A.\ Capano, and P.\ D.\ Ye, J.\  Appl.\ Phys. \textbf{111}, 013716 (2012).

\bibitem{Zirnbauer99} M. R. Zirnbauer, arXiv:hep-th/9905054v2.

\bibitem{tsvelik} M. J. Bhaseen \textit{et al.}, Nucl.\ Phys.\ \textbf{B580}, 688 (2000); A. M. Tsvelik, Phys.\ Rev.\ B \textbf{75}, 184201 (2007).

\bibitem{LeClair} A. LeClair, arXiv:0710.3778v1.

\bibitem{Pruisken} A. M. M. Pruisken and I. S. Burmistrov, Annals  Phys. (N.Y.) {\bf 322}, 1265 (2007); Pisma v ZhETF {\bf 87}, 252 (2008).

\bibitem{Obuse08b}
H.\ Obuse, A.\ R.\ Subramaniam, A.\ Furusaki, I.\ A.\ Gruzberg, and A.\ W.\ W.\ Ludwig, Phys. Rev. Lett.\ \textbf{101}, 116802 (2008).

\bibitem{Evers08b} F.\ Evers, A.\ Mildenberger, and A.\ D.\ Mirlin, Phys.\ Rev.\ Lett.\ \textbf{101}, 116803 (2008).

\bibitem{Slevin09} K. Slevin and T. Ohtsuki, Phys. Rev. B {\bf 80}, 041304 (2009); Int.\ J.\ Mod.\ Phys.\ Conf.\ Ser.\ \textbf{11}, 60 (2012).

\bibitem{Burmistrov10}
I. S. Burmistrov, S.\ Bera, F.\ Evers, I.\ V.\ Gornyi, A.\ D.\ Mirlin, Ann.\ Phys.\ \textbf{326}, 1457 (2011).

\bibitem{Amado11}
M.\ Amado, A.\ V.\ Malyshev, A.\ Sedrakyan, and F.\ Dom\'inguez-Adame, Phys. Rev. Lett. {\bf 107}, 066402 (2011).

\bibitem{stabilitymap}
H.\ Obuse, I.\ A.\ Gruzberg, and F.\ Evers, arXiv:1205.2763.

\bibitem{Kagalovsky99} V. Kagalovsky, B. Horovitz, Y. Avishai, and J. T. Chalker, Phys.~Rev.~Lett.~\textbf{82}, 3516 (1999).

\bibitem{Senthil1999b} \bibinfo{author}{T.~Senthil}, \bibinfo{author}{J.~B. Marston}, and \bibinfo{author}{M.~P.~A. Fisher}, \bibinfo{journal}{Phys. Rev. B}
  \bibinfo{volume}{\textbf{60}}, \bibinfo{pages}{4245} (\bibinfo{date}{1999}).

\bibitem{Gruzberg99} I. A. Gruzberg, A. W. W. Ludwig, and N. Read, Phys.~Rev.~Lett.~\textbf{82}, 4524 (1999).

\bibitem{Cardy00} J. Cardy, Phys.~Rev.~Lett.~\textbf{84}, 3507 (2000).

\bibitem{Beamond02} E. J. Beamond, J. Cardy, and J. T. Chalker, Phys. Rev. B~\textbf{65}, 214301 (2002).

\bibitem{Mirlin03} A. D. Mirlin, F. Evers, and A. Mildenberger, J. Phys. A~\textbf{36}, 3255 (2003).

\bibitem{Subramaniam08} A. R. Subramaniam, I. A. Gruzberg, and A. W. W. Ludwig, Phys. Rev. B {\bf 78}, 245105 (2008).

\bibitem{Bondesan11} R. Bondesan, I. A. Gruzberg, J. L. Jacobsen, H. Obuse, and H. Saleur, Phys. Rev. Lett. {\bf 108}, 126801 (2012).

\bibitem{Chalker88} J. T. Chalker and P. D. Coddington, J.\ Phys.\ C \textbf{21}, 2665 (1988).

\bibitem{Chalker01} J. T. Chalker, N. Read, V. Kagalovsky, B. Horovitz, Y. Avishai, and A. W. W. Ludwig, Phys.~Rev.~B \textbf{65}, 012506 (2001).

\bibitem{Gruzberg01} I. A. Gruzberg, N. Read, and A. W. W. Ludwig, Phys.\ Rev.\ B \textbf{63}, 104422 (2001).

\bibitem{ReadLudwig00} N. Read and A. W. W. Ludwig, Phys.\ Rev.\ B \textbf{63}, 024404 (2000).

\bibitem{Mildenberger07} A. Mildenberger, F. Evers, A. D. Mirlin, and J. T. Chalker, Phys. Rev. B {\bf 75}, 245321 (2007).

\bibitem{Polyakov:1970} A.~M.~Polyakov, JETP Lett.\  {\bf 12}, 381 (1970) [Pisma Zh.\ Eksp.\ Teor.\ Fiz.\ {\bf 12}, 538 (1970)].

\bibitem{BPZ} A.~A.~Belavin, A.~M.~Polyakov, and A.~B.~Zamolodchikov, Nucl.\ Phys.\ B {\bf 241}, 333 (1984).

\bibitem{YellowBook} P. Di Francesco, P. Mathieu, D. Senechal, {\it Conformal field theory}, Springer, 1999.

\bibitem{footnoteFluctuationPartitionFct}
More precisely, it is the disorder average of the logarithm of the partition function, or of the Green's function (quantum mechanical resolvent) which in turn involves the inverse of the partition function, which is often of interest. In either of these quantities the statistical fluctuations of the partition function present a major difficulty.

\bibitem{Obuse07} H. Obuse, A. R. Subramaniam, A. Furusaki, I. A. Gruzberg, and A. W. W. Ludwig, Phys. Rev. Lett. {\bf 98}, 156802 (2007).

\bibitem{Obuse08} H. Obuse, A. R. Subramaniam, A. Furusaki, I. A. Gruzberg, and A. W. W. Ludwig, Physica E {\bf 40}, 1404 (2008).

\bibitem{Obuse10} H. Obuse, A. R. Subramaniam, A. Furusaki, I. A. Gruzberg, and A. W. W. Ludwig, Phys. Rev. B {\bf 82}, 035309 (2010).

\bibitem{GurarieLudwig2005Review} V. Gurarie and A. W. W. Ludwig, in {\it ``From Fields to Strings: Circumnavigating Theoretical Physics''}, Eds. M. Shifman, A. Vainshtein, J. Wheater (World Scientific, 2005); arXiv: hep-th/0409105.

\bibitem{Schramm1999} O. Schramm, Israel J. Math. {\bf 118}, 221 (2000); arXiv: math.PR/9904022.

\bibitem{Werner-review} W. Werner, {\it Random planar curves and Schramm-Loewner evolutions}, Lecture Notes in Mathematics {\bf 1840}, Springer-Verlag (Berlin, 2004); arXiv: math.PR/0303354.

\bibitem{Lawler-book} G. F. Lawler, {\it Conformally invariant processes in the plane}. Mathematical Surveys and Monographs, 114. American Mathematical Society, Providence, RI, 2005.

\bibitem{Kager-Nienhuis-review} W. Kager, B. Nienhuis, J.  Stat. Phys. {\bf 115}, 1149 (2004).

\bibitem{Cardy-review} J. Cardy, Ann. Phys. {\bf 318}, 81 (2005).

\bibitem{BB-review} M. Bauer and D. Bernard, Phys. Rep. {\bf 432}, 115 (2006).

\bibitem{IAG-review} I. A. Gruzberg, J. Phys. A: Math. Gen. {\bf 39}, 12601 (2006).

\bibitem{RBGW07} I. Rushkin, E. Bettelheim, I. A. Gruzberg, and P. Wiegmann, J. Phys. A: Math. Theor. {\bf 40}, 2165 (2007).

\bibitem{Friedrich02} R. Friedrich and W. Werner, C. R. Acad. Sci. Paris, Ser. I {\bf 335}, 947 (2002).

\bibitem{Friedrich03}  R. Friedrich and W. Werner, Comm. Math. Phys. {\bf 243}, 105 (2003).

\bibitem{LSW-conformal-restriction} G. F. Lawler, O. Schramm, and W. Werner, J. Amer. Math. Soc. {\bf 16}, 917 (2003).

\bibitem{Friedrich04} R. Friedrich and J. Kalkkinen, Nucl. Phys. {\bf B687}, 279 (2004).

\bibitem{Werner-restriction-review} W. Werner, Prob. Surveys {\bf 2}, 145 (2005).

\bibitem{BBK2005} M. Bauer, D. Bernard, K. Kyt\"ol\"a, J. Stat. Phys. {\bf 120}, 1125 (2005).

\bibitem{Kytola} K. Kyt\"ol\"a, J. Stat. Phys. {\bf 123}, 1169 (2006).

\bibitem{Graham} K. Graham, J. Stat. Mech. P03008 (2007); arXiv:  math-ph/0511060.

\bibitem{Dubedat-1} J. Dubedat, Ann. Probab. {\bf 33},  223 (2005).

\bibitem{Dubedat-2}  J. Dubedat, Comm. Pure Appl. Math. {\bf 60}, 1792 (2007).

\bibitem{Dubedat-3} J. Dubedat, J. Stat. Phys. {\bf 123}, 1183 (2006).

\bibitem{Janssen99} M. Janssen, M. Metzler, and M. R. Zirnbauer, Phys. Rev. B {\bf 59}, 15\,836 (1999).

\bibitem{RefCommentAbsorbingBoundaries}
Physically, an {\it absorbing boundary} of the sample is one at which the sample is attached to an ideal lead.

\bibitem{footnotePhaseAveraging}
Here we take, as it is common practice, the phases $\phi_j$ to be independent identically distributed random variables with a uniform distribution on each link. (Universal properties will not depend on the details of this probability distribution.)

\bibitem{RG1980} R. W. Rendell and S. M. Girvin, Phys. Rev. B {\bf 23}, 6610 (1981).

\bibitem{ML1993} D. L. Maslov and D. Loss, Phys. Rev. Lett. {\bf 71}, 4222 (1993).

\bibitem{KY1994} D. E. Khmelnitskii and M. Yosefin, Surf. Sci. {\bf 305}, 507 (1994).

\bibitem{XRS1997} S. Xiong, N. Read, and A. D. Stone, Phys. Rev. B {\bf 56}, 3982 (1997).

\bibitem{Senthil2000} T. Senthil and M. P. A. Fisher, Phys. Rev. B {\bf 61}, 9690 (2000).

\bibitem{Read-Green2000} N. Read and D. Green, Phys. Rev. B {\bf 61}, 10\,267 (2000).

\bibitem{Bocquet2000} M. Bocquet, D. Serban, and M. R. Zirnbauer, Nucl. Phys {\bf B578}, 628 (2000).

\bibitem{sets-K}
More precisely, it means that the sets $K$ satisfy the following two properties: (1) $K$ is connected, and ${\overline K} \cap \partial D = \{a, b\}$, (2) ${\mathbb C}\setminus {\overline K}$ is connected.

\bibitem{Coniglio} A. Coniglio, J. Phys. A: Math. Gen. {\bf 15}, 3829 (1982).

\bibitem{Obuse2009} H. Obuse et al. to be published.

\bibitem{Nishimori} I. A. Gruzberg, N. Read, and A. W. W. Ludwig, Phys. Rev. B {\bf  63}, 104422 (2001).

\bibitem{Ikhlef11} Y. Ikhlef, P. Fendley, and J. Cardy, Phys. Rev. B {\bf 84}, 144201 (2011).

\bibitem{Smirnov2001} S. Smirnov, C.R. Acad. Sci. Paris S\'er.
    I Math. {\bf 333}, 239 (2001).

\bibitem{Smirnov2009} S. Smirnov, arXiv: 0909.4499v1 [math.PR].

\bibitem{Binder2007} I. Binder, L. Chayes, and H. K. Lei,
    arXiv:1004.4673v1 [math-ph]; arXiv:1004.4676v1 [math-ph].

\bibitem{Dubedat-ReflBM} J. Dubedat, Ann. I. H. Poincar\'e -- PR {\bf 40}, 539 (2004); arXiv:math/0302250v1 [math.PR].

\bibitem{Cho-Fisher} S. Cho and M. P. A. Fisher, Phys. Rev. B {\bf 55}, 1025 (1997).

\bibitem{Werner-Girsanov} W. Werner, Ann. Fac. Sci. Toulouse Math. (6) 13, no. 1, 121 (2004).

\bibitem{Oksendal} B. {\O}ksendal, {\it Stochastic differential equations}, 6th edition, Springer-Verlag, Berlin, 2003.

\bibitem{Klebaner} F. C. Klebaner, {\it Introduction to stohastic calculus with applications}, 2nd edition, Imperial College Press, London, 2005.

\bibitem{Dotsenko-Fateev} Vl. S. Dotsenko and V. A. Fateev, Nucl. Phys. {\bf B240}, 312 (1984).

\bibitem{Dupa-review} B. Duplantier, in {\it Fractal geometry and applications}, Proc. Symp. Pure Math. vol 72, part 2, Providence, RI, American Mathematical Society, 2004; arXiv: math-ph/0303034.

\bibitem{Bauer-Bernard-2003} M. Bauer and D. Bernard, Phys. Lett {\bf B543}, 135 (2002); Commun. Math. Phys. {\bf 239}, 493 (2003).

\bibitem{footnoteNotationBC}
The subscript on $h_{BC}$ is to indicate that this is, as discussed below in this paragraph,
the  conformal weight of a {\it Boundary Condition Changing} boundary operator.

\bibitem{Cardy1992} J. Cardy, J. Phys. A: Math. Gen. {\bf 25}, L201 (1992).

\bibitem{Watts-formula} G. M. T. Watts, J. Phys. A: Math. Gen. {\bf 29}, L363 (1996); J. J. H. Simmons, P. Kleban, and R. M. Ziff, J. Phys. A: Math. Theor. {\bf 40}, F771 (2007).

\bibitem{Saleur-Bauer-1989} H. Saleur and M. Bauer, Nucl. Phys. B {\bf 320}, 591 (1989).

\bibitem{Aizenman:Duplantier:Aharony} M.~{Aizenman}, B.~{Duplantier}, and A.~{Aharony},  Phys. Rev. Lett. {\bf 83}, 1359 (1999).

\bibitem{Virag2003} B. Virag, Probab. Theory Relat. Fields {\bf 127}, 367 (2003).

\bibitem{Makarov-Smirnov} N. Makarov and S. Smirnov, in the {\it Proceedings of the XVIth International Congress on Mathematical Physics}, edited by P. Exner, pp. 362-371, World Scientific, Singapore (2010); arXiv:0909.5377v1 [math-ph].

\bibitem{Bauer-Bernard-Cantini} M. Bauer, D. Bernard, and L. Cantini, J. Stat. Mech. P07037 (2009).

\bibitem{Tutte} W. T. Tutte. {\it Graph Theory}, Cambridge University Press (2001).

\bibitem{Aigner} M. Aigner. {\it A Course in Enumeration}, Springer-Verlag (2007).

\bibitem{Stanley} R. P. Stanley. {\it Enumerative Combinatorics}, Cambridge University Press (2001).

\bibitem{West} D. B. West. {\it Introduction to Graph Theory}, 2nd ed., Prentice Hall (2000).

\bibitem{Wick} R.~F.~Wick, J.~App.~Phys. {\bf 25}, 741 (1956).

\bibitem{Lippmann-Kuhrt} H.~J.~Lippmann and R.~Kuhrt, Z.~Naturforsch. {\bf 13}, 462 (1958).

\bibitem{Rendell-Girvin} R.~W.~Rendell and S.~M.~Girvin, Phys.~Rev.~B {\bf 23}, 6610 (1981).



\end{thebibliography}
\end{document}